\documentclass[11pt, a4paper]{article}

\usepackage{amsmath, amssymb, amsthm, amsxtra}
\usepackage{hyperref}
\usepackage{algorithm}
\usepackage{algpseudocode}
\usepackage{bbm}
\usepackage{comment}
\usepackage{url}
\usepackage{graphicx}
\usepackage{subfig}
\usepackage{anyfontsize}
\usepackage{multirow}
\usepackage{appendix}
\usepackage{chngcntr}
\usepackage{etoolbox}
\usepackage{lipsum}
\usepackage{pdfpages}
\usepackage{natbib}
\usepackage{tikz}
\usepackage{tkz-graph}
\usepackage{booktabs}
\usepackage{float}
\usepackage{wrapfig}
\usepackage{authblk}
\usetikzlibrary{shapes,arrows.meta,positioning}

\usepackage{breakurl}
\def\UrlBreaks{\do\/\do-}

\title{A Study on Monthly Marine Heatwave Forecasts in New Zealand: An Investigation of Imbalanced Regression Loss Functions with Neural Network Models}
\author[1]{Ding Ning\thanks{Correspondence to: \texttt{ding.ning@pg.canterbury.ac.nz}}}
\author[1]{Varvara Vetrova}
\author[2]{S\'ebastien Delaux}
\author[1]{Rachael Tappenden}
\author[3]{Karin R. Bryan}
\author[4]{Yun Sing Koh}
\date{} 

\affil[1]{School of Mathematics and Statistics, University of Canterbury}
\affil[2]{Oceanum}
\affil[3]{School of Environment, University of Auckland}
\affil[4]{School of Computer Science, University of Auckland}

\begin{document}

\maketitle

\begin{abstract}
Marine heatwaves (MHWs) are extreme ocean-temperature events with significant impacts on marine ecosystems and related industries. Accurate forecasts (one to six months ahead) of MHWs would aid in mitigating these impacts. However, forecasting MHWs presents a challenging imbalanced regression task due to the rarity of extreme temperature anomalies in comparison to more frequent moderate conditions. In this study, we examine monthly MHW forecasts for 12 locations around New Zealand. We use a fully-connected neural network and compare standard and specialized regression loss functions, including the mean squared error (MSE), the mean absolute error (MAE), the Huber, the weighted MSE, the focal-R, the balanced MSE, and a proposed scaling-weighted MSE. Results show that (i) short lead times (one month) are considerably more predictable than three- and six-month leads, (ii) models trained with the standard MSE or MAE losses excel at forecasting average conditions but struggle to capture extremes, and (iii) specialized loss functions such as the balanced MSE and our scaling-weighted MSE substantially improve forecasting of MHW and suspected MHW events. These findings underscore the importance of tailored loss functions for imbalanced regression, particularly in forecasting rare but impactful events such as MHWs.
\end{abstract}

\section{Introduction}

This section provides an overview of marine heatwaves (MHWs), their impacts, and the significance of accurate forecasting methods. It sets the stage for exploring numerical and deep learning (DL) approaches to predict MHWs and related phenomena.

\subsection{Marine Heatwaves}\label{subsec:mhw}

MHWs are observed around the world and have strong impacts on marine ecosystems. Such impacts include shifts in species ranges, local extinctions, and can have a follow-on economic impact on seafood industries \citep{hobday2016hierarchical}. The devastating impact on marine ecosystems caused by MHWs can bring irreversible loss of species or foundation habitats \citep{oliver2019projected}, for example, mass coral bleaching and substantial declines in kelp forests and seagrass meadows \citep{holbrook2020keeping}. MHWs also affect aquaculture businesses and area-restricted fisheries because of the change of the distribution of sea life with follow on effects on production \citep{hobday2018framework}, such as in mussel, oyster and salmon farms. Anthropogenic climate change is expected to cause an increase in both the intensity and frequency of MHWs \citep{hobday2016hierarchical}.

MHWs can be characterized using metrics, such as high maximum temperatures \citep{berkelmans2004comparison}, larger than normal or increasing temperature anomalies \citep{smale2013extreme}, and long degree heating days \citep{maynard2008reeftemp}. Even with the same metrics, different datasets might provide substantially different MHW information \citep{hobday2016hierarchical}. For example, MHWs from daily data show shorter and more intensive events than from monthly data \citep{hobday2016hierarchical}, and MHWs vary by location and season. In some related work \citep{ham2019deep,pravallika2022prediction,taylor2022deep}, the departures from the average sea surface temperature (SST) over a certain period were called ``temperature anomalies'' or ``temperature variability'', and sometimes anomalies were called ``outliers''. In this study, we define sea surface temperature anomalies (SSTAs) as the departures from the average sea surface temperature over a certain period. Spatially, SSTs or SSTAs at a specific location, as a pixel in gridded data, represent the average SSTs or SSTAs within the area that pixel covers. Referring to \citet{hobday2016hierarchical} that used the 90\textsuperscript{th} percentile to define MHWs, we define monthly MHWs as events when the SSTAs greater than the 90\textsuperscript{th} percentile of the SSTAs based on a monthly climatology. For example, a heatwave in January would be defined as an average SSTA for January that is greater than 90\% of previous January SSTAs.

\subsection{Numerical MHW Forecasts}

The ability to accurately predict MHWs, six months in advance for instance, would allow mitigation of their potential impact, for example by collecting healthy samples for repopulation of impacted ecosystems, DNA sampling, and adjusting production beforehand. Dynamical prediction systems have been widely used for MHW forecasts \citep{merryfield2013canadian,saha2014ncep,vecchi2014seasonal}. For example, a case study of the California Current System MHW of 2014-2016 used eight global coupled climate prediction systems to predict the MHW up to eight months ahead \citep{jacox2019predicting}. In this case, two of the four phases were predicted well by dynamic models but two others were missed. Moreover, numerical dynamic models are resource intensive, making it difficult to run enough simulations to perform probabilistic projections. At the same time, there have been more and more observations and completed hindcasts, so statistical explorations of the existing data to make forecasts might be a useful alternative to dynamic modeling. In the last ten years, statistical forecasts have become increasingly sophisticated, with innovations in machine learning (ML) methods, especially DL, being at the forefront.

\subsection{Deep Learning for SST, SSTA, and MHW Forecasts}

A number of ML techniques have been applied to predicting SSTAs and/or SSTA-related climate events. We focus on recent DL approaches to this task. Early work started with convolutional neural networks (CNNs), which have been widely used for spatiotemporal representation learning, and adopted some image and video learning techniques to environmental spatiotemporal data. \citet{ham2019deep,ham2021unified} used a CNN with SSTAs as one of the inputs to forecast the seasonal Nino3.4 SSTA index up to 18 months ahead, which is a climate index based on the SST field over the tropical eastern Pacific that characterizes the El Ni\~{n}o-Southern Oscillation (ENSO) in the equatorial Pacific Ocean. Another distinctive aspect of their work was the application of transfer learning to improve prediction, by leveraging two supplementary datasets. Then \citet{cachay2021world} used a graph neural network (GNN) to tackle the same ENSO forecasting task proposed by \citet{ham2019deep} and outperformed the CNNs for one to six lead month forecasts. For other neural network classes for SSTA-related event forecasts, \citet{ratnam2020machine} proposed a fully-connected neural network (FCN) to forecast SSTAs over the Indian Ocean, which are the indicators of the Indian Ocean Dipole (IOD). Also, the IOD forecasts have been made using an LSTM \citep{pravallika2022prediction} and a CNN \citep{fengpredictability}. Moreover, a CNN was developed to forecast SSTs and MHWs around Australia \citep{boschetti2022sea}. Later work started to address the combination of multiple neural network classes for SST, SSTA, or SSTA-related climate event forecasts. \citet{taylor2022deep} combined the U-Net \citep{ronneberger2015u} with an LSTM to forecast SSTs up to 24 months ahead and validated the forecasts with a focus on specific SST variability-related oscillations (ENSO and IOD) and climate events (the Blob MHW). The U-Net-LSTM used SSTs as one of the inputs to predict SSTs, which then were converted into the Nino3.4 and Nino4 indices and the Blob index. The Blob index is the difference between the monthly average SST climatology (1980-2010) and the monthly average SST computed over the oceanic area off the west coast of the United States, as an indicator of MHWs. In this DL framework, to forecast SSTs, the spatial dependencies and the dependencies between variables were captured by the convolutional autoencoder, the short-term temporal dependencies were captured by the sliding window method, and the long-term temporal dependencies were captured by the LSTM. \citet{taylor2022deep} also mentioned the possible deficiency of transfer learning, which is insufficient to correct the inherent biases in the coupled models \citep{timmermann2018nino}, and highlighted the feasibility of knowledge-based physics-informed ML \citep{karniadakis2021physics}.

In terms of ML for SSTA-related oscillation or climate event forecasts, the majority of the attempts to predict SSTs or SSTAs have focused on reproducing the average conditions but not the extremes, such as the ENSO \citep{ham2019deep,taylor2022deep,boschetti2022sea} and IOD \citep{wu2019seasonal,ratnam2020machine,fengpredictability} forecasts. Nevertheless, the extremes, like MHWs, are much more damaging to ecosystems and productivity, and their forecasts are useful for the climate domain. For example, \citet{ham2019deep,pravallika2022prediction,taylor2022deep} generally formulated the ML tasks for SST or SSTA-related event forecasts as a standard regression or classification task (using a standard loss function) and focused on reproducing the average conditions. However, forecasting SSTA anomalies or extremes calls for a different approach. A label, which is a continuous quantity or a category that a ML model aims to predict, is typically assumed to have a uniform distribution in most ML algorithms. However, label imbalance is common in real-world ML tasks, such as fraud detection, social media spam detection, disease diagnosis, and natural disaster forecasts, where the labels of interest are rare or underrepresented when compared to the other labels. In our proposed ML framework, we view the forecasting tasks as imbalanced regression (for imbalanced labels) to be able to explore the techniques to handle anomalies. We started with imbalanced regression because SSTAs are a continuous variable, and treating different continuous labels as distinct classes is unlikely to yield optimal results because it does not take advantage of the similarity between nearby continuous labels \citep{yang2021delving}. However, \citet{berg2021deep} pointed out that for deep ordinal regression, training a classifier, particularly when utilizing their proposed label devising approach, could outperform training a regressor in terms of accuracy. It is important to note that their approach was not specifically designed for label imbalance tasks. Therefore, in this study, we still place emphasis on imbalanced regression for SSTA and MHW forecasts.

Approaches to tackle imbalanced regression were generally extended from those for imbalanced classification, which can be categorized into two groups: the data-based and the model-based. Data-based approaches to imbalanced classification include oversampling the minority class, undersampling the majority class, and combining both \citep{chawla2002smote,garcia2009evolutionary,he2008adasyn}. These data re-sampling techniques were later introduced to imbalanced regression \citep{torgo2013smote,branco2017smogn,yang2021delving}. Model-based approaches include re-weighting or adjusting the loss function to compensate for class imbalance \citep{cao2019learning,cui2019class,dong2018imbalanced,huang2016learning,huang2019deep}, and applying relevant auxiliary techniques, such as transfer learning \citep{yin2019feature}, metric learning \citep{zhang2017mixup}, meta-learning \citep{shu2019meta}, two-stage training \citep{kang2019decoupling}, and semi-supervised learning and self-supervised learning \citep{yang2020rethinking}. The loss function approaches were later introduced to imbalanced regression \citep{yang2021delving,ren2022balanced}.

In all, we have defined SSTAs and MHWs for this study. We reviewed the numerical approaches and statistical DL approaches for SSTA-related oscillation and MHW forecasts, including different neural network classes with highlighted auxiliary techniques that tackled the forecasts. We then framed the task of forecasting MHWs in New Zealand as an imbalanced regression problem. Subsequently, we reviewed the approaches adapted from tackling imbalanced classification, which were specifically developed for imbalanced regression. In the next subsection, we will present the research questions of this study.

\subsection{Research Questions}

Having conducted a literature review on the DL approaches for SSTA and MHW forecasts, as well as the strategies for tackling imbalanced regression, we are now equipped to establish our research objectives. Alongside this review, we have also defined the essential concepts in this study, including SSTAs, MHWs, and the framed imbalanced regression task for forecasting monthly MHWs in New Zealand. Considering these elements, we propose a two-fold aim for this study.

First, this study aims to investigate the appropriate loss functions for a specific imbalanced regression task: monthly MHW forecasts in New Zealand. A loss function compares observed (target) values with predicted values and measures how well ML models the training data, and the choice of a loss function is a crucial factor that influences the performance of a ML model \citep{nie2018investigation}. A loss function for imbalanced regression accommodates the imbalanced training label distribution and leads to a better predictive performance on rare labels \citep{ren2022balanced}, which are the predicted values of the anomalies or extremes in this case study of MHWs. Related work can be divided into two categories: one is customizing a loss function for specific case studies, such as wind speed forecasts \citep{chen2020new}, COVID-19 diagnosis \citep{khakzar2021towards}, and speech quality evaluation \citep{martin2018deep}. The other is proposing a general loss function, such as the focal-R loss \citep{yang2021delving} and the balanced MSE loss \citep{ren2022balanced}. We synthesized the two approaches to propose a custom scaling-weighted loss function and compared it with the other commonly used regression loss functions for monthly MHW forecasts in New Zealand.

Second, this study aims to identify the degree to which the monthly MHWs at 12 selected New Zealand locations can be predicted by ML models based on regional data, for one, two, three, and six months in advance, especially using the loss functions investigated in the first aim. The formulated imbalanced regression task for forecasting MHWs, which focuses on extreme value or outlier prediction, is different from the balanced regression task for forecasting SSTs or SSTAs. We evaluated the predictions for extreme values and compared the ML models with the persistence models spatially and temporally. The second aim will provide a better understanding of the MHW's predictability with appropriate predictors and models.

\section{Data}

This section outlines the dataset used in this study, detailing the preprocessing steps and the rationale behind the selection of specific data points. The goal is to ensure the data is suitable for developing and testing the proposed ML models for forecasting SSTAs and MHWs.

\subsection{Dataset}

The dataset we used in this study is the Simple Ocean Data Assimilation (SODA) \citep{carton2008reanalysis}, Version 2.2.4. It is global numerical modeling data available from January 1871 to December 2010, with a spatial resolution of $0.5\times0.5$ degrees. SODA is an ocean reanalysis dataset that describes the evolution of a number of ocean variables, including the ocean temperature, over a four dimensional grid (latitude, longitude, depth, and time). We used the version 2.2.4 of the dataset as it extends over a longer period and thus provides more training data than other more recent versions of the reanalysis. MHWs are found at any depth in the water column, but there has been a trend of observing MHWs at the surface because sea temperature measurements at the surface are much more abundant as only the temperature at the surface can be measured by satellites. Thus the MHWs in our research is focused on surface MHWs, although in the future for better prediction it could be useful to explore the use of temperatures at other depths (often in the shape of heat content) and the use of other variables. We considered the temperature in the grid cells that are the nearest to the surface (depth of 5 meters) to represent the sea surface temperature (SST).

SODA 2.2.4 is available on the website of the Asia-Pacific Data-Research Center (APDRC) \citep{apdrcdataset}, a part of the International Pacific Research Center at the University of Hawai'i at Mānoa. The extracted multidimensional grid was of three dimensions (the longitude, 720 degrees from -179.75 to 179.75, the latitude, 330 degrees from -75.25 to 89.25, and the time, 1680 months from January 1871 to December 2010) and one channel (the temperature). We mainly used this SST grid for analysis and experiments.

\subsection{Data Preprocessing}

\begin{figure}[H]
\centering
\includegraphics[scale=0.3]{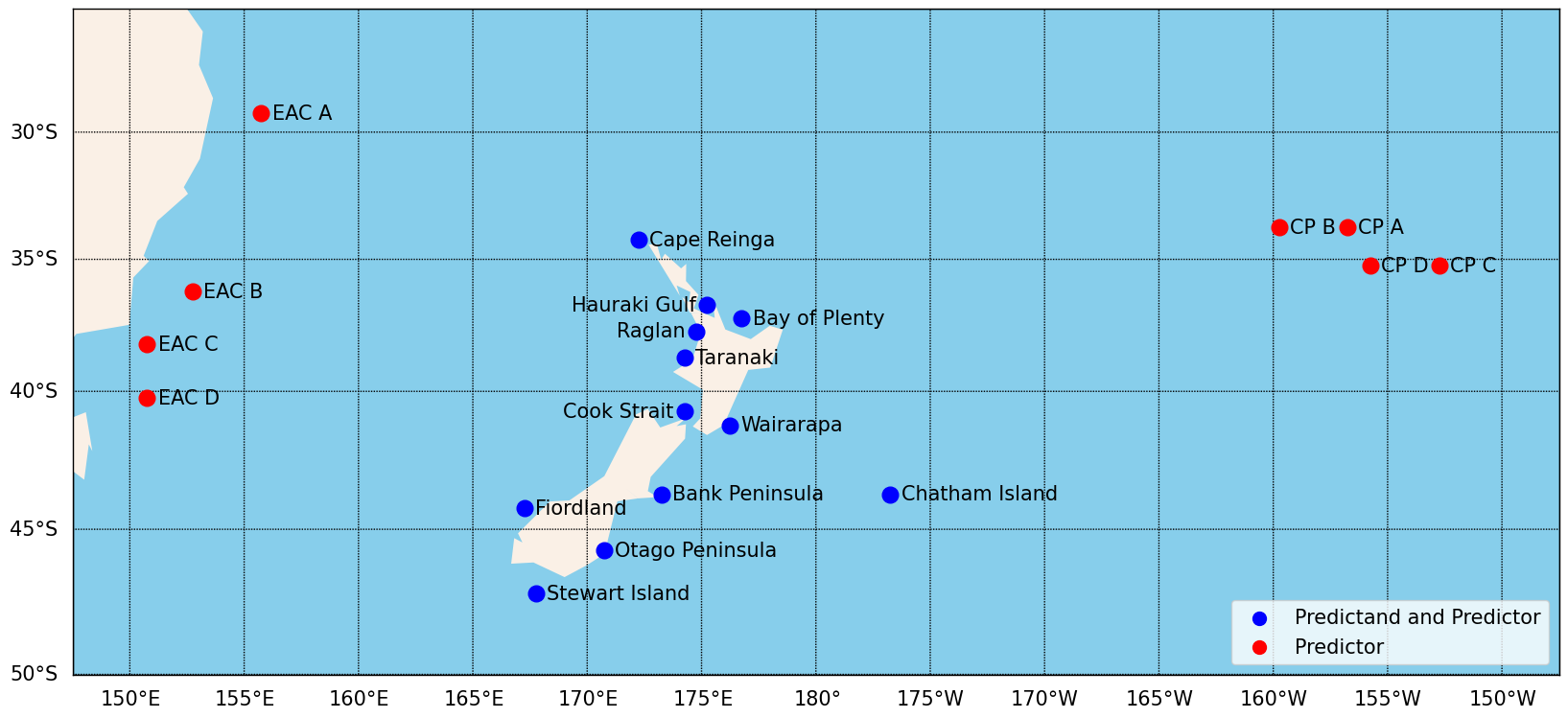}
\caption{20 selected locations of interest for MHW forecasts in New Zealand. The SSTA time series of the locations in blue are both predictands and predictors. The SSTA time series of the locations in red are predictors only.}\label{fg:locs}
\end{figure}

We preprocessed the extracted SST grid into separate sequences that are SST time series at 20 different ocean locations. The 12 locations in New Zealand were referred to the locations of interest for MHW forecasts in the Moana Project \citep{moanaproject}, which uses numerical ocean models to forecast SSTs and MHWs up to seven days in advance. The aim of our research is to extend the MHW forecasts from a daily to a seasonal basis at these selected locations. The 12 locations, which are close to major human settlements and cover most of New Zealand's surrounding oceanic waters, provide MHW forecasts with greater socioeconomic value and representativeness. We also selected eight locations in the two areas where the movement of seawater potentially has an impact on the SSTs in New Zealand. The four locations are in the East Australian Current \citep{sutton2019ocean,elzahaby2021oceanic} and the other four locations are in an area in the Central Pacific, east of North Cape \citep{sutton2019ocean}. The SSTs in these two areas have been demonstrated to be highly correlated with the SSTs in New Zealand. Figure \ref{fg:locs} shows the 20 locations, including 12 New Zealand locations and eight non-New Zealand locations. Tables \ref{tb:label} and \ref{tb:otherfeat} summarize the information of the time series of 20 locations. In addition, it is noticeable that the 20 sequences contain no missing values, because SODA 2.2.4 is modeling data.

\begin{table}[H]
\centering
\scriptsize
\caption{Information and statistics of the 12 label and feature time series of the locations in New Zealand. The test for normal distribution is the Shapiro-Wilk Test \citep{shapiro1965analysis} at a significance level of 0.05, which was applied to each of the 12 complete time series that contains 1680 data points. ``Abbr.'' is the abbreviation and ``Std.'' is the standard deviation, applied to the other tables in this study.}\label{tb:label}
\begin{tabular}{llllll}
\multicolumn{1}{c}{\textbf{Location}} & \multicolumn{1}{c}{\textbf{Abbr.}} & \multicolumn{1}{c}{\textbf{Coordinates}} & \multicolumn{1}{c}{\textbf{Mean}} & \multicolumn{1}{c}{\textbf{Std.}} & \multicolumn{1}{c}{\textbf{Is normally distributed}} \\ \hline
Bay of Plenty                         & BOP                                & 37.25°S, 176.75°E                       & 0.0539                            & 0.8376                            & Yes                                                  \\
Bank Peninsula                        & BP                                 & 43.75°S, 173.25°E                       & -0.0291                           & 0.8573                            & No                                                   \\
Chatham Island                        & CI                                 & 43.75°S, 176.75°W                       & -0.0170                           & 0.9278                            & No                                                   \\
Cape Reinga                           & CR                                 & 34.25°S, 172.25°E                       & 0.0652                            & 0.8237                            & No                                                   \\
Cook Strait                           & CS                                 & 40.75°S, 174.25°E                       & 0.0286                            & 0.6657                            & Yes                                                  \\
Fiordland                             & F                                  & 44.25°S, 167.25°E                       & 0.0519                            & 0.9631                            & No                                                   \\
Hauraki Gulf                          & HG                                 & 36.75°S, 175.25°E                       & 0.0490                            & 0.8536                            & Yes                                                  \\
Otago Peninsula                       & OP                                 & 45.75°S, 170.75°E                       & -0.0091                           & 0.7822                            & No                                                   \\
Raglan                                & R                                  & 37.75°S, 174.75°E                       & 0.0792                            & 0.8447                            & No                                                   \\
Stewart Island                        & SI                                 & 47.25°S, 167.75°E                       & 0.0053                            & 0.8000                            & No                                                   \\
Taranaki                              & T                                  & 38.75°S, 174.25°E                       & 0.1074                            & 0.9864                            & Yes                                                  \\
Wairarapa                             & W                                  & 41.25°S, 176.25°E                       & 0.1052                            & 1.0462                            & No
\end{tabular}
\end{table}

\begin{table}[H]
\centering
\footnotesize
\caption{Information of the other eight feature time series of the locations outside New Zealand.}\label{tb:otherfeat}
\begin{tabular}{lll}
\multicolumn{1}{c}{\textbf{Location}}                & \multicolumn{1}{c}{\textbf{Abbr.}} & \multicolumn{1}{c}{\textbf{Coordinates}} \\ \hline
Location A in the East Australian Current            & EAC A                              & 29.25°S, 155.75°E                        \\
Location B in the East Australian Current            & EAC B                              & 36.25°S, 152.75°E                        \\
Location C in the East Australian Current            & EAC C                              & 38.25°S, 150.75°E                        \\
Location D in the East Australian Current            & EAC D                              & 40.25°S, 150.75°E                        \\
Location A in the Central Pacific east of North Cape & CP A                               & 33.75°S, 156.75°W                        \\
Location B in the Central Pacific east of North Cape & CP B                               & 33.75°S, 159.75°W                        \\
Location C in the Central Pacific east of North Cape & CP C                               & 35.25°S, 152.75°W                        \\
Location D in the Central Pacific east of North Cape & CP D                               & 35.25°S, 155.75°W
\end{tabular}
\end{table}

Each sequence has monthly data points over 140 years, totaling 1680 data points. Therefore, for each month from 1871 to 2010, there are 140 SSTs. To remove seasonality, we split the data temporally into the training (80\%) and test (20\%) sets without shuffling in order to keep temporal dependencies, and used the training set to obtain the mean SSTs by month. Taking January as an example, when we used the 112 years (80\%) SSTs for training, we summed the 112 SSTs and averaged them to get the mean SST for January at each location. This mean SST is a reference value and then was applied to both the training and test sets. We subtracted the mean SST from the 140 January SSTs resulting in 140 SSTAs, where a positive SSTA means the SST was warmer than the reference value and a negative SSTA means the SST was cooler than the reference value. We repeated this step for the other 11 months providing in total 1680 SSTAs for each sequence.

The 12 SSTA time series of the New Zealand locations are the predictands, the labels, or the output of the ML model. Also, the 12 time series together with the eight time series of the non-New Zealand locations are the predictors, the features, or the input of the model. We applied the Shapiro-Wilk Test \citep{shapiro1965analysis} at a significance level of 0.05 to test whether the labels are normally distributed. The normally distributed labels follow a Gaussian distribution, where the majority of the labels are close to the mean and the rest taper off symmetrically towards both tails. If the labels are not normally distributed, the labels' distribution deviates from a normal distribution and the statistical procedures that assume normality may not be appropriate, so additional techniques are required. It is worth noting that more outliers are likely to result in a non-normal distribution. Table \ref{tb:label} includes the basic statistics of the 12 label time series. Appendix \ref{apd:hist} provides the histograms of the 12 time series.

The preprocessed SSTA sequences are stored in the NumPy array files and are available for download. The hyperlink is in Appendix \ref{apd:data}.

\section{Methods}

This section describes the various methods and techniques employed in this study to develop and evaluate the proposed ML models for forecasting SSTAs and MHWs. The following subsections provide detailed explanations of the forecasting task, neural network architecture, loss functions, data re-sampling techniques, evaluation metrics, and model configurations.

\subsection{Forecasting Task}

The forecasting task is an imbalanced regression task performed by ML models and auxiliary computational techniques. The input consists of the SSTA time series of 12 locations in New Zealand and eight locations outside New Zealand. The output is the SSTA time series of the 12 New Zealand locations. The forecasting is conducted for lead times of one, two, three, and six months, resulting in output SSTA time series with one, two, three, and six lags respectively. In accordance with the definition of MHWs in Subsection \ref{subsec:mhw}, the forecasted SSTAs are further classified as MHWs or non-MHWs. The following subsections provide a detailed explanation of the methods used in this analysis.

\subsection{Neural Network Class}

The neural network class we used for this study is fully-connected neural networks (FCNs), a basic form of neural network architectures. Because this study investigates the loss functions for a specific imbalanced regression task, we aimed to focus on experimenting with relevant loss functions and exploring the predictability of monthly MHWs in New Zealand. FCNs can perform various common ML tasks, including regression, classification, detection, recommendation, etc., and FCNs can efficiently learn non-linear hidden representations from a small number of data in sequences, simplifying the hyperparameter experiments. In addition, FCNs have been used for SSTA-related event forecasts \citep{ratnam2020machine}. However, FCNs are slow to train with large amounts of data and data in complex formats, and both conditions of data could be used for improving forecasts, so advanced architectures would be needed.

\subsection{Loss Functions}

The loss functions we selected for normal regression tasks are the mean squared error (MSE), the mean absolute error (MAE), and the Huber loss \citep{huber1964robust}. The Huber loss is expressed as
\begin{align}
&L(y,\hat{y})= \frac{1}{N}\sum_{i=1}^N l_i,\\
&l_i=
  \begin{cases}
  \frac{1}{2}(\hat{y}_i-y_i)^2;
  & if\;|\hat{y}_i-y_i|<\delta,\\
  \delta(|\hat{y}_i-y_i|-\frac{\delta}{2});
  & otherwise,
\end{cases}
\end{align}
where $N$ is the number of data points, $y$ is the observed values, $\hat{y}$ is the predicted values, and $\delta$ is a controllable scaling hyperparameter for the L1 part.

The first loss function for imbalanced regression is the weighted MSE. In the training process, we assigned a larger weight to the mean square errors for the observations greater than the 90\textsuperscript{th} percentile and another larger weight for the mean square errors for the observations greater than the 80\textsuperscript{th} percentile and smaller than or equal to the 90\textsuperscript{th}. The former weight is larger than the latter weight to separate the importance of the MHW predictions. The observations between the 80\textsuperscript{th} and 90\textsuperscript{th} percentiles can be considered as suspected MHWs or MHW warnings. This weighted MSE is expressed as

\begin{align}
&L(y,\hat{y})=\frac{1}{N}\sum_{i=1}^N l_i,\\
&l_i=\begin{cases}
w_{90\%}(\hat{y}_i-y_i)^2;
& y_i>y_{90\%},\\
w_{80\%}(\hat{y}_i-y_i)^2;
& y_{80\%}<y_i\leq y_{90\%},\\
(\hat{y}_i-y_i)^2;
& otherwise,
\end{cases}
\end{align}
where $y_{90\%}$ and $y_{80\%}$ are the 90\textsuperscript{th} and 80\textsuperscript{th} percentiles of $y$, and $w_{90\%}$ and $w_{80\%}$ ($w_{90\%}>w_{80\%}$) are the controllable weight hyperparameters.

In addition to the weighted MSE, we selected the focal-R \citep{yang2021delving} as the second loss function and the balanced MSE \citep{ren2022balanced} as the third loss function for imbalanced regression. The two functions have been explicitly defined in the corresponding literature and are therefore not elaborated upon in this study.

In addition, we proposed a custom scaling-weighted MSE loss function that was designed for this specific imbalanced regression task, based on the distributions of the 12 SSTA time series. The loss function is defined as

\begin{align}
&L(y,\hat{y})=\frac{1}{N}\sum_{i=1}^N l_i,\\
&l_i=\begin{cases}
(w_{90\%}\cdot |y_i|^{\alpha})^{\beta}(\hat{y}_i-y_i)^2;
& y_i>y_{90\%},\\
(w_{80\%}\cdot |y_i|^{\alpha})^{\beta}(\hat{y}_i-y_i)^2;
& y_{80\%}<y_i\leq y_{90\%},\\
(|y_i|^{\alpha})^{\beta}(\hat{y}_i-y_i)^2;
& otherwise,
\end{cases}
\end{align}

where $(w\cdot |y|^{\alpha})^{\beta}$ is a scaling weight term. $\alpha$, $\beta$, $w_{90\%}$, and $w_{80\%}$ are the controllable hyperparameters, where $\alpha\in[1,+\infty]$, $\beta\in(0,1]$, $w_{90\%}\in[1,+\infty]$, $w_{80\%}\in[1,+\infty]$, and $w_{90\%}>w_{80\%}$ are suggested. In this loss function, the weight scales with the observed values $y$ and $\alpha$ controls the degree of scaling. The absolute value of $y$ ensures the term to be a real number. Similar to the weighted MSE, an additional weight $w$ is multiplied to separate the errors between MHW observations and predictions. Furthermore, $\beta$ is added to the scaling weight to control the overestimation of extreme values (the Type I error) that leads to more false positive predictions in the classification context. When $\alpha\cdot\beta=1$, the term scales linearly to $y$ and the computational efficiency increases.

It is possible to simplify the scaling weight term $(w\cdot |y|^{\alpha})^{\beta}$ to $(w'\cdot |y|)^{\gamma}$ with one hyperparameter $\gamma$, where $w'=w^{-\alpha}$ and $\gamma=\alpha\cdot\beta$. However, we kept the two hyperparameters for easy interpretation and convenient control of model performance. In the initial experiments, the weight $w$ could directly use the weight value from the weighted MSE loss. $\beta$ could be set as 1 initially, and depending on performance, $\beta$ could be adjusted to a smaller value to control the number of Type I errors.

The custom scaling-weighted MSE has advantages over the standard weighted MSE by dynamically scaling with anomaly severity via incorporating anomaly magnitude, rather than applying the same values of weights across all cases. This approach allows for better representation of extreme values. Additionally, the controllable hyperparameters offer flexibility to balance the trade-off between false alarms and undetected positives, making the loss function well-suited to MHW prediction. For example, reducing $\alpha$ or $\beta$ lowers the influence of extreme values, leading to more conservative predictions and fewer false positives.

\subsection{Data Re-Sampling}

Besides loss functions, we applied one data re-sampling technique for imbalanced regression: the SMOGN \citep{branco2017smogn}. There is one controllable hyperparameter $k$, the number of nearest neighbors. However, because the dataset has no missing values and the time series are relatively evenly distributed, these techniques designed for situations with substantial missing data in specific ranges may not improve the forecasts in this study.

\subsection{Evaluation Metrics}

We used the MSE to evaluate the overall prediction skill of the predicted SSTA time series.

To evaluate the predictions for MHWs, we used the critical success index (CSI) \citep{schaefer1990critical}. The CSI is also called the threat score, which is equal to the total number of true positive predictions for extremes divided by the sum of the predictions for extremes plus the number of misses. The examples of using the CSI as an evaluation includes rainfall prediction in weather forecasts \citep{pham2020development} and disease diagnosis \citep{larner2021assessing}, where rainfalls and diseases were considered as critical or extreme events. We first used neural network models to predict SSTAs as a regression task. Then we categorized the predicted values by regression into three classes: the MHW ($\hat{y}_i>y_{90\%}$), the suspected MHW ($y_{80\%}<\hat{y}_i\leq y_{90\%}$), and the normal or not a MHW ($\hat{y}_i\leq y_{80\%}$). We used the CSI as the evaluation metrics for the predictions for the first two classes, denoted as ``CSI'' and ``CSI 80'' respectively.

\begin{figure}[H]
\centering
\subfloat{\includegraphics[scale=0.45]{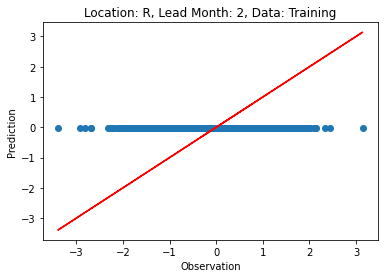}}
\subfloat{\includegraphics[scale=0.45]{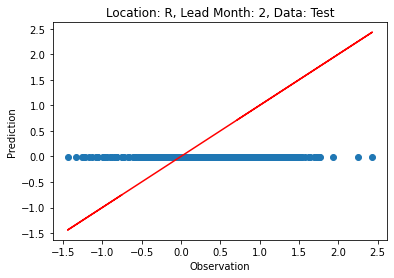}}
\caption{SSTA observation and prediction scatter plots of a model that shows perfect underfitting (PU) on the training and test data respectively, which indicates no ability to predict.}\label{fg:pu}
\end{figure}

The model's training time is another metric of interest to demonstrate the different time efficiency of the loss functions, given the other conditions being the same. Also, we define perfect underfitting (PU) as the event that a model predicts only one value so the model has no ability to predict. We introduce this definition based on the experiments, where we noticed that some models with specific loss functions exhibited the PU phenomenon when performing longer lead forecasts. Therefore, as we trained multiple models using each configuration, we also recorded the ratio of the number of trained models with PU to all models per configuration, denoted as ``PUR". The scatter plots of one example model with PU are shown in Figure \ref{fg:pu}.

In addition, using the MSE and the CSI to compare the performance of models with different loss functions (different regressors) is only appropriate for evaluating on the single dataset. Although the input features are the same, the labels are different, which are time series of different locations. Therefore, the datasets are slightly different and a statistical test that assesses multiple models over multiple datasets is required. We used the non-parametric Friedman Test \citep{friedman1937use,friedman1940comparison} to compare the models and recommend the loss functions for SSTA forecasts, MHW forecasts, and suspected MHW forecasts respectively.

Let $r_i^j$ be the rank of $j$\textsuperscript{th} of $k$ models in terms of an evaluation metric on the $i$\textsuperscript{th} of $N$ datasets (locations). $R_j=\frac{1}{N}\sum_i r_i^j$ is the average ranks of models. The null-hypothesis states that all the models are equivalent so their ranks $R_j$ should be equal. The Friedman Statistic is defined as
\begin{align}
\chi_F^2=\frac{12N}{k(k+1)}[\sum_j R_j^2-\frac{k(k+1)^2}{4}].
\end{align}

\citet{iman1980approximations} later showed that the Friedman Statistic was undesirably conservative. Therefore, a better statistic is defined as
\begin{align}
F_F=\frac{(N-1)\chi_F^2}{N(k-1)-\chi_F^2},
\end{align}
which is distributed according to the F-distribution with $k-1$ and $(k-1)(N-1)$ degrees of freedom. \citet{demvsar2006statistical} introduced the use of the Friedman Test to compare learning algorithms. In this study, we used $F_F$ to assess the comparisons of the models across all locations with regards to SSTA, MHW, and suspected MHW forecasts.

\subsection{Sliding Window}

The sliding window method uses a subset of data points in the time series as inputs rather than all previous data points, and has been used in SST-related forecasts \citep{ham2019deep,taylor2022deep}. The corresponding hyperparameter, window size, determines the number of previous time steps as predictors. The SSTA in the selected sequences have short-term temporal dependencies and we suggests a window size of 3 or larger (three months or more) when using the same sequence as both predictors and preditands. For multiple predictors at monthly time steps, \citet{ham2019deep} used a window size of 3 and \citet{taylor2022deep} used 12 (one year). The window size is a controllable hyperparameter and the values between 3 and 12 were trialed.

\subsection{Model Configurations}\label{subsec:modelconfig}

In ML, a base model refers to the simplest form of a ML model. Because this studies primarily focuses on investigating regression loss functions, we consider the FCN model with the MSE loss as the base model. Therefore, the configuration for the base FCN model: there are one input layer that takes 120 features ($20\times6$, the number of predictors multiplied by the window size) as input, one output layer that outputs one single value, two hidden layers with 100 neurons next to the input layer, and one hidden layer with 50 neurons next to the output layer. The loss function is the MSE. The L1 regularization with a weight of 0.01 is incorporated into the loss function. All activations functions are the hyperbolic tangent. The optimizer is the stochastic gradient descent, with a learning rate of 0.01 and a weight decay of 0.01. The batch size is 64. The forecasting lead times include one, two, three, and six months. The ratio of training data to test data is 4:1. The number of training epochs is 400. We trained five models for the same configuration and we recorded the test MSE, test CSI, and the test CSI 80 of the average predictions by the five models and the average training time. We repeated the process for all 12 locations and obtained a final average result for each metric for all locations.

In designing the FCN architecture, we aimed to balance model complexity with generalization, ensuring it could capture patterns in the SSTA time series data. Preliminary experiments showed that the three-layer structure was sufficient to learn non-linear representations while avoiding overfitting. By incrementally reducing the number of neurons, the model was guided toward abstract features in the third layer, enhancing generalization. This study, however, focuses on evaluating various loss functions rather than refining the architecture. Hence, the chosen architecture was intended to be sufficient to capture non-linear patterns while being computationally efficient, enabling us to train a large number of models quickly for robust experimental comparisons across different loss functions.

In order to investigate the effectiveness of different loss functions, we fixed the other hyperparameters and only modified the loss function from the base model. This process aligns with the goal of ablation experiments, which involve removing or degrading specific components of ML to understand their contribution to the overall system. Because removing the loss function would make neural network training impossible, we replaced the original MSE loss function with the other loss functions to explore their impact. The alternative loss functions for investigation were the MAE, the Huber, the weighted MSE, the focal-R \citep{yang2021delving}, the balanced MSE \citep{ren2022balanced}, and our proposed scaling-weighted MSE. For the hyperparameters associated with some loss functions, in the Huber loss, $\delta=0.5$. In the weighted MSE and the scaling-weighted MSE, $w_{90\%}=1.5$ and $w_{80\%}=1.25$. In the focal-R, the error is the MSE (focal-R-MSE), $\beta=2$, and $\gamma=1$. In the balanced MSE, $\sigma$ was optimized in training. In the scaling-weighted MSE, different sets of $\alpha$ and $\beta$ were attempted. However, there is a potential problem: to reach the best performance, the other hyperparameters should be tuned to suit different loss functions. We will explain this problem with the results in the Discussion. To compare the models across all locations, the Friedman Statistics with regards to the MSE, the CSI, and the CSI 80 were reported.

\section{Results}

This section presents the results of the study, detailing the performance of various models and approaches used for forecasting SSTAs and MHWs. The focus is on evaluating the models based on different metrics to determine their effectiveness in handling the forecasting tasks.

\subsection{One Lead Month Forecasts}\label{subsec:one}

The forecasts started from one month ahead prediction of SSTAs and MHWs, across the 12 selected locations in New Zealand, and the goal was to investigate the effectiveness of different loss functions and different predictions across the selected locations.

We used the MSE, the CSI, and the CSI 80 on the average of five sets of predictions on the test data to evaluate the models' performances. The training time and the PUR of the five trained models further evaluated the performances. We used the Friedman Test to assess these models across all locations.

\subsubsection{Persistence Model}

The persistence model was used for result comparison. The evaluation metrics of the persistence model for each location are in Table \ref{tb:persist}.
\vspace{3mm}

\begin{table}[H]
\centering
\small
\caption{Results of the persistence models for one lead month SSTA and MHW forecasts. A persistence model predicts the next observation the same as the current observation, so the model does not involve a training process.}\label{tb:persist}
\begin{tabular}{llll}
\multicolumn{1}{c}{\textbf{Location}} & \multicolumn{1}{c}{\textbf{MSE↓}} & \multicolumn{1}{c}{\textbf{CSI↑}} & \multicolumn{1}{c}{\textbf{CSI 80↑}} \\ \hline
BOP                         & 0.1690                            & 0.3846                            & 0.4717                               \\
BP                        & 0.4665                            & 0.1875                            & 0.2949                               \\
CI                        & 0.3095                            & 0.2174                            & 0.4211                               \\
CR                           & 0.2468                            & 0.3286                            & 0.4959                               \\
CS                           & 0.1247                            & 0.4043                            & 0.5327                               \\
F                             & 0.3402                            & 0.4699                            & 0.5500                               \\
HG                          & 0.2668                            & 0.4386                            & 0.3750                               \\
OP                       & 0.3155                            & 0.3478                            & 0.4198                               \\
R                                & 0.2377                            & 0.5158                            & 0.6259                               \\
SI                        & 0.2440                            & 0.4545                            & 0.4646                               \\
T                              & 0.3926                            & 0.4595                            & 0.5541                               \\
W                             & 0.4578                            & 0.4607                            & 0.5541                               \\ \hline
Average                               & 0.2976                            & 0.3891                            & 0.4800
\end{tabular}
\end{table}

\subsubsection{Base Model}

The base model was used as a baseline model, whose definition and configurations are in Subsection \ref{subsec:modelconfig}. The evaluation metrics of the base model for each location are in Table \ref{tb:vanilla}. The example observation and prediction scatter plots are in Appendix \ref{apd:scatter}, applied to the models with the other loss functions for one, two, three, and six lead month SSTA and MHW forecasts in this study.
\vspace{3mm}

\begin{table}[H]
\centering
\small
\caption{Results of the base models for one lead month SSTA and MHW forecasts. The results in bold are better than those of the persistence model, applied to the other tables in this study.}\label{tb:vanilla}
\begin{tabular}{llllll}
\multicolumn{1}{c}{\textbf{Location}} & \multicolumn{1}{c}{\textbf{MSE↓}} & \multicolumn{1}{c}{\textbf{CSI↑}} & \multicolumn{1}{c}{\textbf{CSI 80↑}} & \multicolumn{1}{c}{\textbf{Training Time↓}} & \multicolumn{1}{c}{\textbf{PUR↓}} \\ \hline
BOP                         & \textbf{0.1528}                   & 0.0357                            & 0.4211                               & 18.3887                                     &                                   \\
BP                        & \textbf{0.3831}                   & 0.1600                            & 0.2899                               & 18.3435                                     &                                   \\
CI                        & \textbf{0.2592}                   & 0.1765                            & 0.3824                               & 18.6351                                     &                                   \\
CR                           & \textbf{0.2327}                   & 0.1373                            & 0.4224                               & 18.6882                                     &                                   \\
CS                           & \textbf{0.1235}                   & 0.3143                            & 0.4737                               & 18.4263                                     &                                   \\
F                             & \textbf{0.3150}                   & 0.3611                            & \textbf{0.5752}                      & 18.3620                                     & 0\%                               \\
HG                          & \textbf{0.2349}                   & 0.2045                            & \textbf{0.3832}                      & 18.5357                                     &                                   \\
OP                       & \textbf{0.2776}                   & 0.4103                            & \textbf{0.4430}                      & 20.8010                                     &                                   \\
R                                & \textbf{0.2275}                   & 0.2055                            & 0.5859                               & 18.6367                                     &                                   \\
SI                        & \textbf{0.2281}                   & 0.4474                            & 0.4646                               & 18.3489                                     &                                   \\
T                              & 0.3995                            & 0.3735                            & 0.5782                               & 19.4791                                     &                                   \\
W                             & 0.5109                            & 0.2500                            & 0.4789                               & 18.6513                                     &                                   \\ \hline
Average                               & \textbf{0.2787}                   & 0.2563                            & 0.4582                               & 18.7747                                     & 0\%
\end{tabular}
\end{table}

\subsubsection{Models with Different Loss Functions}

To improve the forecasts for MHWs, i.e. improving CSI and CSI 80, we experimented with different loss functions other than the MSE. The evaluation metrics of the models with the MAE loss, the Huber loss, the weighted MSE loss, the focal-R loss, and the balanced MSE loss for each location are respectively in Tables \ref{tb:mae}, \ref{tb:huber}, \ref{tb:wmse}, \ref{tb:focal}, and \ref{tb:bmse}. The evaluation metrics of the models with our proposed scaling-weighted MSE loss with three sets of hyperparameters are respectively in Tables \ref{tb:swmse1}, \ref{tb:swmse2}, and \ref{tb:swmse3}.

\begin{table}[H]
\centering
\small
\caption{Results of the models with the MAE loss for one lead month SSTA and MHW forecasts.}\label{tb:mae}
\begin{tabular}{llllll}
\multicolumn{1}{c}{\textbf{Location}} & \multicolumn{1}{c}{\textbf{MSE↓}} & \multicolumn{1}{c}{\textbf{CSI↑}} & \multicolumn{1}{c}{\textbf{CSI 80↑}} & \multicolumn{1}{c}{\textbf{Training Time↓}} & \multicolumn{1}{c}{\textbf{PUR↓}} \\ \hline
BOP                         & \textbf{0.1551}                   & 0.0357                            & 0.4433                               & 20.0169                                     & 0\%                               \\
BP                        & \textbf{0.3914}                   & 0.0                               & 0.0                                  & 20.4129                                     & 40\%                              \\
CI                        & \textbf{0.2572}                   & 0.0714                            & 0.4028                               & 20.6211                                     & 0\%                               \\
CR                           & \textbf{0.2335}                   & 0.1346                            & 0.4957                               & 19.8763                                     & 0\%                               \\
CS                           & 0.1315                            & 0.2703                            & 0.5000                               & 20.1543                                     & 0\%                               \\
F                             & \textbf{0.3202}                   & 0.2899                            & \textbf{0.5714}                      & 20.0417                                     & 0\%                               \\
HG                          & \textbf{0.2415}                   & 0.0476                            & \textbf{0.3982}                      & 20.4639                                     & 0\%                               \\
\end{tabular}
\end{table}

\begin{table}[H]
\centering
\small
\caption{Results of the models with the Huber loss with $\delta=0.5$ for one lead month SSTA and MHW forecasts.}\label{tb:huber}
\begin{tabular}{llllll}
\multicolumn{1}{c}{\textbf{Location}} & \multicolumn{1}{c}{\textbf{MSE↓}} & \multicolumn{1}{c}{\textbf{CSI↑}} & \multicolumn{1}{c}{\textbf{CSI 80↑}} & \multicolumn{1}{c}{\textbf{Training Time↓}} & \multicolumn{1}{c}{\textbf{PUR↓}} \\ \hline
BOP                         & 0.2137                            & 0.1333                            & 0.3333                               & 18.1167                                     &                                   \\
BP                        & \textbf{0.4538}                   & 0.0400                            & 0.2424                               & 18.5230                                     &                                   \\
CI                        & 0.3455                            & \textbf{0.3889}                   & 0.3286                               & 18.8040                                     &                                   \\
CR                           & 0.2798                            & 0.1724                            & 0.4407                               & 18.7517                                     &                                   \\
CS                           & 0.1678                            & 0.3400                            & 0.4583                               & 18.8105                                     &                                   \\
F                             & 0.4054                            & 0.2778                            & 0.4057                               & 18.3605                                     & 0\%                               \\
HG                          & 0.3089                            & 0.2885                            & \textbf{0.3826}                      & 18.3977                                     &                                   \\
\end{tabular}
\end{table}

\begin{table}[H]
\centering
\small
\caption{Results of the models with the weighted MSE loss with $w_{90\%}=1.5$ and $w_{80\%}=1.25$ for one lead month SSTA and MHW forecasts.}\label{tb:wmse}
\begin{tabular}{llllll}
\multicolumn{1}{c}{\textbf{Location}} & \multicolumn{1}{c}{\textbf{MSE↓}} & \multicolumn{1}{c}{\textbf{CSI↑}} & \multicolumn{1}{c}{\textbf{CSI 80↑}} & \multicolumn{1}{c}{\textbf{Training Time↓}} & \multicolumn{1}{c}{\textbf{PUR↓}} \\ \hline
BOP                         & \textbf{0.1527}                   & 0.0357                            & 0.4316                               & 29.4885                                     &                                   \\
BP                        & \textbf{0.3829}                   & 0.1538                            & 0.2899                               & 30.3551                                     &                                   \\
CI                        & \textbf{0.2612}                   & 0.1765                            & 0.3768                               & 29.8527                                     &                                   \\
CR                           & \textbf{0.2333}                   & 0.1373                            & 0.4397                               & 31.4612                                     &                                   \\
CS                           & \textbf{0.1228}                   & 0.3143                            & 0.4912                               & 30.2711                                     &                                   \\
F                             & \textbf{0.3146}                   & 0.3750                            & \textbf{0.5752}                      & 30.0870                                     & 0\%                               \\
HG                          & \textbf{0.2357}                   & 0.2273                            & \textbf{0.3761}                      & 30.4401                                     &                                   \\
\end{tabular}
\end{table}

\begin{table}[H]
\centering
\small
\caption{Results of the models with the focal-R-MSE loss with $\beta=2$ and $\theta=1$ for one lead month SSTA and MHW forecasts.}\label{tb:focal}
\begin{tabular}{llllll}
\multicolumn{1}{c}{\textbf{Location}} & \multicolumn{1}{c}{\textbf{MSE↓}} & \multicolumn{1}{c}{\textbf{CSI↑}} & \multicolumn{1}{c}{\textbf{CSI 80↑}} & \multicolumn{1}{c}{\textbf{Training Time↓}} & \multicolumn{1}{c}{\textbf{PUR↓}} \\ \hline
BOP                         & \textbf{0.1506}                   & 0.0714                            & 0.4271                               & 24.7791                                     &                                   \\
BP                        & \textbf{0.3848}                   & 0.1600                            & \textbf{0.3000}                      & 25.6488                                     &                                   \\
CI                        & \textbf{0.2587}                   & 0.1765                            & 0.3824                               & 26.4069                                     &                                   \\
CR                           & \textbf{0.2317}                   & 0.0980                            & 0.4174                               & 25.7978                                     &                                   \\
CS                           & \textbf{0.1243}                   & 0.2286                            & 0.4821                               & 25.4496                                     &                                   \\
F                             & \textbf{0.3132}                   & 0.3611                            & \textbf{0.5752}                      & 26.4808                                     & 0\%                               \\
HG                          & \textbf{0.2348}                   & 0.2273                            & \textbf{0.3832}                      & 25.1393                                     &                                   \\
\end{tabular}
\end{table}

\begin{table}[H]
\centering
\small
\caption{Results of the models with the balanced MSE loss with optimized $\sigma$ for one lead month SSTA and MHW forecasts.}\label{tb:bmse}
\begin{tabular}{llllll}
\multicolumn{1}{c}{\textbf{Location}} & \multicolumn{1}{c}{\textbf{MSE↓}} & \multicolumn{1}{c}{\textbf{CSI↑}} & \multicolumn{1}{c}{\textbf{CSI 80↑}} & \multicolumn{1}{c}{\textbf{Training Time↓}} & \multicolumn{1}{c}{\textbf{PUR↓}} \\ \hline
BOP                         & 0.2587                            & 0.2985                            & \textbf{0.4779}                      & 25.9831                                     &                                   \\
BP                        & 0.6484                            & 0.1765                            & \textbf{0.3431}                      & 25.2746                                     &                                   \\
CI                        & 0.3446                            & 0.1944                            & \textbf{0.4524}                      & 24.7637                                     &                                   \\
CR                           & 0.3636                            & \textbf{0.3864}                   & \textbf{0.5102}                      & 26.1010                                     &                                   \\
CS                           & 0.1443                            & 0.3860                            & 0.4692                               & 26.8915                                     &                                   \\
F                             & 0.4006                            & \textbf{0.4949}                   & \textbf{0.5615}                      & 24.6535                                     & 0\%                               \\
HG                          & 0.3419                            & 0.3247                            & \textbf{0.4242}                      & 24.7438                                     &                                   \\
\end{tabular}
\end{table}

\begin{table}[H]
\centering
\small
\caption{Results of the models with the scaling-weighted MSE loss with $\alpha=1.5$, $\beta=0.5$, $w_{90\%}=1.5$, and $w_{80\%}=1.25$ for one lead month SSTA and MHW forecasts.}\label{tb:swmse1}
\begin{tabular}{llllll}
\multicolumn{1}{c}{\textbf{Location}} & \multicolumn{1}{c}{\textbf{MSE↓}} & \multicolumn{1}{c}{\textbf{CSI↑}} & \multicolumn{1}{c}{\textbf{CSI 80↑}} & \multicolumn{1}{c}{\textbf{Training Time↓}} & \multicolumn{1}{c}{\textbf{PUR↓}} \\ \hline
BOP                         & 0.1832                            & 0.3415                            & \textbf{0.4836}                      & 39.5585                                     &                                   \\
BP                        & 0.4882                            & 0.1667                            & \textbf{0.3407}                      & 40.0110                                     &                                   \\
CI                        & \textbf{0.3012}                   & \textbf{0.2222}                   & 0.4177                               & 39.1581                                     &                                   \\
CR                           & 0.2752                            & 0.2879                            & 0.4545                               & 48.1067                                     &                                   \\
CS                           & 0.1409                            & 0.3818                            & 0.4538                               & 41.6232                                     &                                   \\
F                             & 0.3510                            & 0.4598                            & \textbf{0.5772}                      & 40.1473                                     & 0\%                               \\
HG                          & 0.2781                            & 0.3651                            & \textbf{0.4242}                      & 40.2325                                     &                                   \\
\end{tabular}
\end{table}

\begin{table}[H]
\centering
\small
\caption{Results of the models with the scaling-weighted MSE loss with $\alpha=2$, $\beta=0.5$, $w_{90\%}=1.5$, and $w_{80\%}=1.25$ for one lead month SSTA and MHW forecasts.}\label{tb:swmse2}
\begin{tabular}{llllll}
\multicolumn{1}{c}{\textbf{Location}} & \multicolumn{1}{c}{\textbf{MSE↓}} & \multicolumn{1}{c}{\textbf{CSI↑}} & \multicolumn{1}{c}{\textbf{CSI 80↑}} & \multicolumn{1}{c}{\textbf{Training Time↓}} & \multicolumn{1}{c}{\textbf{PUR↓}} \\ \hline
BOP                         & 0.1989                            & 0.3125                            & \textbf{0.4959}                      & 29.1585                                     &                                   \\
BP                        & 0.4959                            & 0.1429                            & \textbf{0.3483}                      & 30.1651                                     &                                   \\
CI                        & 0.3183                            & \textbf{0.2258}                   & \textbf{0.4250}                      & 29.1210                                     &                                   \\
CR                           & 0.3068                            & 0.3151                            & 0.4694                               & 30.3416                                     &                                   \\
CS                           & 0.1471                            & 0.3559                            & 0.4627                               & 29.2815                                     &                                   \\
F                             & 0.4289                            & \textbf{0.4783}                   & \textbf{0.5952}                      & 30.3365                                     & 0\%                               \\
HG                          & 0.2950                            & 0.3636                            & \textbf{0.4058}                      & 29.8237                                     &                                   \\
\end{tabular}
\end{table}

\begin{table}[H]
\centering
\small
\caption{Results of the models with the scaling-weighted MSE loss with $\alpha=2$, $\beta=1$, $w_{90\%}=1.5$, and $w_{80\%}=1.25$ for one lead month SSTA and MHW forecasts.}\label{tb:swmse3}
\begin{tabular}{llllll}
\multicolumn{1}{c}{\textbf{Location}} & \multicolumn{1}{c}{\textbf{MSE↓}} & \multicolumn{1}{c}{\textbf{CSI↑}} & \multicolumn{1}{c}{\textbf{CSI 80↑}} & \multicolumn{1}{c}{\textbf{Training Time↓}} & \multicolumn{1}{c}{\textbf{PUR↓}} \\ \hline
BOP                         & 0.3517                            & 0.2500                            & 0.4207                               & 23.8322                                     &                                   \\
BP                        & 0.6716                            & \textbf{0.2353}                   & 0.2925                               & 27.7756                                     &                                   \\
CI                        & 0.4365                            & 0.2105                            & \textbf{0.4302}                      & 23.5380                                     &                                   \\
CR                           & 0.5111                            & 0.2743                            & 0.4199                               & 23.9643                                     &                                   \\
CS                           & 0.1807                            & 0.3288                            & 0.4748                               & 23.8340                                     &                                   \\
F                             & 0.5689                            & 0.4526                            & \textbf{0.5615}                      & 23.9980                                     & 0\%                               \\
HG                          & 0.3687                            & 0.3059                            & 0.3716                               & 25.1013                                     &                                   \\
\end{tabular}
\end{table}

\subsubsection{Comparison of Multiple Models Across Locations}

We ranked the models based on the MSE, the CSI, and the CSI 80 respectively across all locations. Tables \ref{tb:rankmse}, \ref{tb:rankcsi80}, and \ref{tb:rankcsi80} summarizes the ranks of all models including the persistence model. The Friedman test \citep{friedman1937use,friedman1940comparison,demvsar2006statistical} was used to test the statistical significance of the comparison based on the ranks. The computed values of $F_F$ \citep{demvsar2006statistical} were 65.8918, 11.2928, and 6.9737 for the ranks of the three metrics respectively. Because the number of datasets (locations) is 12 and the number of models including the persistence model is 10, the value of $F$ in the F-distribution with 9 and 99 degrees of freedom denoted as $F(9,99)$ is 1.9758 at a significance of 0.05. $F_F>F(9,99)$ so the null hypothesis is rejected.

We only assessed the models with the Friedman test across all locations for one lead month. Because the PU phenomenon widely existed in the long lead forecasts and the PUR must be combined with the MSE, the CSI, and the CSI 80 for model evaluation, ranking the models solely based on each of the three metrics is inappropriate.

\begin{table}[H]
\centering
\fontsize{5}{6}\selectfont
\caption{Ranks of the models based on the MSE. The three best average ranks are in bold. The acronym ``WMSE'' is ``weighted MSE'', ``FR'' is ``focal-R'', ``BMSE'' is ``balanced MSE'', and ``SWMSE'' is ``scaling-weighted MSE''. The numbers ``1'', ``2'', and ``3'' behind ``SWMSE'' indicate the hyperparameters $\alpha=1.5$, $\beta=0.5$, $\alpha=2$, $\beta=0.5$, and $\alpha=2$, $\beta=1$ respectively, applied to the other tables in this study.}\label{tb:rankmse}
\begin{tabular}{lllllllllll}
\multicolumn{1}{c}{\textbf{Location}} & \multicolumn{1}{c}{\textbf{Persist}} & \multicolumn{1}{c}{\textbf{MSE}} & \multicolumn{1}{c}{\textbf{MAE}} & \multicolumn{1}{c}{\textbf{Huber}} & \multicolumn{1}{c}{\textbf{WMSE}} & \multicolumn{1}{c}{\textbf{FR}} & \multicolumn{1}{c}{\textbf{BMSE}} & \multicolumn{1}{c}{\textbf{SWMSE1}} & \multicolumn{1}{c}{\textbf{SWMSE2}} & \multicolumn{1}{c}{\textbf{SWMSE3}} \\ \hline
BOP                         & 5                                        & 3                                & 4                                & 8                                  & 2                                 & 1                               & 9                                 & 6                                   & 7                                   & 10                                  \\
BP                        & 6                                        & 3                                & 4                                & 5                                  & 1                                 & 2                               & 9                                 & 7                                   & 8                                   & 10                                  \\
CI                        & 6                                        & 3                                & 1                                & 9                                  & 4                                 & 2                               & 8                                 & 5                                   & 7                                   & 10                                  \\
CR                           & 5                                        & 2                                & 4                                & 7                                  & 3                                 & 1                               & 9                                 & 6                                   & 8                                   & 10                                  \\
CS                           & 4                                        & 2                                & 5                                & 9                                  & 1                                 & 3                               & 7                                 & 6                                   & 8                                   & 10                                  \\
F                             & 5                                        & 3                                & 4                                & 8                                  & 2                                 & 1                               & 7                                 & 6                                   & 9                                   & 10                                  \\
HG                          & 5                                        & 2                                & 4                                & 8                                  & 3                                 & 1                               & 9                                 & 6                                   & 7                                   & 10                                  \\
OP                       & 5                                        & 1.5                              & 3                                & 6                                  & 4                                 & 1.5                             & 9                                 & 7                                   & 8                                   & 10                                  \\
R                                & 5                                        & 3                                & 4                                & 9                                  & 2                                 & 1                               & 8                                 & 6                                   & 7                                   & 10                                  \\
SI                        & 5                                        & 2                                & 4                                & 7                                  & 3                                 & 1                               & 9                                 & 6                                   & 8                                   & 10                                  \\
T                              & 1                                        & 5                                & 2.5                              & 9                                  & 4                                 & 2.5                             & 7                                 & 6                                   & 8                                   & 10                                  \\
W                             & 1                                        & 5                                & 2                                & 9                                  & 4                                 & 6                               & 8                                 & 3                                   & 7                                   & 10                                  \\ \hline
Average                               & 4.4167                                   & \textbf{2.875}                   & 3.4583                           & 7.8333                             & \textbf{2.75}                     & \textbf{1.9167}                 & 8.25                              & 5.8333                              & 7.6667                              & 10
\end{tabular}
\end{table}

\begin{table}[H]
\centering
\fontsize{5}{6}\selectfont
\caption{Ranks of the models based on the CSI.}\label{tb:rankcsi}
\begin{tabular}{lllllllllll}
\multicolumn{1}{c}{\textbf{Location}} & \multicolumn{1}{c}{\textbf{Persist}} & \multicolumn{1}{c}{\textbf{MSE}} & \multicolumn{1}{c}{\textbf{MAE}} & \multicolumn{1}{c}{\textbf{Huber}} & \multicolumn{1}{c}{\textbf{WMSE}} & \multicolumn{1}{c}{\textbf{FR}} & \multicolumn{1}{c}{\textbf{BMSE}} & \multicolumn{1}{c}{\textbf{SWMSE1}} & \multicolumn{1}{c}{\textbf{SWMSE2}} & \multicolumn{1}{c}{\textbf{SWMSE3}} \\ \hline
BOP                         & 1                                        & 9                                & 9                                & 6                                  & 9                                 & 7                               & 4                                 & 2                                   & 3                                   & 5                                   \\
BP                        & 2                                        & 5.5                              & 10                               & 9                                  & 7                                 & 5.5                             & 3                                 & 4                                   & 8                                   & 1                                   \\
CI                        & 4                                        & 8                                & 10                               & 1                                  & 8                                 & 8                               & 6                                 & 3                                   & 2                                   & 5                                   \\
CR                           & 2                                        & 7.5                              & 9                                & 6                                  & 7.5                               & 10                              & 1                                 & 4                                   & 3                                   & 5                                   \\
CS                           & 1                                        & 7.5                              & 9                                & 6                                  & 7.5                               & 10                              & 2                                 & 3                                   & 4                                   & 5                                   \\
F                             & 8                                        & 6.5                              & 9                                & 10                                 & 5                                 & 6.5                             & 1                                 & 3                                   & 2                                   & 4                                   \\
HG                          & 1                                        & 9                                & 10                               & 6                                  & 7.5                               & 7.5                             & 4                                 & 2                                   & 3                                   & 5                                   \\
OP                       & 9                                        & 1.5                              & 7.5                              & 10                                 & 3                                 & 1.5                             & 6                                 & 4                                   & 5                                   & 7.5                                 \\
R                                & 1                                        & 7.5                              & 10                               & 9                                  & 7.5                               & 6                               & 2                                 & 5                                   & 3                                   & 4                                   \\
SI                        & 6                                        & 7.5                              & 9                                & 1                                  & 7.5                               & 5                               & 4                                 & 2                                   & 3                                   & 10                                  \\
T                              & 3                                        & 7.5                              & 10                               & 9                                  & 7.5                               & 6                               & 1                                 & 4                                   & 2                                   & 5                                   \\
W                             & 2                                        & 9.5                              & 6                                & 7                                  & 8                                 & 9.5                             & 1                                 & 3                                   & 4                                   & 5                                   \\ \hline
Average                               & \textbf{3.3333}                          & 7.2083                           & 9.0417                           & 6.6667                             & 7.0833                            & 6.875                           & \textbf{2.9167}                   & \textbf{3.25}                       & 3.5                                 & 5.125
\end{tabular}
\end{table}

\begin{table}[H]
\centering
\fontsize{5}{6}\selectfont
\caption{Ranks of the models based on the CSI 80.}\label{tb:rankcsi80}
\begin{tabular}{lllllllllll}
\multicolumn{1}{c}{\textbf{Location}} & \multicolumn{1}{c}{\textbf{Persist}} & \multicolumn{1}{c}{\textbf{MSE}} & \multicolumn{1}{c}{\textbf{MAE}} & \multicolumn{1}{c}{\textbf{Huber}} & \multicolumn{1}{c}{\textbf{WMSE}} & \multicolumn{1}{c}{\textbf{FR}} & \multicolumn{1}{c}{\textbf{BMSE}} & \multicolumn{1}{c}{\textbf{SWMSE1}} & \multicolumn{1}{c}{\textbf{SWMSE2}} & \multicolumn{1}{c}{\textbf{SWMSE3}} \\ \hline
BOP                         & 4                                        & 8                                & 5                                & 10                                 & 6                                 & 7                               & 3                                 & 2                                   & 1                                   & 9                                   \\
BP                        & 5                                        & 7.5                              & 10                               & 9                                  & 7.5                               & 4                               & 2                                 & 3                                   & 1                                   & 6                                   \\
CI                        & 4                                        & 7.5                              & 6                                & 10                                 & 9                                 & 7.5                             & 1                                 & 5                                   & 3                                   & 2                                   \\
CR                           & 2                                        & 8                                & 3                                & 6                                  & 7                                 & 10                              & 1                                 & 5                                   & 4                                   & 9                                   \\
CS                           & 1                                        & 6                                & 2                                & 9                                  & 3                                 & 4                               & 7                                 & 10                                  & 8                                   & 5                                   \\
F                             & 9                                        & 4                                & 6                                & 10                                 & 4                                 & 4                               & 7.5                               & 2                                   & 1                                   & 7.5                                 \\
HG                          & 9                                        & 5.5                              & 4                                & 7                                  & 8                                 & 5.5                             & 1.5                               & 1.5                                 & 3                                   & 10                                  \\
OP                       & 10                                       & 4                                & 5                                & 6.5                                & 6.5                               & 8                               & 1.5                               & 3                                   & 1.5                                 & 9                                   \\
R                                & 1                                        & 6                                & 4                                & 10                                 & 2                                 & 7                               & 8                                 & 3                                   & 5                                   & 9                                   \\
SI                        & 8                                        & 8                                & 4                                & 10                                 & 5                                 & 8                               & 3                                 & 1                                   & 2                                   & 6                                   \\
T                              & 9                                        & 5                                & 7                                & 10                                 & 7                                 & 7                               & 2                                 & 1                                   & 3                                   & 4                                   \\
W                             & 2                                        & 7                                & 6                                & 10                                 & 9                                 & 8                               & 1                                 & 5                                   & 3                                   & 4                                   \\ \hline
Average                               & 5.3333                                   & 6.375                            & 5.1667                           & 8.9583                             & 6.1667                            & 6.6667                          & \textbf{3.2083}                   & \textbf{3.4583}                     & \textbf{2.9583}                     & 6.7083
\end{tabular}
\end{table}

\subsubsection{Data Re-Sampling}

We applied the SMOGN re-sampling \citep{branco2017smogn} to the input data: the 20 predictor time series. However, because there are no missing values in the time series, after the SMOGN was applied, there were new feature vectors generated and some contained missing values. We used the mean to interpolate the missing values. We tested a range of values for the hyperparameter $k$, from 5 to 200. After inputting the re-sampled data to the base model, the performance deteriorated substantially. Therefore, we did not input the re-sampled data to the other models and we do not further report the results of the SMOGN re-sampling.

\subsection{Longer Lead Time Forecasts}

Extending the forecasting horizon, this subsection examines the performance of models predicting SSTAs and MHWs up to six months ahead. The objective is to evaluate the predictability and performance degradation as the lead time increases.

\subsubsection{Fixed-Lag Model}

The SSTA and MHW forecasts were extended up to six months in advance, across all New Zealand locations. We would like to investigate the predictability for longer lead times.

Referring to the persistence model that uses the observation at the last time step as the prediction at the current target time step, we define the model that uses the observation at the second, third, or sixth last time step as the prediction at the current target time step as a fixed-lag model. The persistence model is a one-lag model. We use two-lag, three-lag, and six-lag models to compare with the FCN models in terms of two, three, and six lead month forecasts.

For the longer lead time forecasts, the number of the trained models for each of two, three, and six lead months is the same as the number of the trained models for one lead month. We only report the average results across all 12 locations due to the length constraints of this study. The detailed results, presented in the same format as Subsection \ref{subsec:one}, can be found in Appendix \ref{apd:add}. The code and results are presented in Jupyter Notebooks and available for download. The hyperlink is in Appendix \ref{apd:code}.

\begin{table}[H]
\centering
\scriptsize
\caption{Average results of the fixed-lag models for two, three, and six lead months SSTA and MHW forecasts.}\label{tb:persistall}
\begin{tabular}{llll}
\multicolumn{1}{c}{\textbf{Across All Locations}} & \multicolumn{1}{c}{\textbf{MSE↓}} & \multicolumn{1}{c}{\textbf{CSI↑}} & \multicolumn{1}{c}{\textbf{CSI 80↑}} \\ \hline
Average (two lead months)                         & 0.6240                            & 0.2204                            & 0.3173                               \\
Average (three lead months)                       & 0.8089                            & 0.1459                            & 0.2602                               \\
Average (six lead months)                         & 1.0178                            & 0.1125                            & 0.1882
\end{tabular}
\end{table}

\subsubsection{Base Model}

\begin{table}[H]
\centering
\scriptsize
\caption{Average results of the base models for two, three, and six lead months SSTA and MHW forecasts.}\label{tb:vanillaall}
\begin{tabular}{llllll}
\multicolumn{1}{c}{\textbf{Across All Locations}} & \multicolumn{1}{c}{\textbf{MSE↓}} & \multicolumn{1}{c}{\textbf{CSI↑}} & \multicolumn{1}{c}{\textbf{CSI 80↑}} & \multicolumn{1}{c}{\textbf{Training Time↓}} & \multicolumn{1}{c}{\textbf{PUR↓}} \\ \hline
Average (two lead months)                         & \textbf{0.5246}                   & 0.0                               & 0.2197                               & 22.5095                                     & 0\%                               \\
Average (three lead months)                       & \textbf{0.6370}                   & 0.0                               & 0.0200                               & 22.5024                                     & 38.3\%                            \\
Average (six lead months)                         & \textbf{0.7163}                   & 0.0                               & 0.0                                  & 22.8932                                     & 95\%
\end{tabular}
\end{table}

\subsubsection{Models with Different Loss Functions}

The average evaluation metrics across all locations of the models with the MAE loss, the Huber loss, the weighted MSE loss, the focal-R loss, and the balanced MSE loss for each location are respectively in Tables \ref{tb:maeall}, \ref{tb:huberall}, \ref{tb:wmseall}, \ref{tb:focalall}, and \ref{tb:bmseall}. The average evaluation metrics across all locations of the models with our proposed scaling-weighted MSE loss with three sets of hyperparameters are respectively in Tables \ref{tb:swmse1all}, \ref{tb:swmse2all}, and \ref{tb:swmse3all}.

\begin{table}[H]
\centering
\scriptsize
\caption{Average results of the models with the MAE loss for two, three, and six lead months SSTA and MHW forecasts.}\label{tb:maeall}
\begin{tabular}{llllll}
\multicolumn{1}{c}{\textbf{Across All Locations}} & \multicolumn{1}{c}{\textbf{MSE↓}} & \multicolumn{1}{c}{\textbf{CSI↑}} & \multicolumn{1}{c}{\textbf{CSI 80↑}} & \multicolumn{1}{c}{\textbf{Training Time↓}} & \multicolumn{1}{c}{\textbf{PUR↓}} \\ \hline
Average (two lead months)             & 0.7122                            & 0.0                               & 0.0                                  & 18.5255                                     & 95\%                              \\
Average (three lead months)           & \textbf{0.7376}                   & 0.0                               & 0.0                                  & 18.3525                                     & 100\%                             \\
Average (six lead months)             & \textbf{0.7383}                   & 0.0                               & 0.0                                  & 18.2578                                     & 100\%
\end{tabular}
\end{table}

\begin{table}[H]
\centering
\scriptsize
\caption{Average results of the models with the Huber loss with $\delta=0.5$ for two, three, and six lead months SSTA and MHW forecasts.}\label{tb:huberall}
\begin{tabular}{llllll}
\multicolumn{1}{c}{\textbf{Across All Locations}} & \multicolumn{1}{c}{\textbf{MSE↓}} & \multicolumn{1}{c}{\textbf{CSI↑}} & \multicolumn{1}{c}{\textbf{CSI 80↑}} & \multicolumn{1}{c}{\textbf{Training Time↓}} & \multicolumn{1}{c}{\textbf{PUR↓}} \\ \hline
Average (two lead months)             & \textbf{0.6137}                   & 0.0968                            & 0.1970                               & 21.1989                                     & 0\%                               \\
Average (three lead months)           & \textbf{0.7474}                   & 0.0276                            & 0.1151                               & 21.9763                                     & 0\%                               \\
Average (six lead months)             & \textbf{0.8570}                   & 0.0066                            & 0.0643                               & 21.1274                                     & 0\%
\end{tabular}
\end{table}

\begin{table}[H]
\centering
\scriptsize
\caption{Average results of the models with the weighted MSE loss with $w_{90\%}=1.5$ and $w_{80\%}=1.25$ for two, three, and six lead months SSTA and MHW forecasts.}\label{tb:wmseall}
\begin{tabular}{llllll}
\multicolumn{1}{c}{\textbf{Across All Locations}} & \multicolumn{1}{c}{\textbf{MSE↓}} & \multicolumn{1}{c}{\textbf{CSI↑}} & \multicolumn{1}{c}{\textbf{CSI 80↑}} & \multicolumn{1}{c}{\textbf{Training Time↓}} & \multicolumn{1}{c}{\textbf{PUR↓}} \\ \hline
Average (two lead months)                         & \textbf{0.5257}                   & 0.0                               & 0.2222                               & 30.7924                                     & 0\%                               \\
Average (three lead months)                       & \textbf{0.6323}                   & 0.0                               & 0.0536                               & 31.113                                      & 26.7\%                            \\
Average (six lead months)                         & \textbf{0.7036}                   & 0.0                               & 0.0                                  & 30.1553                                     & 66.7\%
\end{tabular}
\end{table}

\begin{table}[H]
\centering
\scriptsize
\caption{Average results of the models with the focal-R-MSE loss with $\beta=2$ and $\theta=1$ for two, three, and six lead months SSTA and MHW forecasts.}\label{tb:focalall}
\begin{tabular}{llllll}
\multicolumn{1}{c}{\textbf{Across All Locations}} & \multicolumn{1}{c}{\textbf{MSE↓}} & \multicolumn{1}{c}{\textbf{CSI↑}} & \multicolumn{1}{c}{\textbf{CSI 80↑}} & \multicolumn{1}{c}{\textbf{Training Time↓}} & \multicolumn{1}{c}{\textbf{PUR↓}} \\ \hline
Average (two lead months)                         & \textbf{0.5240}                   & 0.0                               & 0.2202                               & 26.128                                      & 0\%                               \\
Average (three lead months)                       & \textbf{0.6484}                   & 0.0                               & 0.0151                               & 26.3625                                     & 48.3\%                            \\
Average (six lead months)                         & \textbf{0.7209}                   & 0.0                               & 0.0                                  & 26.1644                                     & 99.2\%
\end{tabular}
\end{table}

\begin{table}[H]
\centering
\scriptsize
\caption{Results of the models with the balanced MSE loss with optimized $\sigma$ for two, three, and six lead month SSTA and MHW forecasts.}\label{tb:bmseall}
\begin{tabular}{llllll}
\multicolumn{1}{c}{\textbf{Across All Locations}} & \multicolumn{1}{c}{\textbf{MSE↓}} & \multicolumn{1}{c}{\textbf{CSI↑}} & \multicolumn{1}{c}{\textbf{CSI 80↑}} & \multicolumn{1}{c}{\textbf{Training Time↓}} & \multicolumn{1}{c}{\textbf{PUR↓}} \\ \hline
Average (two lead months)             & 1.0370                            & 0.2026                            & 0.3127                               & 25.5092                                     & 1.7\%                             \\
Average (three lead months)           & 1.0233                            & 0.0694                            & 0.0998                               & 25.8331                                     & 58.3\%                            \\
Average (six lead months)             & \textbf{0.7136}                   & 0.0                               & 0.0                                  & 26.0090                                     & 100\%
\end{tabular}
\end{table}

\begin{table}[H]
\centering
\scriptsize
\caption{Results of the models with the scaling-weighted MSE loss with $\alpha=1.5$, $\beta=0.5$, $w_{90\%}=1.5$, and $w_{80\%}=1.25$ for two, three, and six lead month SSTA and MHW forecasts.}\label{tb:swmse1all}
\begin{tabular}{llllll}
\multicolumn{1}{c}{\textbf{Across All Locations}} & \multicolumn{1}{c}{\textbf{MSE↓}} & \multicolumn{1}{c}{\textbf{CSI↑}} & \multicolumn{1}{c}{\textbf{CSI 80↑}} & \multicolumn{1}{c}{\textbf{Training Time↓}} & \multicolumn{1}{c}{\textbf{PUR↓}} \\ \hline
Average (two lead months)             & 0.6418                            & 0.1926                            & 0.3161                               & 40.4254                                     & 0\%                               \\
Average (three lead months)           & \textbf{0.7360}                   & 0.0866                            & 0.2297                               & 38.9710                                     & 0\%                               \\
Average (six lead months)             & \textbf{0.8411}                   & 0.0223                            & 0.1187                               & 38.0058                                     & 15\%
\end{tabular}
\end{table}

\begin{table}[H]
\centering
\scriptsize
\caption{Results of the models with the scaling-weighted MSE loss with $\alpha=2$, $\beta=0.5$, $w_{90\%}=1.5$, and $w_{80\%}=1.25$ for two, three, and six lead month SSTA and MHW forecasts.}\label{tb:swmse2all}
\begin{tabular}{llllll}
\multicolumn{1}{c}{\textbf{Across All Locations}} & \multicolumn{1}{c}{\textbf{MSE↓}} & \multicolumn{1}{c}{\textbf{CSI↑}} & \multicolumn{1}{c}{\textbf{CSI 80↑}} & \multicolumn{1}{c}{\textbf{Training Time↓}} & \multicolumn{1}{c}{\textbf{PUR↓}} \\ \hline
Average (two lead months)             & 0.6988                            & 0.1976                            & \textbf{0.3213}                      & 29.6779                                     & 0\%                               \\
Average (three lead months)           & 0.8176                            & 0.1170                            & 0.2300                               & 30.1820                                     & 0\%                               \\
Average (six lead months)             & 1.0395                            & 0.0682                            & \textbf{0.1913}                      & 30.3764                                     & 0\%
\end{tabular}
\end{table}

\begin{table}[H]
\centering
\scriptsize
\caption{Results of the models with the scaling-weighted MSE loss with $\alpha=2$, $\beta=1$, $w_{90\%}=1.5$, and $w_{80\%}=1.25$ for two, three, and six lead month SSTA and MHW forecasts.}\label{tb:swmse3all}
\begin{tabular}{llllll}
\multicolumn{1}{c}{\textbf{Across All Locations}} & \multicolumn{1}{c}{\textbf{MSE↓}} & \multicolumn{1}{c}{\textbf{CSI↑}} & \multicolumn{1}{c}{\textbf{CSI 80↑}} & \multicolumn{1}{c}{\textbf{Training Time↓}} & \multicolumn{1}{c}{\textbf{PUR↓}} \\ \hline
Average (two lead months)             & 1.0222                            & 0.1761                            & 0.2978                               & 24.6314                                     & 0\%                               \\
Average (three lead months)           & 1.2485                            & 0.1356                            & 0.2310                               & 24.5649                                     & 0\%                               \\
Average (six lead months)             & 1.5275                            & 0.0970                            & 0.1695                               & 25.0025                                     & 0\%
\end{tabular}
\end{table}

\section{Discussion}

This section provides an of the results, focusing on the performance of different loss functions and their impact on the predictive accuracy of SSTAs and MHWs.

\subsection{Loss Functions}

Evaluating the effectiveness of various loss functions is crucial for understanding their impact on forecasting performance, especially for tasks involving extreme events like MHWs.

\subsubsection{Average MSE, CSI and CSI 80}

The base model with the MSE loss generally returned small test MSEs across most locations. The training time was short and the PUR was low. However, according to the CSI and the scatter plots, the MSE loss did not accurately predict the extreme values. The MSE loss was unable to learn hidden patterns when forecasting six months ahead using our FCN configuration.

The MAE loss returned slightly larger test MSEs than the MSE loss. When predicting suspected MHWs, the MAE loss performed almost the same as the MSE loss, but when predicting MHWs, the MAE loss demonstrated its lower sensitivity to the extreme values than the MSE loss. Also, the PUR was substantially high and the MAE loss was unable to forecast SSTAs and MHWs two, three, and six months ahead using our FCN configuration. Given these problems, we deduce that the MAE loss encouraged the models to make more conservative predictions and was unable to capture the complexity of the data, by being less sensitive to individual errors. For long lead time forecasts, the models had difficulties learning the hidden patterns from the current input data.

The Huber loss was generally outperformed by the MSE loss for one lead month forecasts in terms of the MSE and the CSI. However, the Huber loss had shorter training time and no underfitted models for all lead month forecasts. Therefore, the Huber loss could be used for two, three, and six lead month SSTA forecasts. As combing properties of the MSE and the MAE, the Huber loss is less influenced by outliers like the MAE while still providing meaningful gradients for backpropagation like the MSE.

The weighted MSE loss returned similar MSEs to the MSE loss but slightly better CSIs. The PURs of the weighted MSE were generally lower than the MSE loss. Compared with the Huber loss, the weighted MSE had better forecasts for both SSTAs and MHWs for one and two months ahead, which implies the effectiveness of the use of large weights for extreme values. However, considering the high PURs for three and six months, we do not suggest the weighted MSE loss for long lead forecasts. In addition, the weighted MSE was slow in training.

The performance of the focal-R loss was quite similar to the weighted MSE, referring to the evaluation metrics and scatter plots. The focal-R was slightly faster than the weighted MSE in training, but for three and six lead months the focal-R had higher PURs. The focal-R was proposed for the situation where there are significant missing values in specific value ranges. The techniques introduced with the focal-R \citep{yang2021delving} attempts to amend the substantial data imbalance to the normal distribution and then perform learning. In this specific case of SSTA and MHW forecasts, the distributions of predictands and predictors are normally distributed or nearly normally distributed, with no missing values and limited outliers. Therefore, the focal-R loss performed similar to the standard MSE loss and the weighted MSE loss.

The balanced MSE loss outperformed the other loss functions and the persistence model on the one lead month MHW forecast with generally better CSIs. The trade-off was the high MSEs for SSTA forecasts. Compared with the fixed-lag models, the balanced MSE predicted suspected MHWs generally better and predicted MHWs slightly better at some locations. However, the PURs for three and six lead months were high, so we do not suggest using the balanced MSE for long lead forecasts, where the patterns learned from features are limited for prediction.

Our proposed scaling-weighted MSE loss was able to perform similarly to the balanced MSE, in terms of the MSE and the CSI. There was almost no perfect underfitting in training for all lead months. We noticed that the training time was long when the exponential hyperparameter $\alpha=1.5$. When $\beta=1$, there were more Type I errors. Also, our proposed loss function was able to balance the trade-off between the MSE and the CSI, returning similar CSIs without making MSEs larger. Our proposed loss was the only loss function that was able to forecast MHWs for longer lead times, although it has not generally outperformed the fixed-lag models.

In all, the results of this imbalanced task imply a trade-off between accurate prediction for normal values and accurate prediction for anomalies or extreme values. The regressors trained with the MSE loss and the focal-R endow larger weights to the normal, while the regressors trained with the balanced MSE and our scaling-weighted MSE adjust larger weights to anomalies. We suggest utilizing different loss functions to predict distinct classes of labels in the imbalanced regression task of SSTA and MHW forecasts.

\subsubsection{Average Rank}

Based on the ranks and the Friedman Test for all models across all locations for one lead month forecasts, we recommend the three most highly-ranked models: the MSE, the weighted MSE, and the focal-R for forecasting SSTAs, the persistence model, the balanced MSE, and our scaling-weighted MSE for forecasting MHWs, and the balanced MSE and our scaling-weighted MSE for forecasting suspected MHWs. For longer lead time forecasts, the situations are more complicated and require evaluation with the PUR case by case. For forecasts at different locations, Figure \ref{fg:locs_results_losses} summarizes the three highest-ranked loss functions for each location.

\subsection{Locations}

We mainly discuss the one lead month forecasts by location.

In terms of the SSTA forecasts, the FCNs outperformed the persistence models at ten New Zealand locations, besides Taranaki and Wairarapa. According to the exploratory data analysis, the means of the SSTA time series at Taranaki and Wairarapa are around 0.1 instead of 0 and they have the two largest standard deviations, which might be the reasons for FCNs being unable to outperform the persistence model.

The forecasts of MHWs are generally more challenging than the forecasts of suspected MHWs. For MHW forecasts, the FCN models could outperform the persistence model at Bank Peninsula, Chatham Island, Cape Reinga, Fiordland, Otago Peninsula, Stewart Island, Taranaki, and Wairarapa. We noticed that among the four locations with the SSTA time series normally distributed, the FCN models did not outperform the persistence model at Bay of Plenty, Cook Strait, and Hauraki Gulf. For suspected MHW forecasts, only at two locations: Cook Strait and Raglan, the FCNs models did not outperform the persistence model. The balanced MSE and our scaling-weighted MSE generally had larger CSIs and CSI 80s than the persistence model at more locations than the other loss functions.

To further investigate the regional difference in predictability of the FCN models, we performed a Monte Carlo test. Cape Reinga, Hauraki Gulf, Bay of Plenty, Raglan, Taranaki, Cook Strait, and Wairarapa were the seven locations in the northern group. Bank Peninsula, Chatham Island, Fiordland, Otago Peninsula, and Stewart Island were the five locations in the southern group. We selected the regressors with the MSE loss and the balanced MSE loss for one lead month SSTA and MHW forecasts as examples. We calculated the differences between the average MSEs and the average CSIs of the two groups respectively. Then, we conducted a Monte Carlo test by randomly dividing the 12 locations into two groups of equal size and comparing the differences in the two metrics with the differences observed in the original two groups. After running 10000 simulations for each case, we observed that the probability of encountering counterexamples in all four cases exceeded 0.1. Considering a significance level of 0.01, the simulation results indicate that there is insufficient evidence to reject the null hypothesis, suggesting no significant difference in predictability between the two regions. The regressors exhibited similar performances in both regions, and it is unlikely that training different regressors for each region would result in a substantial improvement in predictability.

The results of the longer lead time forecasts by location were in the Jupyter Notebooks accessible in Appendix \ref{apd:code}.

In general, the forecasts for southern locations were more accurate than those for northern locations. We deduce the reasons could be that SSTs in colder regions are more stable than those in warmer regions and especially in New Zealand, the southern oceanic part is less influenced by the East Australian Current.

\subsection{Lead Times}

The one lead month SSTA and suspected MHW forecasts were generally good, where the FCNs with the MSE, the MAE, the weighted MSE, and the focal-R loss mostly outperformed the persistence model in terms of the MSE, and the FCNs with the balanced MSE and our scaling-weighted MSE loss mostly outperformed the persistence model in terms of the CSI 80. For MHW forecasts, the persistence models mostly outperformed the FCNs. Typically, the occurrence of consecutive extreme values tends to result in good performance for persistence models in forecasting extremes, and vice versa. Also, there was almost no perfect underfitting in training.

The two lead month SSTA forecasts exhibit moderate performances, where the FCNs with the MSE, the Huber, the weighted MSE, and the focal-R loss mostly outperformed the fixed-lag model in terms of the MSE. Comparatively, the two lead month suspected MHW forecasts were not good, where only the FCNs with our scaling-weighted MSE mostly outperformed the fixed-lag model in terms of the CSI 80 metric. For MHW forecasts, the fixed-lag models outperformed the FCNs. Also, only the FCNs with the MAE loss had a large number of underfitted models.

For the three and six lead month SSTA forecasts, the models might simply predict one value to reach lower MSEs than the fixed-lag model. The FCNs with the MSE, the MAE, the weighted MSE, the focal-R, and the balanced MSE loss had large PURs and were unable to effectively learn hidden representations to make long lead forecasts given the current inputs and configurations. The FCNs with the Huber loss and our proposed scaling-weighted MSE loss were able to make forecasts, where the Huber loss generally returned lower MSEs and the scaling-weighted MSE generally returned lower CSI 80s than the fixed-lag model. The long lead SSTA and MHW forecasts are challenging.

\begin{figure}[H]
\centering
\includegraphics[scale=0.3]{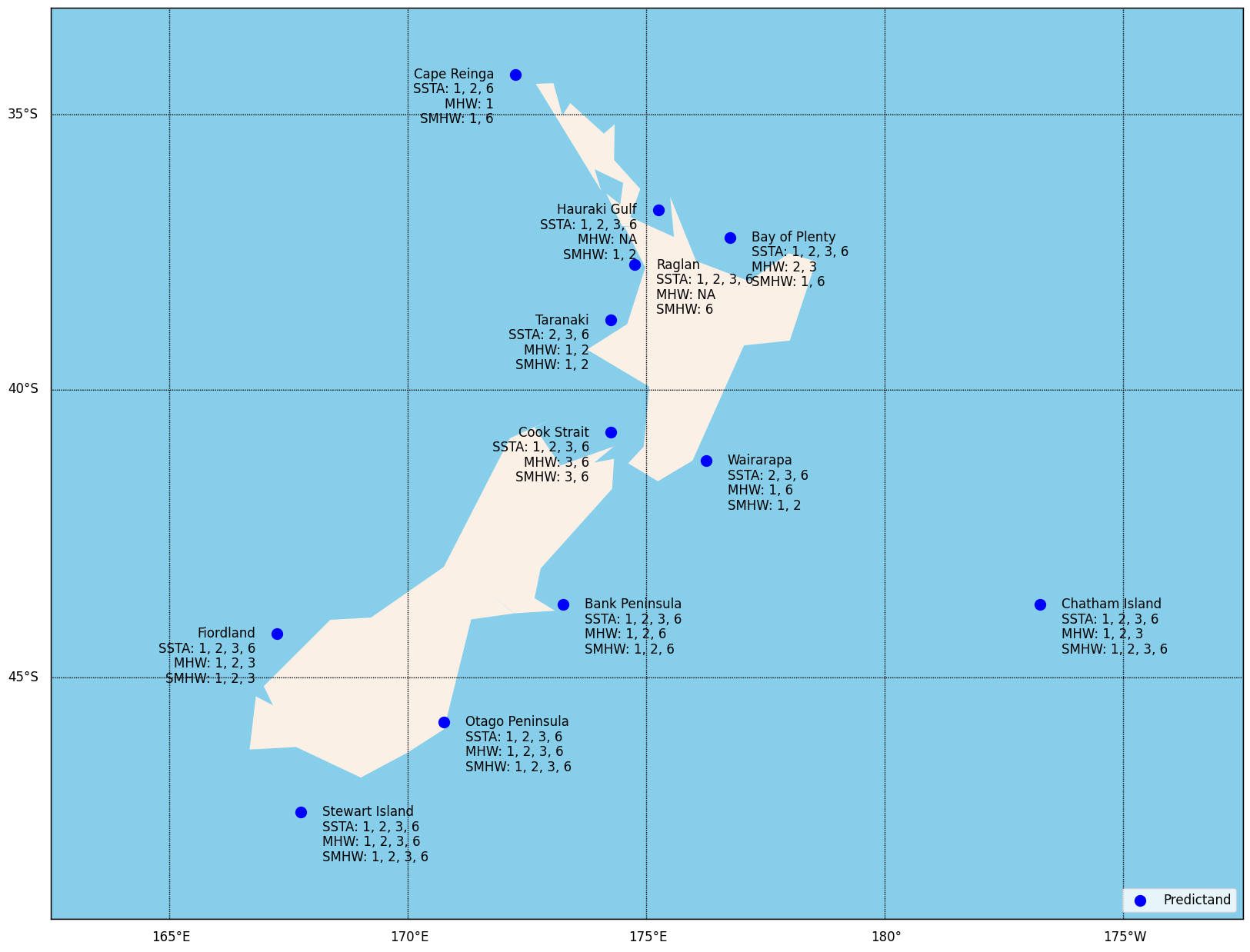}
\caption{12 selected locations with the corresponding results of monthly SSTA and MHW forecasts. The acronym ``MHW'' is ``MHW'' and ``SMHW'' is ``suspected MHW'', applied to the other figures in this study. The numbers following each type of forecasts are the numbers of lead months. If at least one of the model outperformed the persistence or fixed-lag model at one location in terms of the MSE for SSTAs, the CSI for MHWs, and the CSI 80 for suspected MHWs with the PUR not greater than 20\%, the corresponding type of forecasts and the lead month are added to the location.}\label{fg:locs_results}
\end{figure}

\begin{figure}[H]
\centering
\includegraphics[scale=0.3]{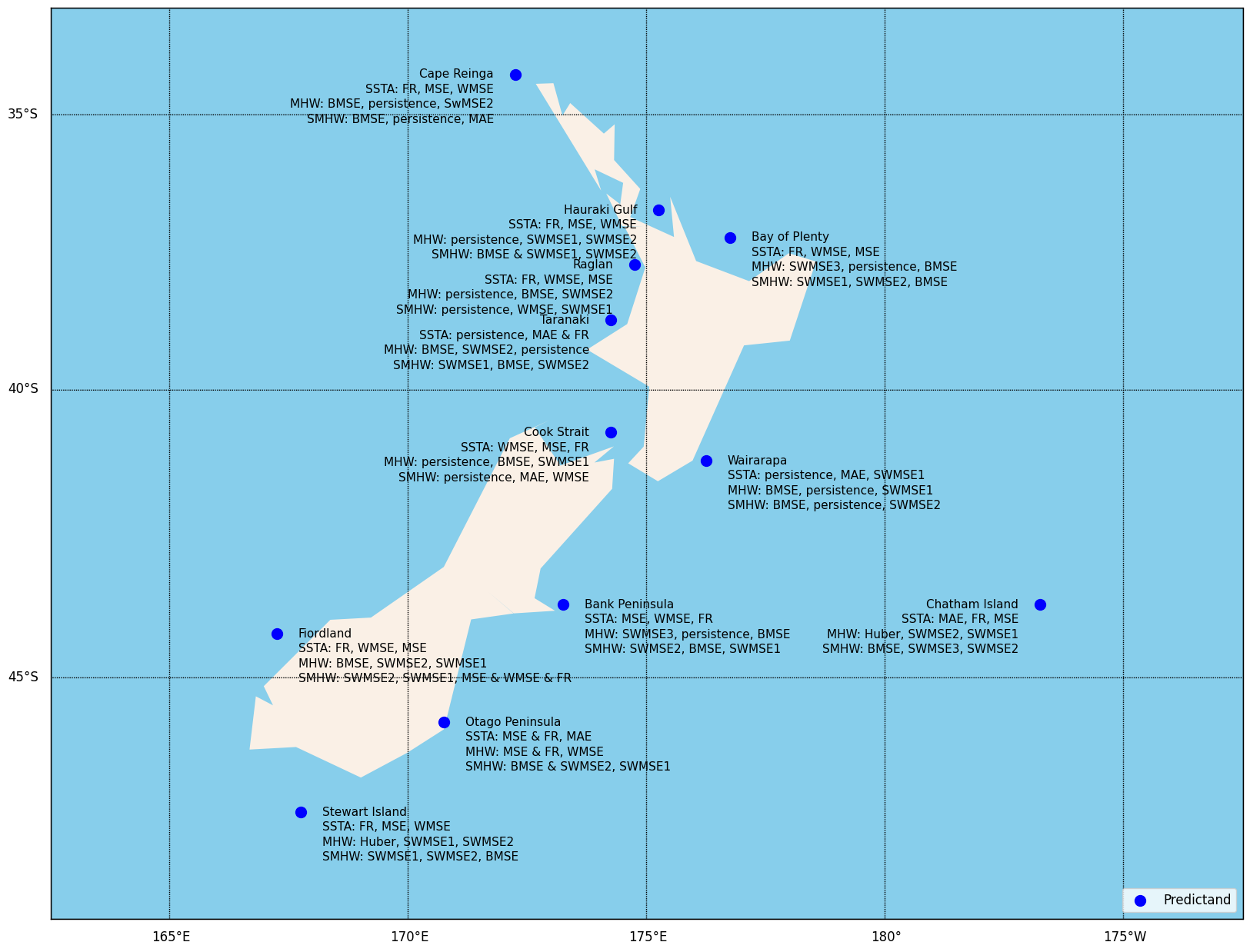}
\caption{12 selected locations with the corresponding ranks of the loss functions based on the Friedman test with regards to monthly SSTA and MHW forecasts. Only the top three ranked predictors for one lead month SSTA, MHW, and SMHW at each location are selected. ``\&'' denotes the same rank.}\label{fg:locs_results_losses}
\end{figure}

Figure \ref{fg:locs_results} summarizes the results by location and by lead time. Figure \ref{fg:locs_results_losses} summarizes the ranks of the loss functions in terms of one lead month forecasts only. In Appendix \ref{apd:add}, Table \ref{tb:all_results} summarizes the results by loss function, by location, and by lead time.

\section{Conclusion}

This section summarizes the key contributions of this study and outlines directions for future work.

\subsection{Contributions}

This study makes three major contributions.

First, we assessed the predictability of SSTAs and MHWs using ML models for 12 New Zealand locations and for one, two, three, and six lead months. We analyzed the results by location and lead time, respectively. Generally, the forecasts for southern locations exhibit better performance compared to those for northern locations, possibly due to more stable changes in SSTAs in colder water. Six-month lead MHW forecasts remain challenging because the current data inputs are insufficient to learn the complex patterns for long lead time forecasts, and the randomness in the climate system might make long-lead statistical forecasts unattainable.

Second, we identified the influence of regression loss functions for monthly SSTA and MHW forecasts in New Zealand. For SSTA forecasts, we suggest using the standard MSE loss, the weighted MSE, and the focal-R loss. For MHW forecasts, we suggest using the persistence or fixed-lag model, the balanced MSE loss, and our proposed scaling-weighted MSE loss. For suspected MHW forecasts, we suggest using the balanced MSE loss and our scaling-weighted MSE loss. This finding can be extended to experiments on global SSTA and MHW forecasting.

Third, we proposed a scaling-weighted MSE loss function specifically designed for the imbalanced regression task of monthly MHW forecasts in New Zealand. The scaling-weighted loss function aims to achieve a balance between predictions for normal values and extreme values, offering flexibility in controlling predictions for extreme values and Type I errors, and preventing underfitting. Our proposed scaling-weighted MSE loss has demonstrated its effectiveness in forecasting MHWs at specific locations, including Chatham Island, Fiordland, Otago Peninsula, Stewart Island, and Taranaki, by outperforming the persistence or fixed-lag model.

\subsection{Further Work}

We plan to address the following aspects in subsequent studies.

First, the regional data inputs might not be sufficient for SSTA and MHW forecasts in New Zealand. We recommend increasing the number of inputs by expanding to global sources.

Second, given the substantial increase in inputs, we suggest using an alternative neural network class to perform representation learning on the large amount of data. While CNNs have been used for SST forecasts \citep{taylor2022deep}, we propose experimenting with GNNs. Graph re-sampling provides flexibility in adjusting the number of inputs, and GNNs could potentially model climate teleconnections.

Third, relying on SSTAs as the only variable might not be sufficient for SSTA and MHW forecasts. Incorporating other ocean or climate variables could potentially enhance the forecasts by including them as either node attributes or edge attributes in graph re-sampling.

Fourth, the generalization of our proposed scaling-weighted MSE loss should be further justified through global SSTA and MHW forecasts and other imbalanced regression tasks. There is still room for improvement and generalization in the loss function.

\bibliographystyle{icml2024}
\bibliography{paper}

\begin{thebibliography}{59}
\providecommand{\natexlab}[1]{#1}
\providecommand{\url}[1]{\texttt{#1}}
\expandafter\ifx\csname urlstyle\endcsname\relax
  \providecommand{\doi}[1]{doi: #1}\else
  \providecommand{\doi}{doi: \begingroup \urlstyle{rm}\Url}\fi

\bibitem[{Asia-Pacific Data Research Center}(accessed 2022)]{apdrcdataset}
{Asia-Pacific Data Research Center}.
\newblock {AVHRR Sea Surface Temperature Data}.
\newblock \url{http://apdrc.soest.hawaii.edu/las/v6/dataset?catitem=4781},
  accessed 2022.
\newblock Accessed on August 11, 2022.

\bibitem[Berg et~al.(2021)Berg, Oskarsson, and O'Connor]{berg2021deep}
Berg, A., Oskarsson, M., and O'Connor, M.
\newblock Deep ordinal regression with label diversity.
\newblock In \emph{2020 25th International Conference on Pattern Recognition
  (ICPR)}, pp.\  2740--2747. IEEE, 2021.

\bibitem[Berkelmans et~al.(2004)Berkelmans, De’ath, Kininmonth, and
  Skirving]{berkelmans2004comparison}
Berkelmans, R., De’ath, G., Kininmonth, S., and Skirving, W.~J.
\newblock A comparison of the 1998 and 2002 coral bleaching events on the great
  barrier reef: spatial correlation, patterns, and predictions.
\newblock \emph{Coral reefs}, 23\penalty0 (1):\penalty0 74--83, 2004.

\bibitem[Boschetti et~al.(2022)Boschetti, Feng, Hartog, Hobday, and
  Zhang]{boschetti2022sea}
Boschetti, F., Feng, M., Hartog, J.~R., Hobday, A.~J., and Zhang, X.
\newblock Sea surface temperature predictability at the interface between
  oceanographic modelling and machine learning.
\newblock 2022.

\bibitem[Branco et~al.(2017)Branco, Torgo, and Ribeiro]{branco2017smogn}
Branco, P., Torgo, L., and Ribeiro, R.~P.
\newblock Smogn: A pre-processing approach for imbalanced regression.
\newblock In \emph{First International Workshop on Learning with Imbalanced
  Domains: Theory and Applications}, pp.\  36--50. PMLR, 2017.

\bibitem[Cachay et~al.(2021)Cachay, Erickson, Bucker, Pokropek, Potosnak, Bire,
  Osei, and L{\"u}tjens]{cachay2021world}
Cachay, S.~R., Erickson, E., Bucker, A. F.~C., Pokropek, E., Potosnak, W.,
  Bire, S., Osei, S., and L{\"u}tjens, B.
\newblock The world as a graph: Improving el niño forecasts with graph neural
  networks.
\newblock \emph{arXiv preprint arXiv:2104.05089}, 2021.

\bibitem[Cao et~al.(2019)Cao, Wei, Gaidon, Arechiga, and Ma]{cao2019learning}
Cao, K., Wei, C., Gaidon, A., Arechiga, N., and Ma, T.
\newblock Learning imbalanced datasets with label-distribution-aware margin
  loss.
\newblock \emph{Advances in neural information processing systems}, 32, 2019.

\bibitem[Carton \& Giese(2008)Carton and Giese]{carton2008reanalysis}
Carton, J.~A. and Giese, B.~S.
\newblock A reanalysis of ocean climate using simple ocean data assimilation
  (soda).
\newblock \emph{Monthly weather review}, 136\penalty0 (8):\penalty0 2999--3017,
  2008.

\bibitem[Chawla et~al.(2002)Chawla, Bowyer, Hall, and
  Kegelmeyer]{chawla2002smote}
Chawla, N.~V., Bowyer, K.~W., Hall, L.~O., and Kegelmeyer, W.~P.
\newblock {SMOTE}: synthetic minority over-sampling technique.
\newblock \emph{Journal of artificial intelligence research}, 16:\penalty0
  321--357, 2002.

\bibitem[Chen et~al.(2020)Chen, Yu, Ullah, Wu, Liu, Huang, Gao, Jiang, and
  Nie]{chen2020new}
Chen, X., Yu, R., Ullah, S., Wu, D., Liu, M., Huang, Y., Gao, H., Jiang, J.,
  and Nie, N.
\newblock A new weighted mse loss for wind speed forecasting based on deep
  learning models.
\newblock In \emph{EGU General Assembly Conference Abstracts}, pp.\  12899,
  2020.

\bibitem[Cui et~al.(2019)Cui, Jia, Lin, Song, and Belongie]{cui2019class}
Cui, Y., Jia, M., Lin, T.-Y., Song, Y., and Belongie, S.
\newblock Class-balanced loss based on effective number of samples.
\newblock In \emph{Proceedings of the IEEE/CVF Conference on Computer Vision
  and Pattern Recognition}, pp.\  9268--9277, 2019.

\bibitem[Dem{\v{s}}ar(2006)]{demvsar2006statistical}
Dem{\v{s}}ar, J.
\newblock Statistical comparisons of classifiers over multiple data sets.
\newblock \emph{The Journal of Machine learning research}, 7:\penalty0 1--30,
  2006.

\bibitem[Dong et~al.(2018)Dong, Gong, and Zhu]{dong2018imbalanced}
Dong, Q., Gong, S., and Zhu, X.
\newblock Imbalanced deep learning by minority class incremental rectification.
\newblock \emph{IEEE transactions on pattern analysis and machine
  intelligence}, 41\penalty0 (6):\penalty0 1367--1381, 2018.

\bibitem[Elzahaby et~al.(2021)Elzahaby, Schaeffer, Roughan, and
  Delaux]{elzahaby2021oceanic}
Elzahaby, Y., Schaeffer, A., Roughan, M., and Delaux, S.
\newblock Oceanic circulation drives the deepest and longest marine heatwaves
  in the east australian current system.
\newblock \emph{Geophysical Research Letters}, 48\penalty0 (17):\penalty0
  e2021GL094785, 2021.

\bibitem[Feng et~al.(2022)Feng, Boschetti, Ling, Zhang, Hartog, Akhtar, Shi,
  Luo, and Hobday]{fengpredictability}
Feng, M., Boschetti, F., Ling, F., Zhang, X., Hartog, J.~R., Akhtar, M., Shi,
  L., Luo, J.-J., and Hobday, A.~J.
\newblock Predictability of sea surface temperature anomalies at the eastern
  pole of the indian ocean dipole-using a convolutional neural network model.
\newblock \emph{Frontiers in Climate}, pp.\  143, 2022.

\bibitem[Friedman(1937)]{friedman1937use}
Friedman, M.
\newblock The use of ranks to avoid the assumption of normality implicit in the
  analysis of variance.
\newblock \emph{Journal of the american statistical association}, 32\penalty0
  (200):\penalty0 675--701, 1937.

\bibitem[Friedman(1940)]{friedman1940comparison}
Friedman, M.
\newblock A comparison of alternative tests of significance for the problem of
  m rankings.
\newblock \emph{The Annals of Mathematical Statistics}, 11\penalty0
  (1):\penalty0 86--92, 1940.

\bibitem[Garc{\'\i}a \& Herrera(2009)Garc{\'\i}a and
  Herrera]{garcia2009evolutionary}
Garc{\'\i}a, S. and Herrera, F.
\newblock Evolutionary undersampling for classification with imbalanced
  datasets: Proposals and taxonomy.
\newblock \emph{Evolutionary computation}, 17\penalty0 (3):\penalty0 275--306,
  2009.

\bibitem[Ham et~al.(2019)Ham, Kim, and Luo]{ham2019deep}
Ham, Y.-G., Kim, J.-H., and Luo, J.-J.
\newblock Deep learning for multi-year enso forecasts.
\newblock \emph{Nature}, 573\penalty0 (7775):\penalty0 568--572, 2019.

\bibitem[Ham et~al.(2021)Ham, Kim, Kim, and On]{ham2021unified}
Ham, Y.-G., Kim, J.-H., Kim, E.-S., and On, K.-W.
\newblock Unified deep learning model for el ni{\~n}o/southern oscillation
  forecasts by incorporating seasonality in climate data.
\newblock \emph{Science Bulletin}, 66\penalty0 (13):\penalty0 1358--1366, 2021.

\bibitem[He et~al.(2008)He, Bai, Garcia, and Li]{he2008adasyn}
He, H., Bai, Y., Garcia, E.~A., and Li, S.
\newblock Adasyn: Adaptive synthetic sampling approach for imbalanced learning.
\newblock In \emph{Proceedings of the IEEE International Joint Conference on
  Neural Networks (IEEE World Congress on Computational Intelligence)}, pp.\
  1322--1328. IEEE, 2008.

\bibitem[Hobday et~al.(2016)Hobday, Alexander, Perkins, Smale, Straub, Oliver,
  Benthuysen, Burrows, Donat, Feng, et~al.]{hobday2016hierarchical}
Hobday, A.~J., Alexander, L.~V., Perkins, S.~E., Smale, D.~A., Straub, S.~C.,
  Oliver, E.~C., Benthuysen, J.~A., Burrows, M.~T., Donat, M.~G., Feng, M.,
  et~al.
\newblock A hierarchical approach to defining marine heatwaves.
\newblock \emph{Progress in Oceanography}, 141:\penalty0 227--238, 2016.

\bibitem[Hobday et~al.(2018)Hobday, Spillman, Eveson, Hartog, Zhang, and
  Brodie]{hobday2018framework}
Hobday, A.~J., Spillman, C.~M., Eveson, J.~P., Hartog, J.~R., Zhang, X., and
  Brodie, S.
\newblock A framework for combining seasonal forecasts and climate projections
  to aid risk management for fisheries and aquaculture.
\newblock \emph{Frontiers in Marine Science}, pp.\  137, 2018.

\bibitem[Holbrook et~al.(2020)Holbrook, Sen~Gupta, Oliver, Hobday, Benthuysen,
  Scannell, Smale, and Wernberg]{holbrook2020keeping}
Holbrook, N.~J., Sen~Gupta, A., Oliver, E.~C., Hobday, A.~J., Benthuysen,
  J.~A., Scannell, H.~A., Smale, D.~A., and Wernberg, T.
\newblock Keeping pace with marine heatwaves.
\newblock \emph{Nature Reviews Earth \& Environment}, 1\penalty0 (9):\penalty0
  482--493, 2020.

\bibitem[Huang et~al.(2016)Huang, Li, Loy, and Tang]{huang2016learning}
Huang, C., Li, Y., Loy, C.~C., and Tang, X.
\newblock Learning deep representation for imbalanced classification.
\newblock In \emph{Proceedings of the IEEE Conference on Computer Vision and
  Pattern Recognition}, pp.\  5375--5384, 2016.

\bibitem[Huang et~al.(2019)Huang, Li, Loy, and Tang]{huang2019deep}
Huang, C., Li, Y., Loy, C.~C., and Tang, X.
\newblock Deep imbalanced learning for face recognition and attribute
  prediction.
\newblock \emph{IEEE transactions on pattern analysis and machine
  intelligence}, 42\penalty0 (11):\penalty0 2781--2794, 2019.

\bibitem[Huber(1964)]{huber1964robust}
Huber, P.~J.
\newblock Robust estimation of a location parameter.
\newblock \emph{The Annals of Mathematical Statistics}, pp.\  73--101, 1964.

\bibitem[Iman \& Davenport(1980)Iman and Davenport]{iman1980approximations}
Iman, R.~L. and Davenport, J.~M.
\newblock Approximations of the critical region of the fbietkan statistic.
\newblock \emph{Communications in Statistics-Theory and Methods}, 9\penalty0
  (6):\penalty0 571--595, 1980.

\bibitem[Jacox et~al.(2019)Jacox, Tommasi, Alexander, Hervieux, and
  Stock]{jacox2019predicting}
Jacox, M.~G., Tommasi, D., Alexander, M.~A., Hervieux, G., and Stock, C.~A.
\newblock Predicting the evolution of the 2014--2016 {California} current
  system marine heatwave from an ensemble of coupled global climate forecasts.
\newblock \emph{Frontiers in Marine Science}, 6:\penalty0 497, 2019.

\bibitem[Kang et~al.(2019)Kang, Xie, Rohrbach, Yan, Gordo, Feng, and
  Kalantidis]{kang2019decoupling}
Kang, B., Xie, S., Rohrbach, M., Yan, Z., Gordo, A., Feng, J., and Kalantidis,
  Y.
\newblock Decoupling representation and classifier for long-tailed recognition.
\newblock In \emph{International Conference on Learning Representations}, 2019.

\bibitem[Karniadakis et~al.(2021)Karniadakis, Kevrekidis, Lu, Perdikaris, Wang,
  and Yang]{karniadakis2021physics}
Karniadakis, G.~E., Kevrekidis, I.~G., Lu, L., Perdikaris, P., Wang, S., and
  Yang, L.
\newblock Physics-informed machine learning.
\newblock \emph{Nature Reviews Physics}, 3\penalty0 (6):\penalty0 422--440,
  2021.

\bibitem[Khakzar et~al.(2021)Khakzar, Musatian, Buchberger, Valeriano~Quiroz,
  Pinger, Baselizadeh, Kim, and Navab]{khakzar2021towards}
Khakzar, A., Musatian, S., Buchberger, J., Valeriano~Quiroz, I., Pinger, N.,
  Baselizadeh, S., Kim, S.~T., and Navab, N.
\newblock Towards semantic interpretation of thoracic disease and covid-19
  diagnosis models.
\newblock In \emph{Proceedings of the International Conference on Medical Image
  Computing and Computer-Assisted Intervention}, pp.\  499--508. Springer,
  2021.

\bibitem[Larner(2021)]{larner2021assessing}
Larner, A.
\newblock Assessing cognitive screeners with the critical success index.
\newblock \emph{Progress in Neurology and Psychiatry}, 25\penalty0
  (3):\penalty0 33--37, 2021.

\bibitem[Martin-Donas et~al.(2018)Martin-Donas, Gomez, Gonzalez, and
  Peinado]{martin2018deep}
Martin-Donas, J.~M., Gomez, A.~M., Gonzalez, J.~A., and Peinado, A.~M.
\newblock A deep learning loss function based on the perceptual evaluation of
  the speech quality.
\newblock \emph{IEEE Signal processing letters}, 25\penalty0 (11):\penalty0
  1680--1684, 2018.

\bibitem[Maynard et~al.(2008)Maynard, Turner, Anthony, Baird, Berkelmans,
  Eakin, Johnson, Marshall, Packer, Rea, et~al.]{maynard2008reeftemp}
Maynard, J.~A., Turner, P.~J., Anthony, K.~R., Baird, A.~H., Berkelmans, R.,
  Eakin, C.~M., Johnson, J., Marshall, P.~A., Packer, G.~R., Rea, A., et~al.
\newblock Reeftemp: An interactive monitoring system for coral bleaching using
  high-resolution sst and improved stress predictors.
\newblock \emph{Geophysical Research Letters}, 35\penalty0 (5), 2008.

\bibitem[Merryfield et~al.(2013)Merryfield, Lee, Boer, Kharin, Scinocca, Flato,
  Ajayamohan, Fyfe, Tang, and Polavarapu]{merryfield2013canadian}
Merryfield, W.~J., Lee, W.-S., Boer, G.~J., Kharin, V.~V., Scinocca, J.~F.,
  Flato, G.~M., Ajayamohan, R., Fyfe, J.~C., Tang, Y., and Polavarapu, S.
\newblock The canadian seasonal to interannual prediction system. part i:
  Models and initialization.
\newblock \emph{Monthly weather review}, 141\penalty0 (8):\penalty0 2910--2945,
  2013.

\bibitem[{MetOcean Solutions}(accessed 2023)]{moanaproject}
{MetOcean Solutions}.
\newblock {Marine Heatwave Forecast}.
\newblock \url{https://www.moanaproject.org/marine-heatwave-forecast}, accessed
  2023.
\newblock Accessed on February 18, 2023.

\bibitem[Nie et~al.(2018)Nie, Hu, and Li]{nie2018investigation}
Nie, F., Hu, Z., and Li, X.
\newblock An investigation for loss functions widely used in machine learning.
\newblock \emph{Communications in Information and Systems}, 18\penalty0
  (1):\penalty0 37--52, 2018.

\bibitem[Oliver et~al.(2019)Oliver, Burrows, Donat, Sen~Gupta, Alexander,
  Perkins-Kirkpatrick, Benthuysen, Hobday, Holbrook, Moore,
  et~al.]{oliver2019projected}
Oliver, E.~C., Burrows, M.~T., Donat, M.~G., Sen~Gupta, A., Alexander, L.~V.,
  Perkins-Kirkpatrick, S.~E., Benthuysen, J.~A., Hobday, A.~J., Holbrook,
  N.~J., Moore, P.~J., et~al.
\newblock Projected marine heatwaves in the 21st century and the potential for
  ecological impact.
\newblock \emph{Frontiers in Marine Science}, 6:\penalty0 734, 2019.

\bibitem[Pham et~al.(2020)Pham, Le, Le, Bui, Le, Ly, and
  Prakash]{pham2020development}
Pham, B.~T., Le, L.~M., Le, T.-T., Bui, K.-T.~T., Le, V.~M., Ly, H.-B., and
  Prakash, I.
\newblock Development of advanced artificial intelligence models for daily
  rainfall prediction.
\newblock \emph{Atmospheric Research}, 237:\penalty0 104845, 2020.

\bibitem[Pravallika et~al.(2022)Pravallika, Vasavi, and
  Vighneshwar]{pravallika2022prediction}
Pravallika, M.~S., Vasavi, S., and Vighneshwar, S.
\newblock Prediction of temperature anomaly in indian ocean based on
  autoregressive long short-term memory neural network.
\newblock \emph{Neural Computing and Applications}, 34\penalty0 (10):\penalty0
  7537--7545, 2022.

\bibitem[Ratnam et~al.(2020)Ratnam, Dijkstra, and Behera]{ratnam2020machine}
Ratnam, J., Dijkstra, H., and Behera, S.~K.
\newblock A machine learning based prediction system for the indian ocean
  dipole.
\newblock \emph{Scientific reports}, 10\penalty0 (1):\penalty0 1--11, 2020.

\bibitem[Ren et~al.(2022)Ren, Zhang, Yu, and Liu]{ren2022balanced}
Ren, J., Zhang, M., Yu, C., and Liu, Z.
\newblock Balanced mse for imbalanced visual regression.
\newblock pp.\  7926--7935, 2022.

\bibitem[Ronneberger et~al.(2015)Ronneberger, Fischer, and
  Brox]{ronneberger2015u}
Ronneberger, O., Fischer, P., and Brox, T.
\newblock U-net: Convolutional networks for biomedical image segmentation.
\newblock In \emph{Proceedings of the International Conference on Medical Image
  Computing and Computer-Assisted Intervention}, pp.\  234--241. Springer,
  2015.

\bibitem[Saha et~al.(2014)Saha, Moorthi, Wu, Wang, Nadiga, Tripp, Behringer,
  Hou, Chuang, Iredell, et~al.]{saha2014ncep}
Saha, S., Moorthi, S., Wu, X., Wang, J., Nadiga, S., Tripp, P., Behringer, D.,
  Hou, Y.-T., Chuang, H.-y., Iredell, M., et~al.
\newblock The ncep climate forecast system version 2.
\newblock \emph{Journal of climate}, 27\penalty0 (6):\penalty0 2185--2208,
  2014.

\bibitem[Schaefer(1990)]{schaefer1990critical}
Schaefer, J.~T.
\newblock The critical success index as an indicator of warning skill.
\newblock \emph{Weather and forecasting}, 5\penalty0 (4):\penalty0 570--575,
  1990.

\bibitem[Shapiro \& Wilk(1965)Shapiro and Wilk]{shapiro1965analysis}
Shapiro, S.~S. and Wilk, M.~B.
\newblock An analysis of variance test for normality (complete samples).
\newblock \emph{Biometrika}, 52\penalty0 (3/4):\penalty0 591--611, 1965.

\bibitem[Shu et~al.(2019)Shu, Xie, Yi, Zhao, Zhou, Xu, and Meng]{shu2019meta}
Shu, J., Xie, Q., Yi, L., Zhao, Q., Zhou, S., Xu, Z., and Meng, D.
\newblock Meta-weight-net: Learning an explicit mapping for sample weighting.
\newblock \emph{Advances in neural information processing systems}, 32, 2019.

\bibitem[Smale \& Wernberg(2013)Smale and Wernberg]{smale2013extreme}
Smale, D.~A. and Wernberg, T.
\newblock Extreme climatic event drives range contraction of a habitat-forming
  species.
\newblock \emph{Proceedings of the Royal Society B: Biological Sciences},
  280\penalty0 (1754):\penalty0 20122829, 2013.

\bibitem[Sutton \& Bowen(2019)Sutton and Bowen]{sutton2019ocean}
Sutton, P.~J. and Bowen, M.
\newblock Ocean temperature change around new zealand over the last 36 years.
\newblock \emph{New Zealand Journal of Marine and Freshwater Research},
  53\penalty0 (3):\penalty0 305--326, 2019.

\bibitem[Taylor \& Feng(2022)Taylor and Feng]{taylor2022deep}
Taylor, J. and Feng, M.
\newblock A deep learning model for forecasting global monthly mean sea surface
  temperature anomalies.
\newblock \emph{Frontiers in Climate}, 4:\penalty0 178, 2022.

\bibitem[Timmermann et~al.(2018)Timmermann, An, Kug, Jin, Cai, Capotondi, Cobb,
  Lengaigne, McPhaden, Stuecker, et~al.]{timmermann2018nino}
Timmermann, A., An, S.-I., Kug, J.-S., Jin, F.-F., Cai, W., Capotondi, A.,
  Cobb, K.~M., Lengaigne, M., McPhaden, M.~J., Stuecker, M.~F., et~al.
\newblock El ni{\~n}o--southern oscillation complexity.
\newblock \emph{Nature}, 559\penalty0 (7715):\penalty0 535--545, 2018.

\bibitem[Torgo et~al.(2013)Torgo, Ribeiro, Pfahringer, and
  Branco]{torgo2013smote}
Torgo, L., Ribeiro, R.~P., Pfahringer, B., and Branco, P.
\newblock {SMOTE} for regression.
\newblock In \emph{Proceedings of the Portuguese Conference on Artificial
  Intelligence}, pp.\  378--389. Springer, 2013.

\bibitem[Vecchi et~al.(2014)Vecchi, Delworth, Gudgel, Kapnick, Rosati,
  Wittenberg, Zeng, Anderson, Balaji, Dixon, et~al.]{vecchi2014seasonal}
Vecchi, G.~A., Delworth, T., Gudgel, R., Kapnick, S., Rosati, A., Wittenberg,
  A.~T., Zeng, F., Anderson, W., Balaji, V., Dixon, K., et~al.
\newblock On the seasonal forecasting of regional tropical cyclone activity.
\newblock \emph{Journal of Climate}, 27\penalty0 (21):\penalty0 7994--8016,
  2014.

\bibitem[Wu \& Tang(2019)Wu and Tang]{wu2019seasonal}
Wu, Y. and Tang, Y.
\newblock Seasonal predictability of the tropical indian ocean sst in the north
  american multimodel ensemble.
\newblock \emph{Climate Dynamics}, 53\penalty0 (5):\penalty0 3361--3372, 2019.

\bibitem[Yang \& Xu(2020)Yang and Xu]{yang2020rethinking}
Yang, Y. and Xu, Z.
\newblock Rethinking the value of labels for improving class-imbalanced
  learning.
\newblock \emph{Advances in neural information processing systems},
  33:\penalty0 19290--19301, 2020.

\bibitem[Yang et~al.(2021)Yang, Zha, Chen, Wang, and Katabi]{yang2021delving}
Yang, Y., Zha, K., Chen, Y., Wang, H., and Katabi, D.
\newblock Delving into deep imbalanced regression.
\newblock In \emph{Proceedings of the 38th International Conference on Machine
  Learning}, pp.\  11842--11851. PMLR, 2021.

\bibitem[Yin et~al.(2019)Yin, Yu, Sohn, Liu, and Chandraker]{yin2019feature}
Yin, X., Yu, X., Sohn, K., Liu, X., and Chandraker, M.
\newblock Feature transfer learning for face recognition with under-represented
  data.
\newblock In \emph{Proceedings of the IEEE/CVF Conference on Computer Vision
  and Pattern Recognition}, pp.\  5704--5713, 2019.

\bibitem[Zhang et~al.(2018)Zhang, Cisse, Dauphin, and
  Lopez-Paz]{zhang2017mixup}
Zhang, H., Cisse, M., Dauphin, Y.~N., and Lopez-Paz, D.
\newblock mixup: Beyond empirical risk minimization.
\newblock In \emph{International Conference on Learning Representations}, 2018.

\end{thebibliography}

\appendix

\section{Appendix}

\subsection{Histograms of SSTA Time Series}\label{apd:hist}

\begin{figure}[H]
\centering
\subfloat{\includegraphics[scale=0.45]{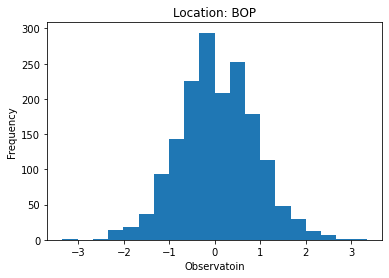}}
\subfloat{\includegraphics[scale=0.45]{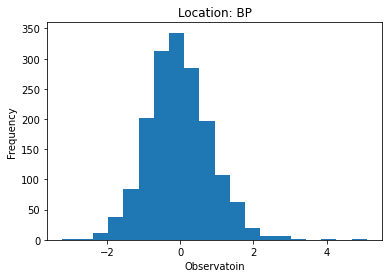}}\\
\subfloat{\includegraphics[scale=0.45]{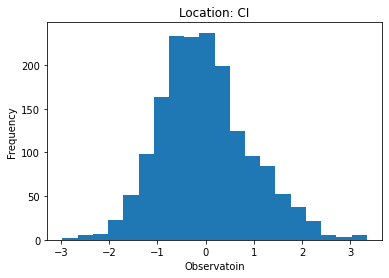}}
\subfloat{\includegraphics[scale=0.45]{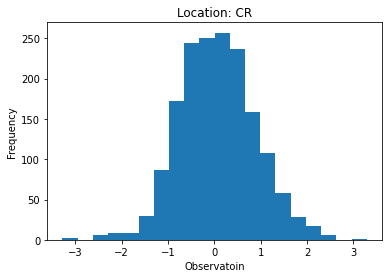}}\\
\subfloat{\includegraphics[scale=0.45]{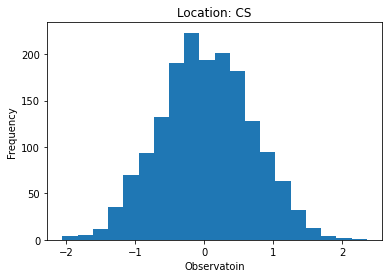}}
\subfloat{\includegraphics[scale=0.45]{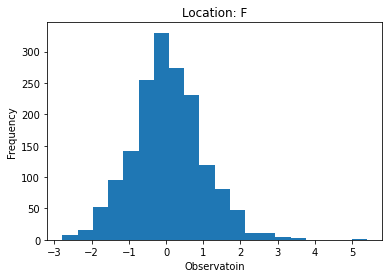}}
\caption{Histograms of the SSTA time series of the six selected NZ locations: Bay of Plenty, Bank Peninsula, Chatham Island, Cape Reinga, Cook Strait, and Fiordland.}
\end{figure}

\begin{figure}[H]
\centering
\subfloat{\includegraphics[scale=0.45]{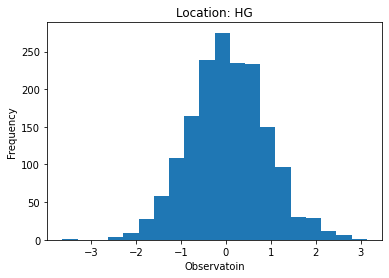}}
\subfloat{\includegraphics[scale=0.45]{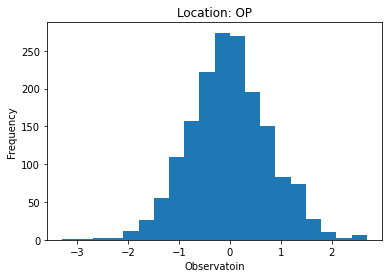}}\\
\subfloat{\includegraphics[scale=0.45]{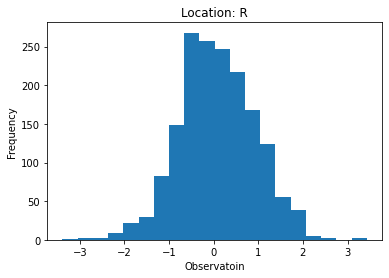}}
\subfloat{\includegraphics[scale=0.45]{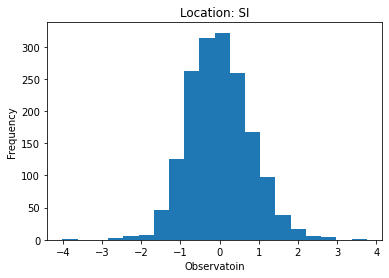}}\\
\subfloat{\includegraphics[scale=0.45]{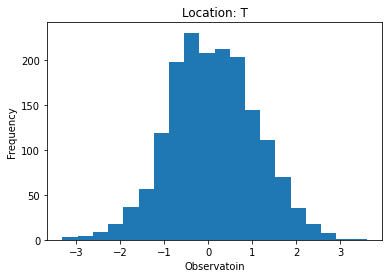}}
\subfloat{\includegraphics[scale=0.45]{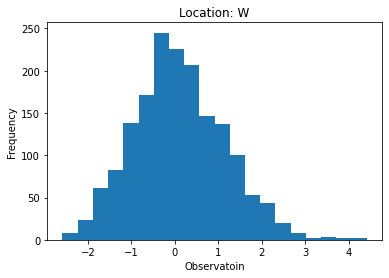}}
\caption{Histograms of the SSTA time series of the six selected NZ locations: Hauraki Gulf, Otago Peninsula, Raglan, Stewart Island, Taranaki, and Wairarapa.}
\end{figure}

\subsection{Additional Results}\label{apd:add}

\subsubsection{Two Lead Month Forecasts}\label{apd:two}

\begin{table}[H]
\centering
\small
\caption{Results of the two-lag models for two lead month SSTA and MHW forecasts.}
\begin{tabular}{llll}
\multicolumn{1}{c}{\textbf{Location}} & \multicolumn{1}{c}{\textbf{MSE↓}} & \multicolumn{1}{c}{\textbf{CSI↑}} & \multicolumn{1}{c}{\textbf{CSI 80↑}} \\ \hline
BOP                         & 0.3339                            & 0.2000                            & 0.3684                               \\
BP                        & 0.8608                            & 0.0556                            & 0.1477                               \\
CI                        & 0.7036                            & 0.0370                            & 0.1868                               \\
CR                           & 0.4255                            & 0.2078                            & 0.3636                               \\
CS                           & 0.2569                            & 0.2692                            & 0.3898                               \\
F                             & 0.7640                            & 0.3407                            & 0.4198                               \\
HG                          & 0.5289                            & 0.2812                            & 0.2581                               \\
\end{tabular}
\end{table}

\begin{table}[H]
\centering
\small
\caption{Results of the base models for two lead month SSTA and MHW forecasts. The results in bold are better than those of the corresponding fixed-lag model, applied to the
other tables in this study.}
\begin{tabular}{llllll}
\multicolumn{1}{c}{\textbf{Location}} & \multicolumn{1}{c}{\textbf{MSE↓}} & \multicolumn{1}{c}{\textbf{CSI↑}} & \multicolumn{1}{c}{\textbf{CSI 80↑}} & \multicolumn{1}{c}{\textbf{Training Time↓}} & \multicolumn{1}{c}{\textbf{PUR↓}} \\ \hline
BOP                         & \textbf{0.2997}                   & 0.0                               & 0.1818                               & 22.3693                            &                          \\
BP                        & \textbf{0.5798}                   & 0.0                               & 0.1324                               & 22.2763                            &                          \\
CI                        & \textbf{0.5285}                   & 0.0                               & 0.1406                               & 22.5064                            &                          \\
CR                           & \textbf{0.3740}                   & 0.0                               & 0.1600                               & 22.3860                            &                          \\
CS                           & \textbf{0.2327}                   & 0.0                               & 0.3426                               & 22.4155                            &                          \\
F                             & \textbf{0.6498}                   & 0.0                               & 0.3084                               & 22.3326                            & 0\%                      \\
HG                          & \textbf{0.4216}                   & 0.0                               & 0.2258                               & 22.2642                            &                          \\
\end{tabular}
\end{table}

\begin{table}[H]
\centering
\small
\caption{Results of the models with the MAE loss for two lead month SSTA and MHW forecasts.}
\begin{tabular}{llllll}
\multicolumn{1}{c}{\textbf{Location}} & \multicolumn{1}{c}{\textbf{MSE↓}} & \multicolumn{1}{c}{\textbf{CSI↑}} & \multicolumn{1}{c}{\textbf{CSI 80↑}} & \multicolumn{1}{c}{\textbf{Training Time↓}} & \multicolumn{1}{c}{\textbf{PUR↓}} \\ \hline
BOP                         & 0.4893                            & 0.0                               & 0.0                                  & 18.2881                                     & 0\%                               \\
BP                        & \textbf{0.6074}                   & 0.0                               & 0.0                                  & 18.4212                                     & 100\%                             \\
CI                        & \textbf{0.6224}                   & 0.0                               & 0.0                                  & 18.3987                                     & 100\%                             \\
CR                           & 0.5773                            & 0.0                               & 0.0                                  & 18.6549                                     & 100\%                             \\
CS                           & 0.3282                            & 0.0                               & 0.0                                  & 18.3463                                     & 80\%                              \\
F                             & 0.8867                            & 0.0                               & 0.0                                  & 18.3205                                     & 100\%                             \\
HG                          & 0.5620                            & 0.0                               & 0.0                                  & 18.1188                                     & 100\%                             \\
\end{tabular}
\end{table}

\begin{table}[H]
\centering
\small
\caption{Results of the models with the Huber loss with $\delta=0.5$ for two lead month SSTA and MHW forecasts.}
\begin{tabular}{llllll}
\multicolumn{1}{c}{\textbf{Location}} & \multicolumn{1}{c}{\textbf{MSE↓}} & \multicolumn{1}{c}{\textbf{CSI↑}} & \multicolumn{1}{c}{\textbf{CSI 80↑}} & \multicolumn{1}{c}{\textbf{Training Time↓}} & \multicolumn{1}{c}{\textbf{PUR↓}} \\ \hline
BOP                         & 0.3595                            & 0.0312                            & 0.1900                               & 20.7981                                     &                                   \\
BP                        & \textbf{0.6079}                   & 0.0500                            & 0.1186                               & 20.6532                                     &                                   \\
CI                        & \textbf{0.6373}                   & \textbf{0.0667}                   & 0.0690                               & 21.5118                                     &                                   \\
CR                           & 0.5063                            & 0.0213                            & 0.1364                               & 20.7767                                     &                                   \\
CS                           & 0.3087                            & 0.2045                            & 0.3008                               & 20.9256                                     &                                   \\
F                             & 0.8751                            & 0.1077                            & 0.1321                               & 22.2434                                     & 0\%                               \\
HG                          & \textbf{0.4807}                   & 0.1489                            & 0.2566                               & 21.5380                                     &                                   \\
\end{tabular}
\end{table}

\begin{table}[H]
\centering
\small
\caption{Results of the models with the weighted MSE loss with $w_{90\%}=1.5$ and $w_{80\%}=1.25$ for two lead month SSTA and MHW forecasts.}
\begin{tabular}{llllll}
\multicolumn{1}{c}{\textbf{Location}} & \multicolumn{1}{c}{\textbf{MSE↓}} & \multicolumn{1}{c}{\textbf{CSI↑}} & \multicolumn{1}{c}{\textbf{CSI 80↑}} & \multicolumn{1}{c}{\textbf{Training Time↓}} & \multicolumn{1}{c}{\textbf{PUR↓}} \\ \hline
BOP                         & \textbf{0.3087}                   & 0.0                               & 0.1889                               & 30.1683                            &                          \\
BP                        & \textbf{0.5794}                   & 0.0                               & 0.1343                               & 30.7600                            &                          \\
CI                        & \textbf{0.5264}                   & 0.0                               & 0.1406                               & 30.9739                            &                          \\
CR                           & \textbf{0.3774}                   & 0.0                               & 0.1782                               & 30.7878                            &                          \\
CS                           & \textbf{0.2322}                   & 0.0                               & 0.3426                               & 30.8247                            &                          \\
F                             & \textbf{0.6497}                   & 0.0                               & 0.3178                               & 30.3270                            & 0\%                      \\
HG                          & \textbf{0.4262}                   & 0.0                               & 0.2083                               & 30.7889                            &                          \\
\end{tabular}
\end{table}

\begin{table}[H]
\centering
\small
\caption{Results of the models with the focal-R-MSE loss with $\beta=2$ and $\theta=1$ for two lead month SSTA and MHW forecasts.}
\begin{tabular}{llllll}
\multicolumn{1}{c}{\textbf{Location}} & \multicolumn{1}{c}{\textbf{MSE↓}} & \multicolumn{1}{c}{\textbf{CSI↑}} & \multicolumn{1}{c}{\textbf{CSI 80↑}} & \multicolumn{1}{c}{\textbf{Training Time↓}} & \multicolumn{1}{c}{\textbf{PUR↓}} \\ \hline
BOP                         & \textbf{0.2952}                   & 0.0                               & 0.1818                               & 26.9496                                     &                                   \\
BP                        & \textbf{0.5826}                   & 0.0                               & 0.1343                               & 27.1529                                     &                                   \\
CI                        & \textbf{0.5315}                   & 0.0                               & 0.1343                               & 26.2288                                     &                                   \\
CR                           & \textbf{0.3713}                   & 0.0                               & 0.1531                               & 25.1065                                     &                                   \\
CS                           & \textbf{0.2322}                   & 0.0                               & 0.3333                               & 26.9141                                     &                                   \\
F                             & \textbf{0.6489}                   & 0.0                               & 0.3271                               & 26.5709                                     & 0\%                               \\
HG                          & \textbf{0.4210}                   & 0.0                               & 0.2022                               & 26.2504                                     &                                   \\
\end{tabular}
\end{table}

\begin{table}[H]
\centering
\small
\caption{Results of the models with the balanced MSE loss with optimized $\sigma$ for two lead month SSTA and MHW forecasts.}
\begin{tabular}{llllll}
\multicolumn{1}{c}{\textbf{Location}} & \multicolumn{1}{c}{\textbf{MSE↓}} & \multicolumn{1}{c}{\textbf{CSI↑}} & \multicolumn{1}{c}{\textbf{CSI 80↑}} & \multicolumn{1}{c}{\textbf{Training Time↓}} & \multicolumn{1}{c}{\textbf{PUR↓}} \\ \hline
BOP                         & 0.8127                            & 0.1429                            & 0.2744                               & 25.3428                                     & 0\%                               \\
BP                        & 1.4049                            & \textbf{0.1404}                   & \textbf{0.1765}                      & 25.9395                                     & 0\%                               \\
CI                        & 1.1446                            & \textbf{0.0727}                   & \textbf{0.2371}                      & 24.3099                                     & 0\%                               \\
CR                           & 0.6960                            & 0.1786                            & 0.2877                               & 24.9177                                     & 20\%                              \\
CS                           & 0.3438                            & 0.2152                            & 0.3566                               & 26.4798                                     & 0\%                               \\
F                             & 1.3362                            & 0.3265                            & \textbf{0.4275}                      & 25.0495                                     & 0\%                               \\
HG                          & 1.0137                            & 0.1604                            & 0.2455                               & 24.8199                                     & 0\%                               \\
\end{tabular}
\end{table}

\begin{table}[H]
\centering
\small
\caption{Results of the models with the scaling-weighted MSE loss with $\alpha=1.5$, $\beta=0.5$, $w_{90\%}=1.5$, and $w_{80\%}=1.25$ for two lead month SSTA and MHW forecasts.}
\begin{tabular}{llllll}
\multicolumn{1}{c}{\textbf{Location}} & \multicolumn{1}{c}{\textbf{MSE↓}} & \multicolumn{1}{c}{\textbf{CSI↑}} & \multicolumn{1}{c}{\textbf{CSI 80↑}} & \multicolumn{1}{c}{\textbf{Training Time↓}} & \multicolumn{1}{c}{\textbf{PUR↓}} \\ \hline
BOP                         & 0.4530                            & \textbf{0.2143}                   & 0.2699                               & 40.2966                                     &                                   \\
BP                        & \textbf{0.7674}                   & \textbf{0.0930}                   & \textbf{0.1724}                      & 40.3910                                     &                                   \\
CI                        & \textbf{0.6338}                   & \textbf{0.0769}                   & \textbf{0.2048}                      & 40.6418                                     &                                   \\
CR                           & 0.4785                            & 0.1167                            & 0.2938                               & 39.6368                                     &                                   \\
CS                           & 0.2821                            & 0.2143                            & 0.3592                               & 41.0330                                     &                                   \\
F                             & 0.8256                            & 0.2796                            & \textbf{0.4462}                      & 40.0886                                     & 0\%                               \\
HG                          & 0.5925                            & \textbf{0.2051}                   & \textbf{0.2629}                      & 41.3833                                     &                                   \\
\end{tabular}
\end{table}

\begin{table}[H]
\centering
\small
\caption{Results of the models with the scaling-weighted MSE loss with $\alpha=2$, $\beta=0.5$, $w_{90\%}=1.5$, and $w_{80\%}=1.25$ for two lead month SSTA and MHW forecasts.}
\begin{tabular}{llllll}
\multicolumn{1}{c}{\textbf{Location}} & \multicolumn{1}{c}{\textbf{MSE↓}} & \multicolumn{1}{c}{\textbf{CSI↑}} & \multicolumn{1}{c}{\textbf{CSI 80↑}} & \multicolumn{1}{c}{\textbf{Training Time↓}} & \multicolumn{1}{c}{\textbf{PUR↓}} \\ \hline
BOP                         & 0.5167                            & \textbf{0.2241}                   & 0.2743                               & 29.4121                                     &                                   \\
BP                        & \textbf{0.7908}                   & 0.0263                            & \textbf{0.1538}                      & 29.7531                                     &                                   \\
CI                        & \textbf{0.6844}                   & \textbf{0.1200}                   & \textbf{0.2195}                      & 29.5277                                     &                                   \\
CR                           & 0.5194                            & 0.1757                            & 0.3214                               & 29.7048                                     &                                   \\
CS                           & 0.3011                            & 0.2121                            & 0.3517                               & 29.9427                                     &                                   \\
F                             & 0.9411                            & \textbf{0.3535}                   & \textbf{0.4538}                      & 29.6311                                     & 0\%                               \\
HG                          & 0.6719                            & 0.1765                            & \textbf{0.2713}                      & 29.9946                                     &                                   \\
\end{tabular}
\end{table}

\begin{table}[H]
\centering
\small
\caption{Results of the models with the scaling-weighted MSE loss with $\alpha=2$, $\beta=1$, $w_{90\%}=1.5$, and $w_{80\%}=1.25$ for two lead month SSTA and MHW forecasts.}
\begin{tabular}{llllll}
\multicolumn{1}{c}{\textbf{Location}} & \multicolumn{1}{c}{\textbf{MSE↓}} & \multicolumn{1}{c}{\textbf{CSI↑}} & \multicolumn{1}{c}{\textbf{CSI 80↑}} & \multicolumn{1}{c}{\textbf{Training Time↓}} & \multicolumn{1}{c}{\textbf{PUR↓}} \\ \hline
BOP                         & 0.8973                            & 0.1452                            & 0.2714                               & 24.3832                                     &                                   \\
BP                        & 1.2872                            & \textbf{0.0769}                   & 0.1441                               & 24.7838                                     &                                   \\
CI                        & 0.9448                            & \textbf{0.0833}                   & \textbf{0.2381}                      & 24.5354                                     &                                   \\
CR                           & 0.9383                            & 0.1304                            & 0.2857                               & 24.4787                                     &                                   \\
CS                           & 0.4038                            & 0.2062                            & 0.3354                               & 24.7088                                     &                                   \\
F                             & 1.3063                            & 0.2581                            & 0.4154                               & 24.7834                                     & 0\%                               \\
HG                          & 0.8998                            & 0.1610                            & 0.2316                               & 24.5456                                     &                                   \\
\end{tabular}
\end{table}

\subsubsection{Three Lead Month Forecasts}\label{apd:three}

\begin{table}[H]
\centering
\footnotesize
\caption{Results of the three-lag models for three lead month SSTA and MHW forecasts.}
\begin{tabular}{llll}
\multicolumn{1}{c}{\textbf{Location}} & \multicolumn{1}{c}{\textbf{MSE↓}} & \multicolumn{1}{c}{\textbf{CSI↑}} & \multicolumn{1}{c}{\textbf{CSI 80↑}} \\ \hline
BOP                         & 0.4076                            & 0.1250                            & 0.3109                               \\
BP                        & 1.0517                            & 0.0                               & 0.0860                               \\
CI                        & 0.9345                            & 0.0370                            & 0.1134                               \\
CR                           & 0.5004                            & 0.1625                            & 0.3358                               \\
CS                           & 0.3439                            & 0.1379                            & 0.3016                               \\
F                             & 1.0635                            & 0.2079                            & 0.3191                               \\
HG                          & 0.6400                            & 0.2239                            & 0.2480                               \\
\end{tabular}
\end{table}

\begin{table}[H]
\centering
\small
\caption{Results of the base models for three lead month SSTA and MHW forecasts.}
\begin{tabular}{llllll}
\multicolumn{1}{c}{\textbf{Location}} & \multicolumn{1}{c}{\textbf{MSE↓}} & \multicolumn{1}{c}{\textbf{CSI↑}} & \multicolumn{1}{c}{\textbf{CSI 80↑}} & \multicolumn{1}{c}{\textbf{Training Time↓}} & \multicolumn{1}{c}{\textbf{PUR↓}} \\ \hline
BOP                         & \textbf{0.1528}                   & 0.0357                            & 0.4211                               & 18.3887                                     &                                   \\
BP                        & \textbf{0.3831}                   & 0.1600                            & 0.2899                               & 18.3435                                     &                                   \\
CI                        & \textbf{0.2592}                   & 0.1765                            & 0.3824                               & 18.6351                                     &                                   \\
CR                           & \textbf{0.2327}                   & 0.1373                            & 0.4224                               & 18.6882                                     &                                   \\
CS                           & \textbf{0.1235}                   & 0.3143                            & 0.4737                               & 18.4263                                     &                                   \\
F                             & \textbf{0.3150}                   & 0.3611                            & \textbf{0.5752}                      & 18.3620                                     & 0\%                               \\
HG                          & \textbf{0.2349}                   & 0.2045                            & \textbf{0.3832}                      & 18.5357                                     &                                   \\
\end{tabular}
\end{table}

\begin{table}[H]
\centering
\small
\caption{Results of the models with the MAE loss for three lead month SSTA and MHW forecasts.}
\begin{tabular}{llllll}
\multicolumn{1}{c}{\textbf{Location}} & \multicolumn{1}{c}{\textbf{MSE↓}} & \multicolumn{1}{c}{\textbf{CSI↑}} & \multicolumn{1}{c}{\textbf{CSI 80↑}} & \multicolumn{1}{c}{\textbf{Training Time↓}} & \multicolumn{1}{c}{\textbf{PUR↓}} \\ \hline
BOP                         & 0.4894                            & 0.0                               & 0.0                                  & 18.4441                                     &                                   \\
BP                        & \textbf{0.6071}                   & 0.0                               & 0.0                                  & 18.1606                                     &                                   \\
CI                        & \textbf{0.6225}                   & 0.0                               & 0.0                                  & 19.0742                                     &                                   \\
CR                           & 0.5772                            & 0.0                               & 0.0                                  & 18.1971                                     &                                   \\
CS                           & 0.3861                            & 0.0                               & 0.0                                  & 18.1512                                     &                                   \\
F                             & \textbf{0.8866}                   & 0.0                               & 0.0                                  & 18.7150                                     & 100\%                             \\
HG                          & \textbf{0.5616}                   & 0.0                               & 0.0                                  & 18.6446                                     &                                   \\
\end{tabular}
\end{table}

\begin{table}[H]
\centering
\small
\caption{Results of the models with the Huber loss with $\delta=0.5$ for three lead month SSTA and MHW forecasts.}
\begin{tabular}{llllll}
\multicolumn{1}{c}{\textbf{Location}} & \multicolumn{1}{c}{\textbf{MSE↓}} & \multicolumn{1}{c}{\textbf{CSI↑}} & \multicolumn{1}{c}{\textbf{CSI 80↑}} & \multicolumn{1}{c}{\textbf{Training Time↓}} & \multicolumn{1}{c}{\textbf{PUR↓}} \\ \hline
BOP                         & 0.4743                            & 0.0                               & 0.0957                               & 23.1707                                     &                                   \\
BP                        & \textbf{0.7335}                   & 0.0                               & 0.0149                               & 21.8572                                     &                                   \\
CI                        & \textbf{0.6773}                   & 0.0                               & 0.0308                               & 22.2779                                     &                                   \\
CR                           & 0.5270                            & 0.0185                            & 0.1565                               & 21.9849                                     &                                   \\
CS                           & 0.3624                            & 0.0                               & 0.1453                               & 20.8197                                     &                                   \\
F                             & \textbf{1.0469}                   & 0.0492                            & 0.1031                               & 22.3711                                     & 0\%                               \\
HG                          & \textbf{0.6138}                   & 0.0                               & 0.1402                               & 22.2079                                     &                                   \\
\end{tabular}
\end{table}

\begin{table}[H]
\centering
\small
\caption{Results of the models with the weighted MSE loss with $w_{90\%}=1.5$ and $w_{80\%}=1.25$ for three lead month SSTA and MHW forecasts.}
\begin{tabular}{llllll}
\multicolumn{1}{c}{\textbf{Location}} & \multicolumn{1}{c}{\textbf{MSE↓}} & \multicolumn{1}{c}{\textbf{CSI↑}} & \multicolumn{1}{c}{\textbf{CSI 80↑}} & \multicolumn{1}{c}{\textbf{Training Time↓}} & \multicolumn{1}{c}{\textbf{PUR↓}} \\ \hline
BOP                         & \textbf{0.3554}                   & 0.0                               & 0.0588                               & 31.1222                            & 0\%                      \\
BP                        & \textbf{0.6231}                   & 0.0                               & 0.0                                  & 30.7661                            & 100\%                    \\
CI                        & \textbf{0.6123}                   & 0.0                               & 0.0161                               & 31.1717                            & 0\%                      \\
CR                           & 0.5545                            & 0.0                               & 0.0                                  & 30.8064                            & 100\%                    \\
CS                           & \textbf{0.2825}                   & 0.0                               & 0.1923                               & 30.7800                            & 0\%                      \\
F                             & \textbf{0.8189}                   & 0.0                               & 0.1188                               & 31.3309                            & 0\%                      \\
HG                          & \textbf{0.4695}                   & 0.0                               & 0.1124                               & 31.5299                            & 0\%                      \\
\end{tabular}
\end{table}

\begin{table}[H]
\centering
\small
\caption{Results of the models with the focal-R-MSE loss with $\beta=2$ and $\theta=1$ for three lead month SSTA and MHW forecasts.}
\begin{tabular}{llllll}
\multicolumn{1}{c}{\textbf{Location}} & \multicolumn{1}{c}{\textbf{MSE↓}} & \multicolumn{1}{c}{\textbf{CSI↑}} & \multicolumn{1}{c}{\textbf{CSI 80↑}} & \multicolumn{1}{c}{\textbf{Training Time↓}} & \multicolumn{1}{c}{\textbf{PUR↓}} \\ \hline
BOP                         & \textbf{0.3450}                   & 0.0                               & 0.0                                  & 26.2805                                     & 20\%                              \\
BP                        & \textbf{0.6225}                   & 0.0                               & 0.0                                  & 25.0737                                     & 100\%                             \\
CI                        & \textbf{0.6128}                   & 0.0                               & 0.0333                               & 26.5760                                     & 0\%                               \\
CR                           & 0.5608                            & 0.0                               & 0.0                                  & 26.6553                                     & 100\%                             \\
CS                           & \textbf{0.3387}                   & 0.0                               & 0.0                                  & 25.3281                                     & 80\%                              \\
F                             & \textbf{0.8349}                   & 0.0                               & 0.1000                               & 27.4149                                     & 0\%                               \\
HG                          & \textbf{0.4613}                   & 0.0                               & 0.0                                  & 23.9966                                     & 40\%                              \\
\end{tabular}
\end{table}

\begin{table}[H]
\centering
\small
\caption{Results of the models with the balanced MSE loss with optimized $\sigma$ for three lead month SSTA and MHW forecasts.}
\begin{tabular}{llllll}
\multicolumn{1}{c}{\textbf{Location}} & \multicolumn{1}{c}{\textbf{MSE↓}} & \multicolumn{1}{c}{\textbf{CSI↑}} & \multicolumn{1}{c}{\textbf{CSI 80↑}} & \multicolumn{1}{c}{\textbf{Training Time↓}} & \multicolumn{1}{c}{\textbf{PUR↓}} \\ \hline
BOP                         & 0.4532                            & 0.0                               & 0.0                                  & 25.4122                                     & 100\%                             \\
BP                        & \textbf{0.6744}                   & 0.0                               & 0.0                                  & 25.8747                                     & 100\%                             \\
CI                        & 1.0443                            & \textbf{0.0417}                   & \textbf{0.1489}                      & 26.6697                                     & 0\%                               \\
CR                           & 0.5555                            & 0.0                               & 0.0                                  & 25.2941                                     & 100\%                             \\
CS                           & 0.3751                            & 0.0                               & 0.0                                  & 25.7201                                     & 100\%                             \\
F                             & 2.0951                            & \textbf{0.2128}                   & 0.2920                               & 26.5994                                     & 0\%                               \\
HG                          & \textbf{0.5513}                   & 0.0                               & 0.0                                  & 26.8769                                     & 100\%                             \\
\end{tabular}
\end{table}

\begin{table}[H]
\centering
\small
\caption{Results of the models with the scaling-weighted MSE loss with $\alpha=1.5$, $\beta=0.5$, $w_{90\%}=1.5$, and $w_{80\%}=1.25$ for three lead month SSTA and MHW forecasts.}
\begin{tabular}{llllll}
\multicolumn{1}{c}{\textbf{Location}} & \multicolumn{1}{c}{\textbf{MSE↓}} & \multicolumn{1}{c}{\textbf{CSI↑}} & \multicolumn{1}{c}{\textbf{CSI 80↑}} & \multicolumn{1}{c}{\textbf{Training Time↓}} & \multicolumn{1}{c}{\textbf{PUR↓}} \\ \hline
BOP                         & 0.5440                            & \textbf{0.1286}                   & 0.2590                               & 39.5022                                     &                                   \\
BP                        & \textbf{0.7200}                   & 0.0                               & 0.0714                               & 38.9915                                     &                                   \\
CI                        & \textbf{0.6768}                   & 0.0                               & 0.0588                               & 40.6756                                     &                                   \\
CR                           & 0.5420                            & 0.0                               & 0.2877                               & 38.4012                                     &                                   \\
CS                           & \textbf{0.3256}                   & 0.0652                            & 0.2721                               & 38.1835                                     &                                   \\
F                             & 1.1251                            & 0.1358                            & 0.2595                               & 38.7145                                     & 0\%                               \\
HG                          & 0.6555                            & 0.0921                            & 0.2460                               & 38.9647                                     &                                   \\
\end{tabular}
\end{table}

\begin{table}[H]
\centering
\small
\caption{Results of the models with the scaling-weighted MSE loss with $\alpha=2$, $\beta=0.5$, $w_{90\%}=1.5$, and $w_{80\%}=1.25$ for three lead month SSTA and MHW forecasts.}
\begin{tabular}{llllll}
\multicolumn{1}{c}{\textbf{Location}} & \multicolumn{1}{c}{\textbf{MSE↓}} & \multicolumn{1}{c}{\textbf{CSI↑}} & \multicolumn{1}{c}{\textbf{CSI 80↑}} & \multicolumn{1}{c}{\textbf{Training Time↓}} & \multicolumn{1}{c}{\textbf{PUR↓}} \\ \hline
BOP                         & 0.5167                            & \textbf{0.2241}                   & 0.2743                               & 29.4121                                     &                                   \\
BP                        & \textbf{0.7908}                   & 0.0263                            & \textbf{0.1538}                      & 29.7531                                     &                                   \\
CI                        & \textbf{0.6844}                   & \textbf{0.1200}                   & \textbf{0.2195}                      & 29.5277                                     &                                   \\
CR                           & 0.5194                            & 0.1757                            & 0.3214                               & 29.7048                                     &                                   \\
CS                           & 0.3011                            & 0.2121                            & 0.3517                               & 29.9427                                     &                                   \\
F                             & 0.9411                            & \textbf{0.3535}                   & \textbf{0.4538}                      & 29.6311                                     & 0\%                               \\
HG                          & 0.6719                            & 0.1765                            & \textbf{0.2713}                      & 29.9946                                     &                                   \\
\end{tabular}
\end{table}

\begin{table}[H]
\centering
\small
\caption{Results of the models with the scaling-weighted MSE loss with $\alpha=2$, $\beta=1$, $w_{90\%}=1.5$, and $w_{80\%}=1.25$ for three lead month SSTA and MHW forecasts.}
\begin{tabular}{llllll}
\multicolumn{1}{c}{\textbf{Location}} & \multicolumn{1}{c}{\textbf{MSE↓}} & \multicolumn{1}{c}{\textbf{CSI↑}} & \multicolumn{1}{c}{\textbf{CSI 80↑}} & \multicolumn{1}{c}{\textbf{Training Time↓}} & \multicolumn{1}{c}{\textbf{PUR↓}} \\ \hline
BOP                         & 0.8963                            & \textbf{0.1429}                   & 0.2771                               & 24.4695                                     &                                   \\
BP                        & 1.4682                            & \textbf{0.0714}                   & 0.0806                               & 24.3445                                     &                                   \\
CI                        & 1.1323                            & \textbf{0.0455}                   & 0.0957                               & 24.4202                                     &                                   \\
CR                           & 1.0557                            & 0.1224                            & 0.2593                               & 24.5512                                     &                                   \\
CS                           & 0.4999                            & \textbf{0.1712}                   & \textbf{0.3254}                      & 24.4688                                     &                                   \\
F                             & 1.4537                            & \textbf{0.2447}                   & \textbf{0.3623}                      & 24.7630                                     & 0\%                               \\
HG                          & 1.0358                            & 0.1074                            & 0.1949                               & 24.5973                                     &                                   \\
\end{tabular}
\end{table}

\subsubsection{Six Lead Month Forecasts}\label{apd:six}

\begin{table}[H]
\centering
\small
\caption{Results of the six-lag models for six lead month SSTA and MHW forecasts.}
%
\end{table}

\begin{table}[H]
\centering
\small
\caption{Results of the base models for six lead month SSTA and MHW forecasts.}
%
\end{table}

\begin{table}[H]
\centering
\small
\caption{Results of the models with the MAE loss for six lead month SSTA and MHW forecasts.}
%
\end{table}

\begin{table}[H]
\centering
\small
\caption{Results of the models with the Huber loss with $\delta=0.5$ for six lead month SSTA and MHW forecasts.}
%
\end{table}

\begin{table}[H]
\centering
\small
\caption{Results of the models with the weighted MSE loss with $w_{90\%}=1.5$ and $w_{80\%}=1.25$ for six lead month SSTA and MHW forecasts.}
%
\end{table}

\begin{table}[H]
\centering
\small
\caption{Results of the models with the focal-R-MSE loss with $\beta=2$ and $\theta=1$ for six lead month SSTA and MHW forecasts.}
%
\end{table}

\begin{table}[H]
\centering
\small
\caption{Results of the models with the balanced MSE loss with optimized $\sigma$ for six lead month SSTA and MHW forecasts.}
%
\end{table}

\begin{table}[H]
\centering
\small
\caption{Results of the models with the scaling-weighted MSE loss with $\alpha=1.5$, $\beta=0.5$, $w_{90\%}=1.5$, and $w_{80\%}=1.25$ for six lead month SSTA and MHW forecasts.}
%
\end{table}

\begin{table}[H]
\centering
\small
\caption{Results of the models with the scaling-weighted MSE loss with $\alpha=2$, $\beta=0.5$, $w_{90\%}=1.5$, and $w_{80\%}=1.25$ for six lead month SSTA and MHW forecasts.}
%
\end{table}

\begin{table}[H]
\centering
\small
\caption{Results of the models with the scaling-weighted MSE loss with $\alpha=2$, $\beta=1$, $w_{90\%}=1.5$, and $w_{80\%}=1.25$ for six lead month SSTA and MHW forecasts.}
%
\end{table}

\subsubsection{Summary of the Results}\label{apd:summary}

\begin{table}[H]
\centering
\scriptsize
\caption{12 selected locations with the corresponding results of monthly SSTA and MHW forecasts (Part 1). In each cell for a location and a lead time, the three lines show the models with different loss functions that outperformed the persistence or fixed-lag model in terms of the MSE for SSTA forecasts, the CSI for MHW forecasts, and the CSI 80 for suspected MHW forecasts, with the PUR not greater than 20\%.}\label{tb:all_results}
%
\end{table}

\begin{table}[H]
\centering
\footnotesize
\caption{12 selected locations with the corresponding results of monthly SSTA and MHW forecasts (Part 2).}
%
\end{table}

\begin{table}[H]
\centering
\footnotesize
\caption{12 selected locations with the corresponding results of monthly SSTA and MHW forecasts (Part 3).}
%
\end{table}

\begin{table}[H]
\centering
\footnotesize
\caption{12 selected locations with the corresponding results of monthly SSTA and MHW forecasts (Part 4).}
%
\end{table}

\subsection{Example Scatter Plots}\label{apd:scatter}

\subsubsection{One Lead Month Forecasts}

\begin{figure}[H]
\centering
\subfloat{\includegraphics[scale=0.45]{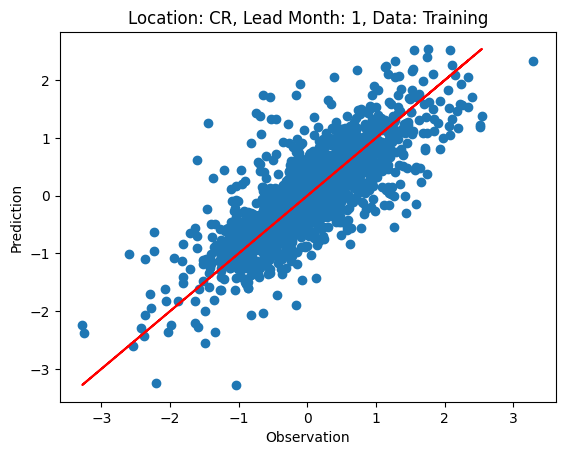}}
\subfloat{\includegraphics[scale=0.45]{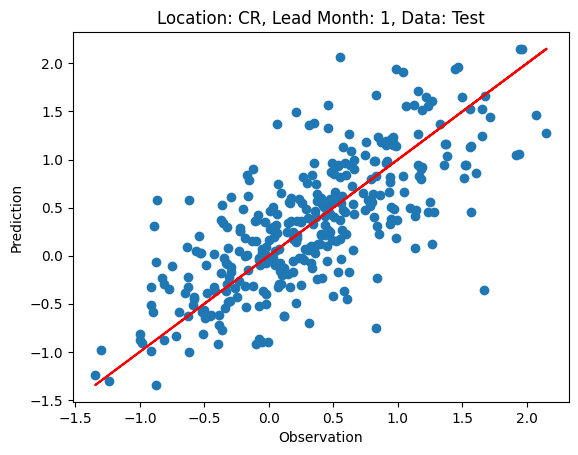}}
\caption{SSTA observation and prediction scatter plots of the persistence model for one lead month forecasts at Cape Reinga, on the training and test data respectively.}
\end{figure}

\begin{figure}[H]
\centering
\subfloat{\includegraphics[scale=0.45]{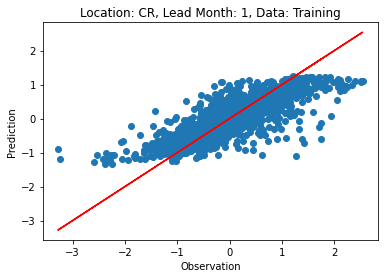}}
\subfloat{\includegraphics[scale=0.45]{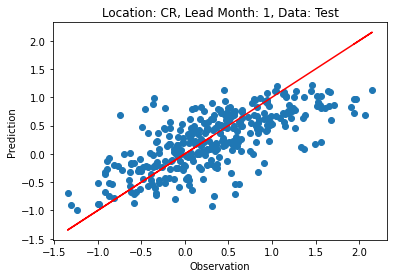}}\\
\subfloat{\includegraphics[scale=0.45]{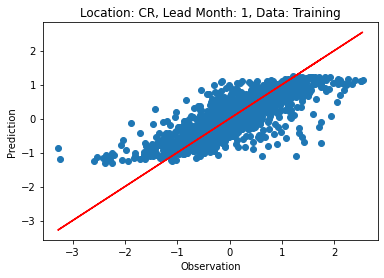}}
\subfloat{\includegraphics[scale=0.45]{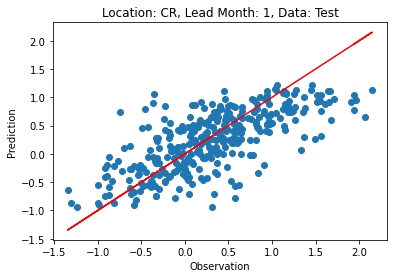}}\\
\subfloat{\includegraphics[scale=0.45]{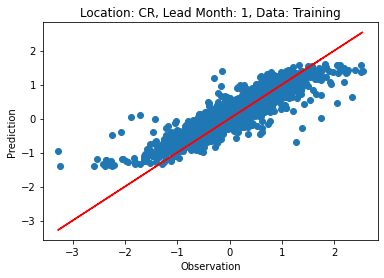}}
\subfloat{\includegraphics[scale=0.45]{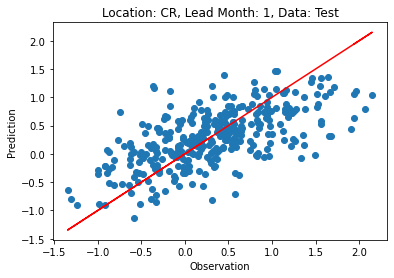}}
\caption{SSTA observation and prediction scatter plots of the last model for one lead month forecasts at Cape Reinga, on the training and test data respectively. The corresponding loss functions from top to bottom are the MSE, the MAE, and the Huber.}
\end{figure}

\begin{figure}[H]
\centering
\subfloat{\includegraphics[scale=0.45]{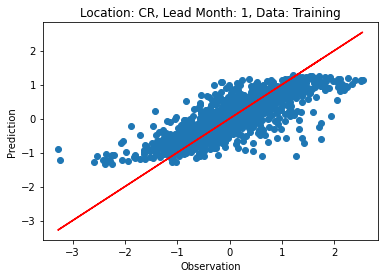}}
\subfloat{\includegraphics[scale=0.45]{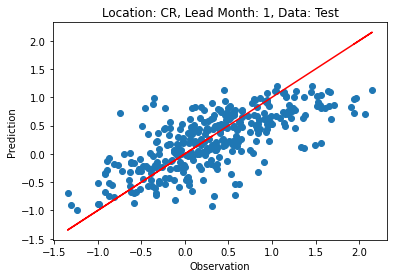}}\\
\subfloat{\includegraphics[scale=0.45]{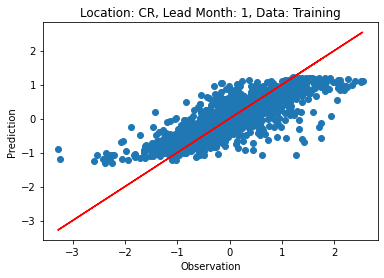}}
\subfloat{\includegraphics[scale=0.45]{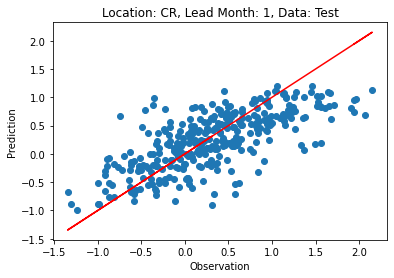}}\\
\subfloat{\includegraphics[scale=0.45]{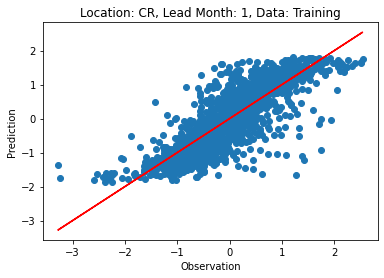}}
\subfloat{\includegraphics[scale=0.45]{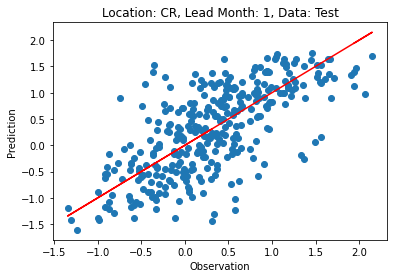}}
\caption{SSTA observation and prediction scatter plots of the last model for one lead month forecasts at Cape Reinga, on the training and test data respectively. The corresponding loss functions from top to bottom are the weighted MSE, the focal-R, and the balanced MSE.}
\end{figure}

\begin{figure}[H]
\centering
\subfloat{\includegraphics[scale=0.45]{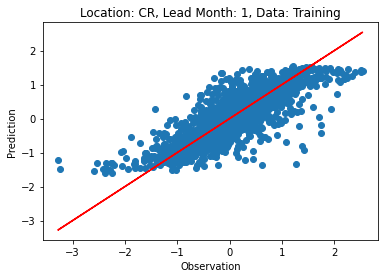}}
\subfloat{\includegraphics[scale=0.45]{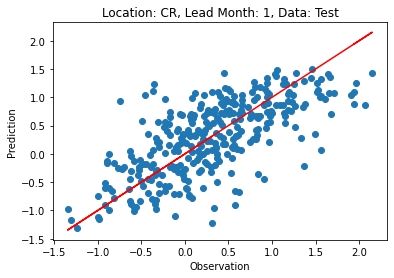}}\\
\subfloat{\includegraphics[scale=0.45]{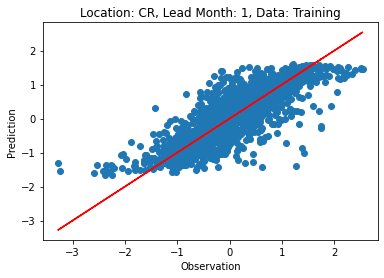}}
\subfloat{\includegraphics[scale=0.45]{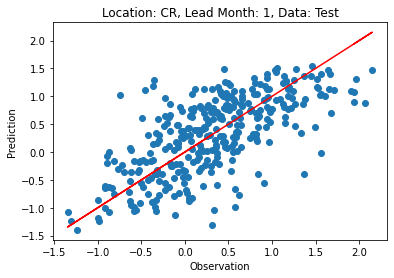}}\\
\subfloat{\includegraphics[scale=0.45]{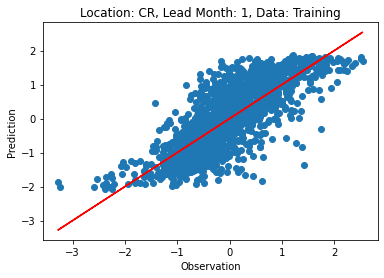}}
\subfloat{\includegraphics[scale=0.45]{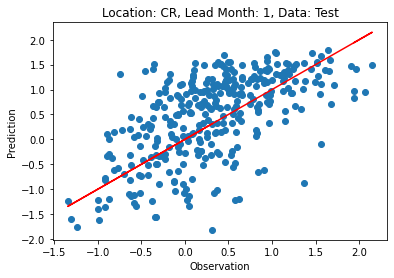}}
\caption{SSTA observation and prediction scatter plots of the last model for one lead month forecasts at Cape Reinga, on the training and test data respectively. The corresponding loss function is our proposed scaling-weighted MSE, from top to bottom with the hyperparameters $\alpha=1.5$, $\beta=0.5$, $\alpha=2$, $\beta=0.5$, and $\alpha=2$, $\beta=1$ respectively.}
\end{figure}

\begin{figure}[H]
\centering
\subfloat{\includegraphics[scale=0.45]{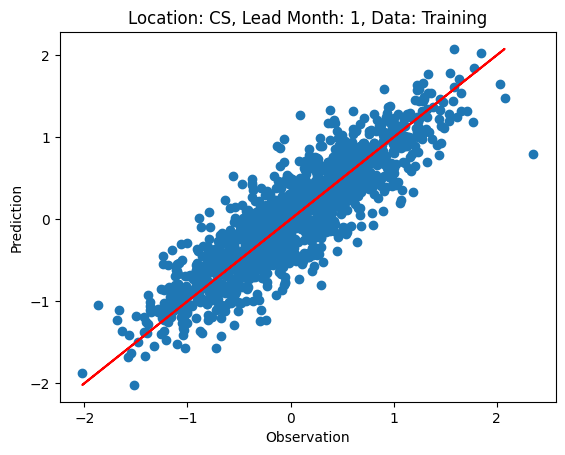}}
\subfloat{\includegraphics[scale=0.45]{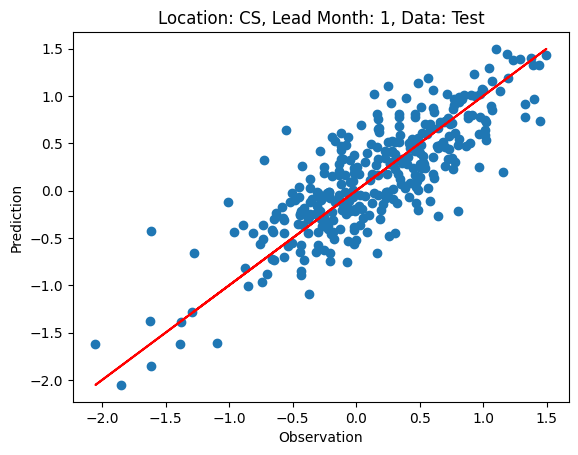}}
\caption{SSTA observation and prediction scatter plots of the persistence model for one lead month forecasts at Cook Strait, on the training and test data respectively.}
\end{figure}

\begin{figure}[H]
\centering
\subfloat{\includegraphics[scale=0.45]{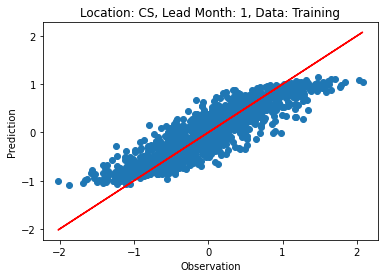}}
\subfloat{\includegraphics[scale=0.45]{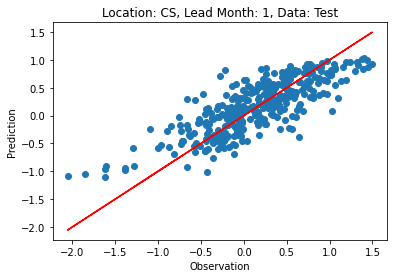}}\\
\subfloat{\includegraphics[scale=0.45]{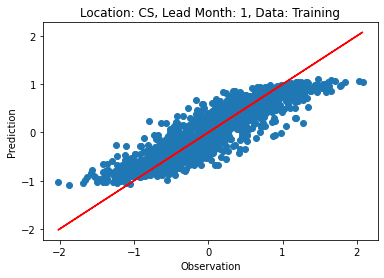}}
\subfloat{\includegraphics[scale=0.45]{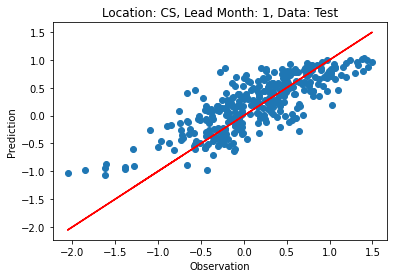}}\\
\subfloat{\includegraphics[scale=0.45]{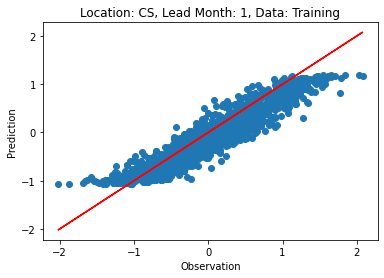}}
\subfloat{\includegraphics[scale=0.45]{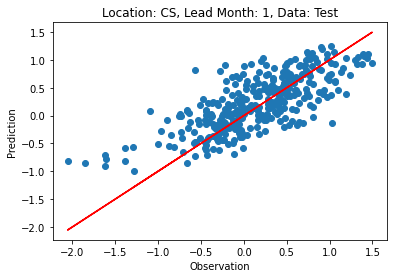}}
\caption{SSTA observation and prediction scatter plots of the last model for one lead month forecasts at Cook Strait, on the training and test data respectively. The corresponding loss functions from top to bottom are the MSE, the MAE, and the Huber.}
\end{figure}

\begin{figure}[H]
\centering
\subfloat{\includegraphics[scale=0.45]{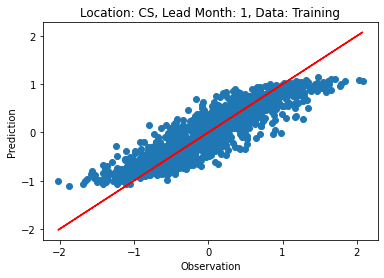}}
\subfloat{\includegraphics[scale=0.45]{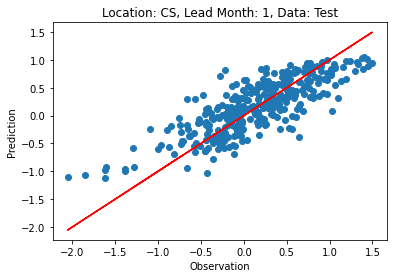}}\\
\subfloat{\includegraphics[scale=0.45]{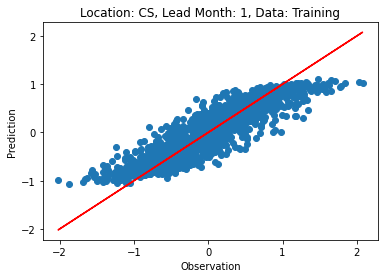}}
\subfloat{\includegraphics[scale=0.45]{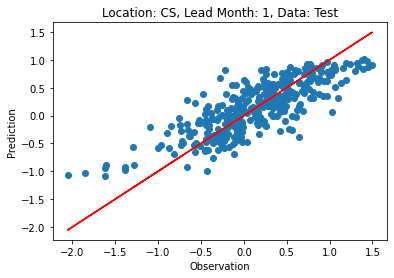}}\\
\subfloat{\includegraphics[scale=0.45]{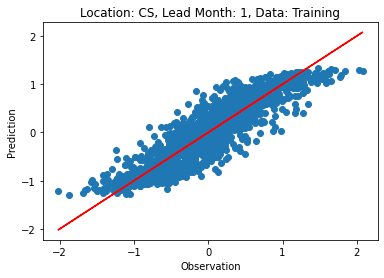}}
\subfloat{\includegraphics[scale=0.45]{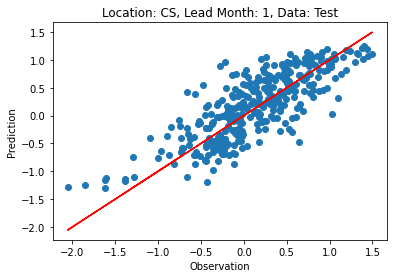}}
\caption{SSTA observation and prediction scatter plots of the last model for one lead month forecasts at Cook Strait, on the training and test data respectively. The corresponding loss functions from top to bottom are the weighted MSE, the focal-R, and the balanced MSE.}
\end{figure}

\begin{figure}[H]
\centering
\subfloat{\includegraphics[scale=0.45]{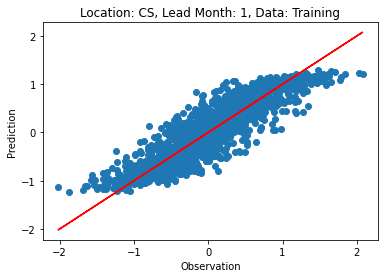}}
\subfloat{\includegraphics[scale=0.45]{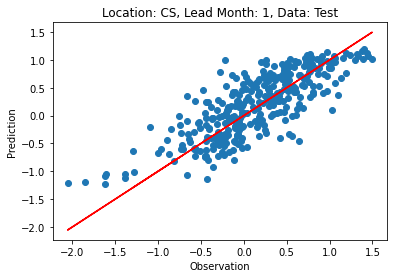}}\\
\subfloat{\includegraphics[scale=0.45]{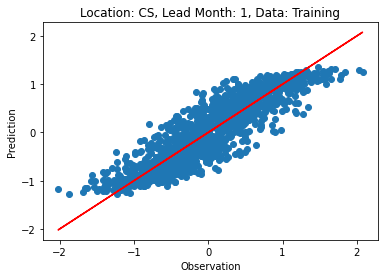}}
\subfloat{\includegraphics[scale=0.45]{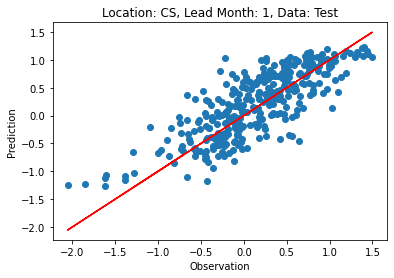}}\\
\subfloat{\includegraphics[scale=0.45]{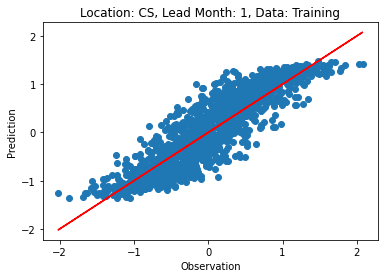}}
\subfloat{\includegraphics[scale=0.45]{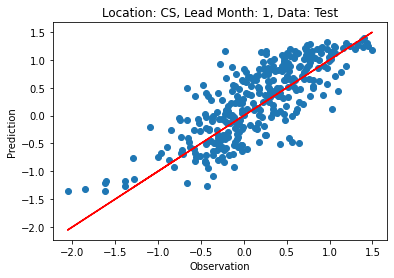}}
\caption{SSTA observation and prediction scatter plots of the last model for one lead month forecasts at Cook Strait, on the training and test data respectively. The corresponding loss function is our proposed scaling-weighted MSE, from top to bottom with the hyperparameters $\alpha=1.5$, $\beta=0.5$, $\alpha=2$, $\beta=0.5$, and $\alpha=2$, $\beta=1$ respectively.}
\end{figure}

\begin{figure}[H]
\centering
\subfloat{\includegraphics[scale=0.45]{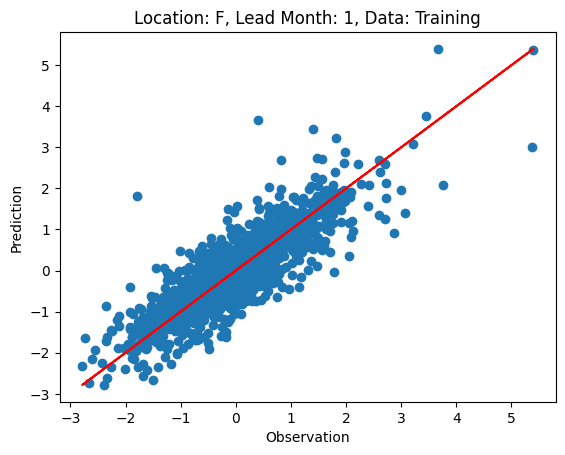}}
\subfloat{\includegraphics[scale=0.45]{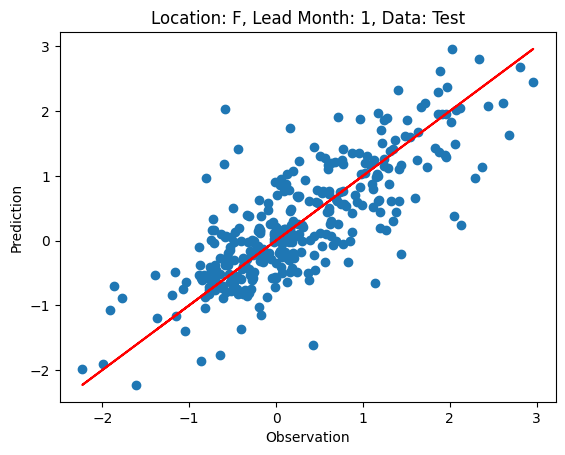}}
\caption{SSTA observation and prediction scatter plots of the persistence model for one lead month forecasts at Fiordland, on the training and test data respectively.}
\end{figure}

\begin{figure}[H]
\centering
\subfloat{\includegraphics[scale=0.45]{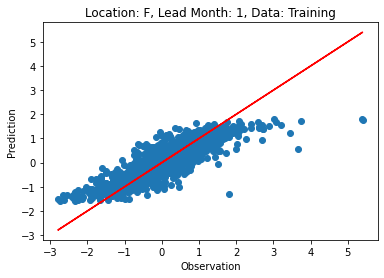}}
\subfloat{\includegraphics[scale=0.45]{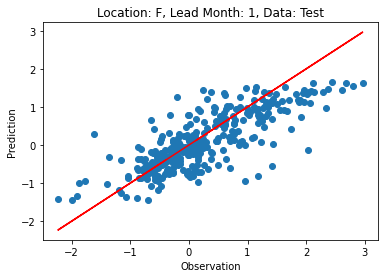}}\\
\subfloat{\includegraphics[scale=0.45]{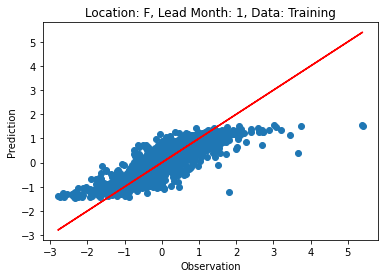}}
\subfloat{\includegraphics[scale=0.45]{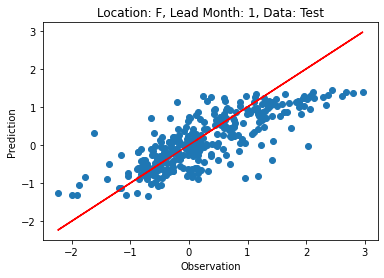}}\\
\subfloat{\includegraphics[scale=0.45]{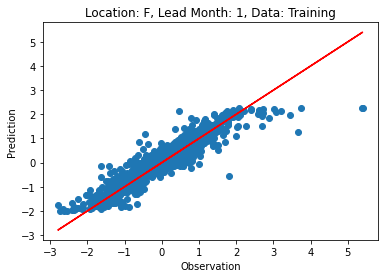}}
\subfloat{\includegraphics[scale=0.45]{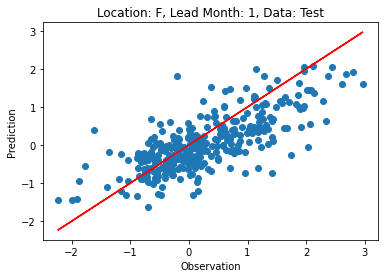}}
\caption{SSTA observation and prediction scatter plots of the last model for one lead month forecasts at Fiordland, on the training and test data respectively. The corresponding loss functions from top to bottom are the MSE, the MAE, and the Huber.}
\end{figure}

\begin{figure}[H]
\centering
\subfloat{\includegraphics[scale=0.45]{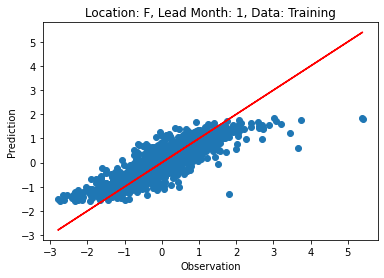}}
\subfloat{\includegraphics[scale=0.45]{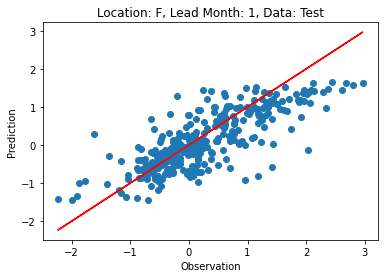}}\\
\subfloat{\includegraphics[scale=0.45]{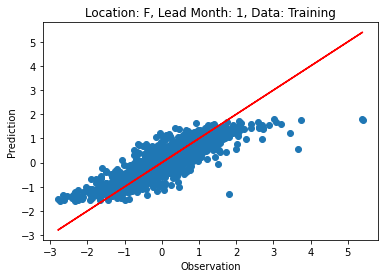}}
\subfloat{\includegraphics[scale=0.45]{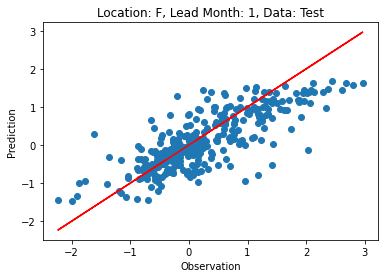}}\\
\subfloat{\includegraphics[scale=0.45]{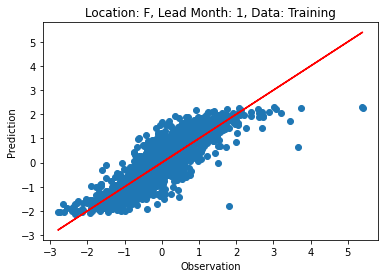}}
\subfloat{\includegraphics[scale=0.45]{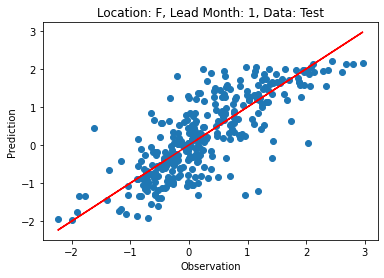}}
\caption{SSTA observation and prediction scatter plots of the last model for one lead month forecasts at Fiordland, on the training and test data respectively. The corresponding loss functions from top to bottom are the weighted MSE, the focal-R, and the balanced MSE.}
\end{figure}

\begin{figure}[H]
\centering
\subfloat{\includegraphics[scale=0.45]{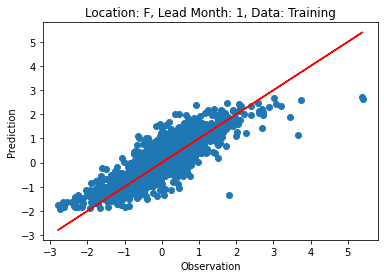}}
\subfloat{\includegraphics[scale=0.45]{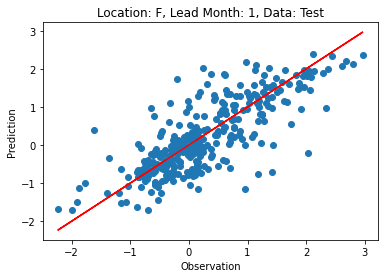}}\\
\subfloat{\includegraphics[scale=0.45]{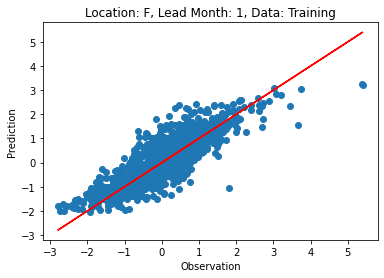}}
\subfloat{\includegraphics[scale=0.45]{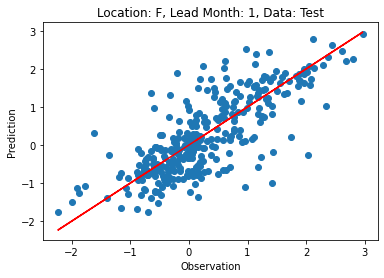}}\\
\subfloat{\includegraphics[scale=0.45]{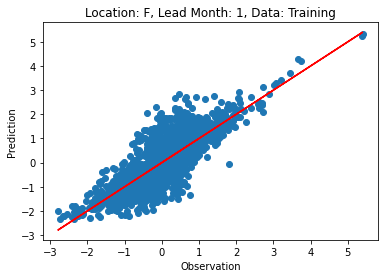}}
\subfloat{\includegraphics[scale=0.45]{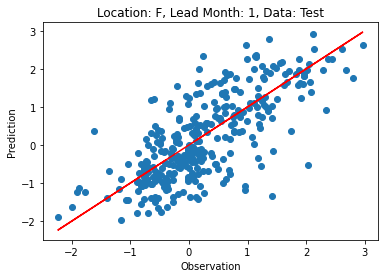}}
\caption{SSTA observation and prediction scatter plots of the last model for one lead month forecasts at Fiordland, on the training and test data respectively. The corresponding loss function is our proposed scaling-weighted MSE, from top to bottom with the hyperparameters $\alpha=1.5$, $\beta=0.5$, $\alpha=2$, $\beta=0.5$, and $\alpha=2$, $\beta=1$ respectively.}
\end{figure}

\subsubsection{Two Lead Month Forecasts}

\begin{figure}[H]
\centering
\subfloat{\includegraphics[scale=0.45]{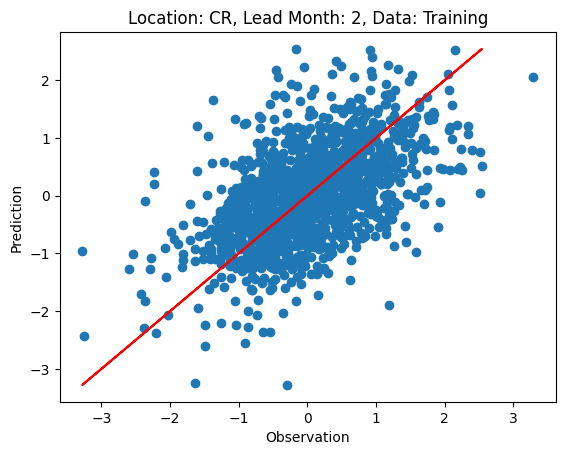}}
\subfloat{\includegraphics[scale=0.45]{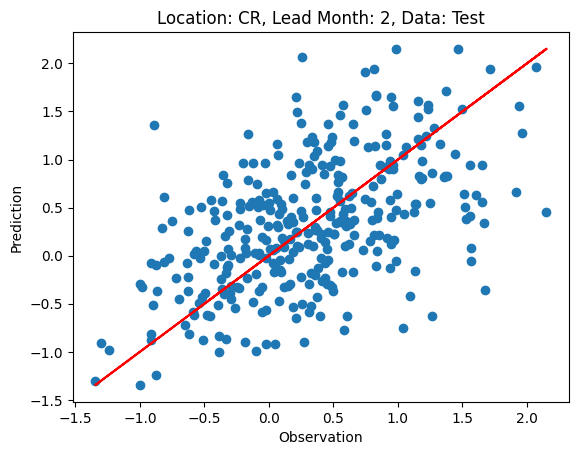}}
\caption{SSTA observation and prediction scatter plots of the two-lag model for two lead month forecasts at Cape Reinga, on the training and test data respectively.}
\end{figure}

\begin{figure}[H]
\centering
\subfloat{\includegraphics[scale=0.45]{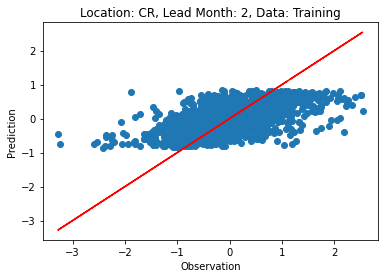}}
\subfloat{\includegraphics[scale=0.45]{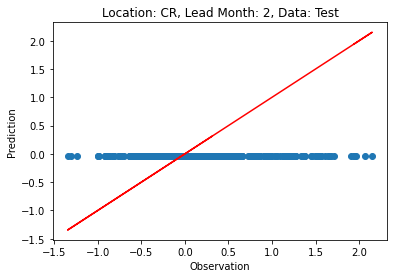}}\\
\subfloat{\includegraphics[scale=0.45]{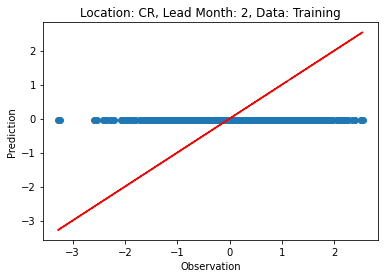}}
\subfloat{\includegraphics[scale=0.45]{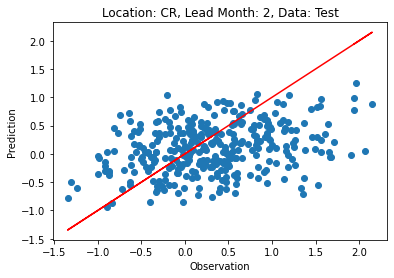}}\\
\subfloat{\includegraphics[scale=0.45]{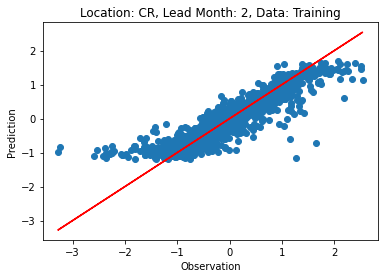}}
\subfloat{\includegraphics[scale=0.45]{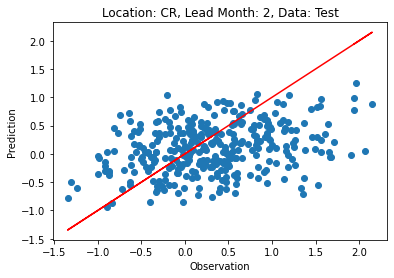}}
\caption{SSTA observation and prediction scatter plots of the last model for two lead month forecasts at Cape Reinga, on the training and test data respectively. The corresponding loss functions from top to bottom are the MSE, the MAE, and the Huber.}
\end{figure}

\begin{figure}[H]
\centering
\subfloat{\includegraphics[scale=0.45]{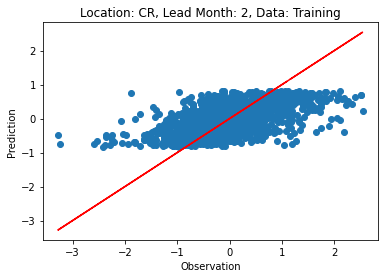}}
\subfloat{\includegraphics[scale=0.45]{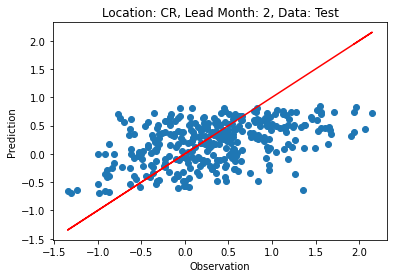}}\\
\subfloat{\includegraphics[scale=0.45]{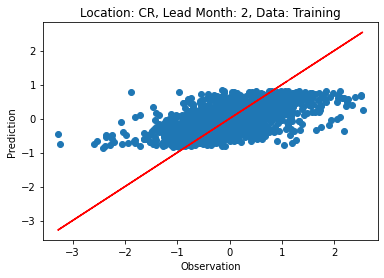}}
\subfloat{\includegraphics[scale=0.45]{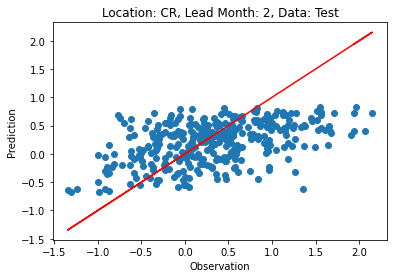}}\\
\subfloat{\includegraphics[scale=0.45]{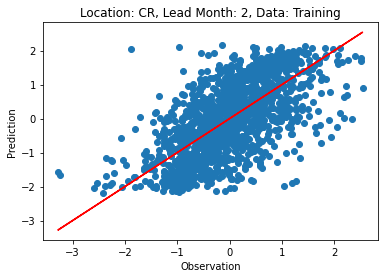}}
\subfloat{\includegraphics[scale=0.45]{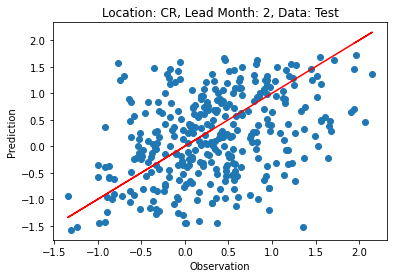}}
\caption{SSTA observation and prediction scatter plots of the last model for two lead month forecasts at Cape Reinga, on the training and test data respectively. The corresponding loss functions from top to bottom are the weighted MSE, the focal-R, and the balanced MSE.}
\end{figure}

\begin{figure}[H]
\centering
\subfloat{\includegraphics[scale=0.45]{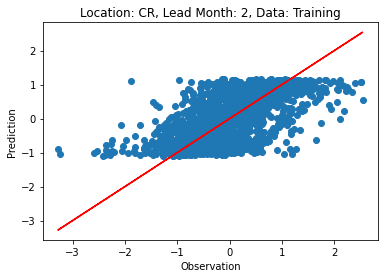}}
\subfloat{\includegraphics[scale=0.45]{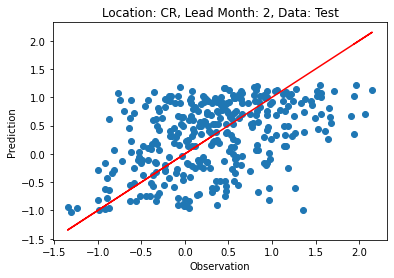}}\\
\subfloat{\includegraphics[scale=0.45]{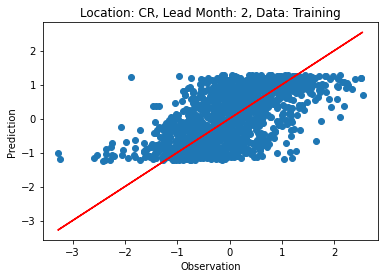}}
\subfloat{\includegraphics[scale=0.45]{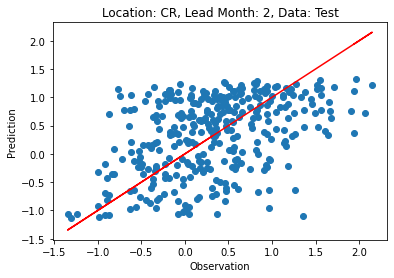}}\\
\subfloat{\includegraphics[scale=0.45]{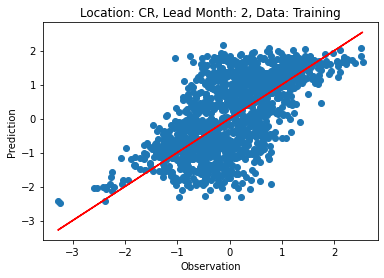}}
\subfloat{\includegraphics[scale=0.45]{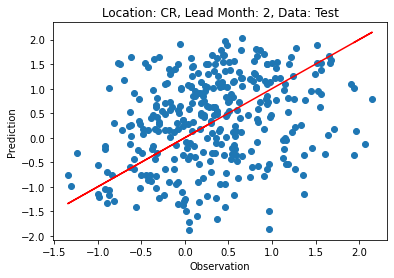}}
\caption{SSTA observation and prediction scatter plots of the last model for two lead month forecasts at Cape Reinga, on the training and test data respectively. The corresponding loss function is our proposed scaling-weighted MSE, from top to bottom with the hyperparameters $\alpha=1.5$, $\beta=0.5$, $\alpha=2$, $\beta=0.5$, and $\alpha=2$, $\beta=1$ respectively.}
\end{figure}

\begin{figure}[H]
\centering
\subfloat{\includegraphics[scale=0.45]{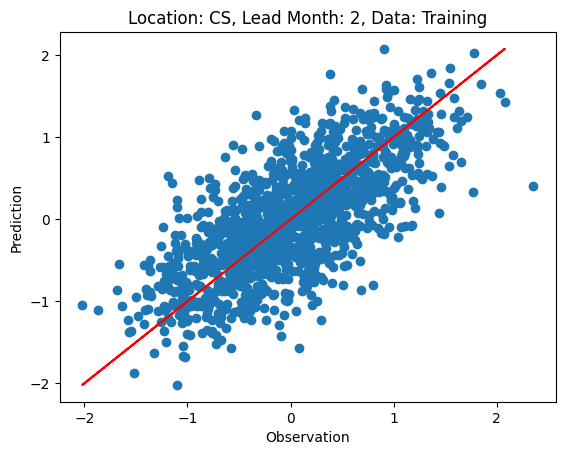}}
\subfloat{\includegraphics[scale=0.45]{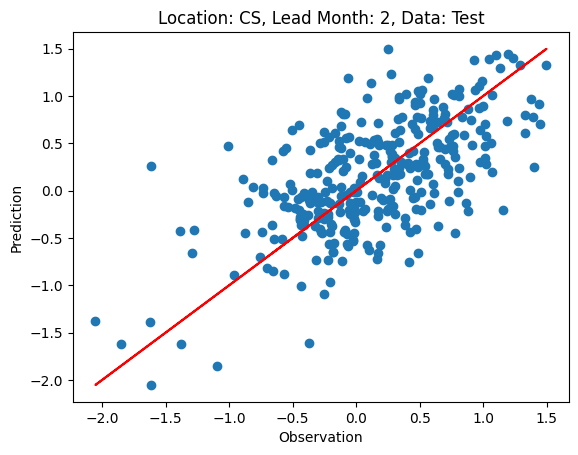}}
\caption{SSTA observation and prediction scatter plots of the two-lag model for two lead month forecasts at Cook Strait, on the training and test data respectively.}
\end{figure}

\begin{figure}[H]
\centering
\subfloat{\includegraphics[scale=0.45]{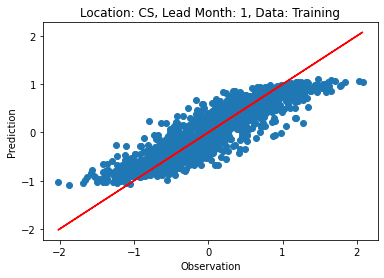}}
\subfloat{\includegraphics[scale=0.45]{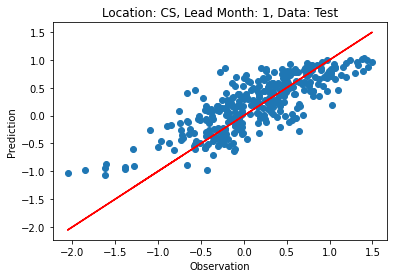}}\\
\subfloat{\includegraphics[scale=0.45]{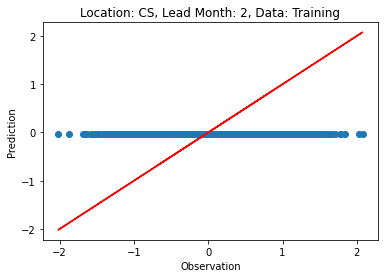}}
\subfloat{\includegraphics[scale=0.45]{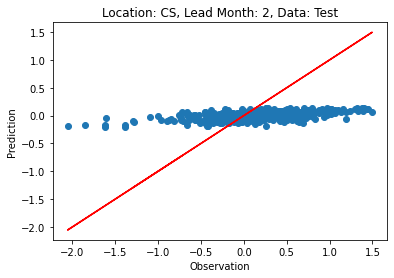}}\\
\subfloat{\includegraphics[scale=0.45]{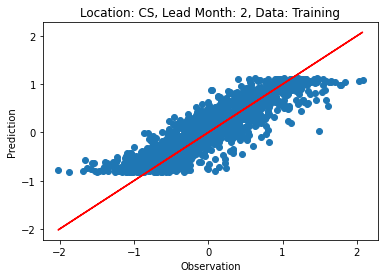}}
\subfloat{\includegraphics[scale=0.45]{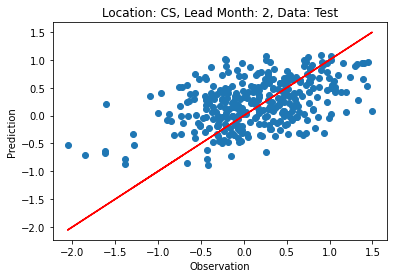}}
\caption{SSTA observation and prediction scatter plots of the last model for two lead month forecasts at Cook Strait, on the training and test data respectively. The corresponding loss functions from top to bottom are the MSE, the MAE, and the Huber.}
\end{figure}

\begin{figure}[H]
\centering
\subfloat{\includegraphics[scale=0.45]{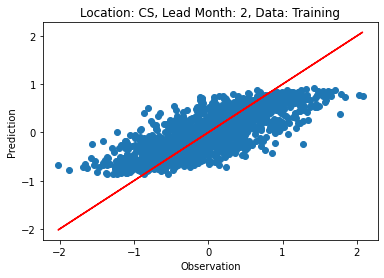}}
\subfloat{\includegraphics[scale=0.45]{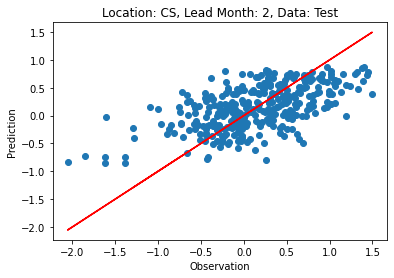}}\\
\subfloat{\includegraphics[scale=0.45]{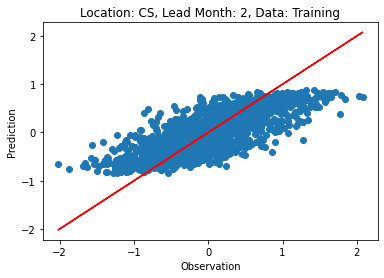}}
\subfloat{\includegraphics[scale=0.45]{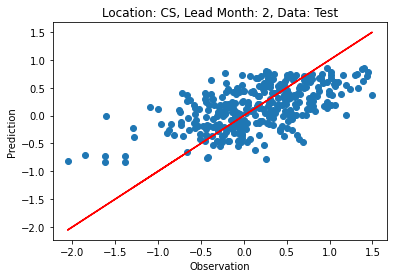}}\\
\subfloat{\includegraphics[scale=0.45]{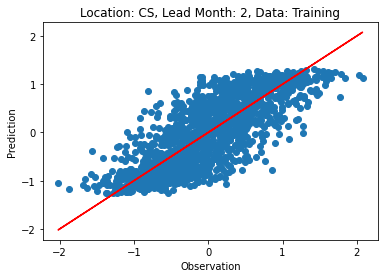}}
\subfloat{\includegraphics[scale=0.45]{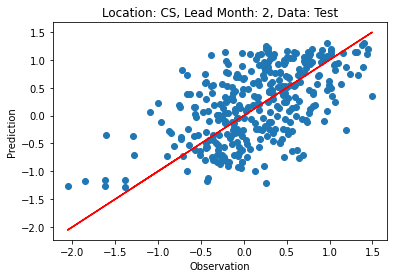}}
\caption{SSTA observation and prediction scatter plots of the last model for two lead month forecasts at Cook Strait, on the training and test data respectively. The corresponding loss functions from top to bottom are the weighted MSE, the focal-R, and the balanced MSE.}
\end{figure}

\begin{figure}[H]
\centering
\subfloat{\includegraphics[scale=0.45]{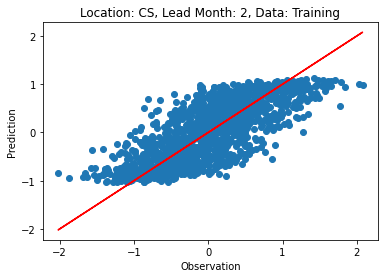}}
\subfloat{\includegraphics[scale=0.45]{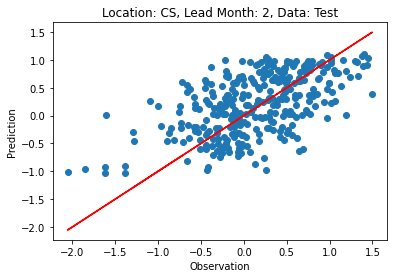}}\\
\subfloat{\includegraphics[scale=0.45]{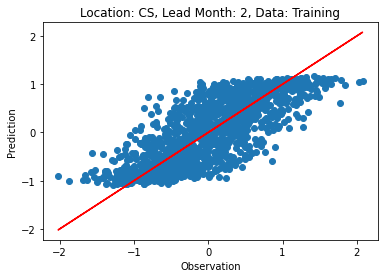}}
\subfloat{\includegraphics[scale=0.45]{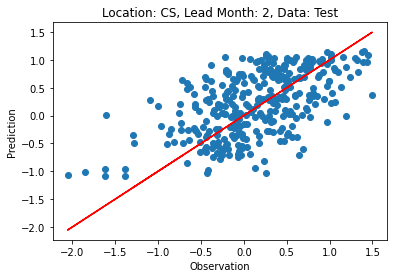}}\\
\subfloat{\includegraphics[scale=0.45]{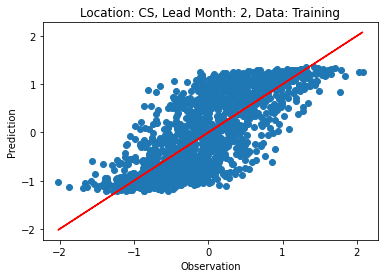}}
\subfloat{\includegraphics[scale=0.45]{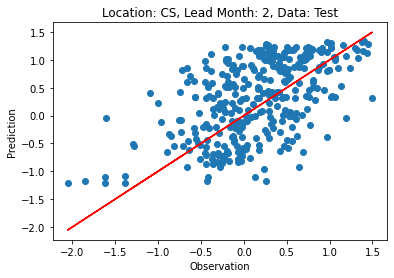}}
\caption{SSTA observation and prediction scatter plots of the last model for two lead month forecasts at Cook Strait, on the training and test data respectively. The corresponding loss function is our proposed scaling-weighted MSE, from top to bottom with the hyperparameters $\alpha=1.5$, $\beta=0.5$, $\alpha=2$, $\beta=0.5$, and $\alpha=2$, $\beta=1$ respectively.}
\end{figure}

\begin{figure}[H]
\centering
\subfloat{\includegraphics[scale=0.45]{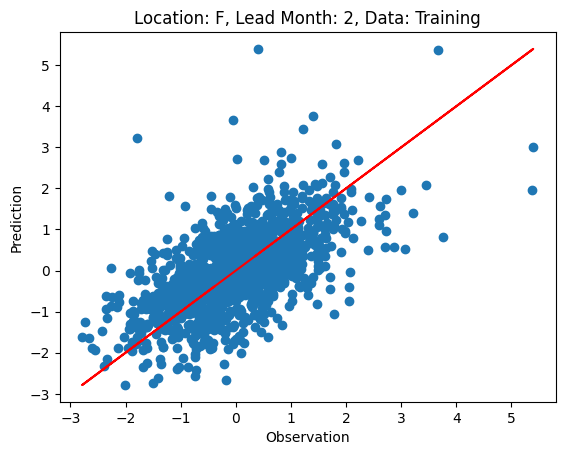}}
\subfloat{\includegraphics[scale=0.45]{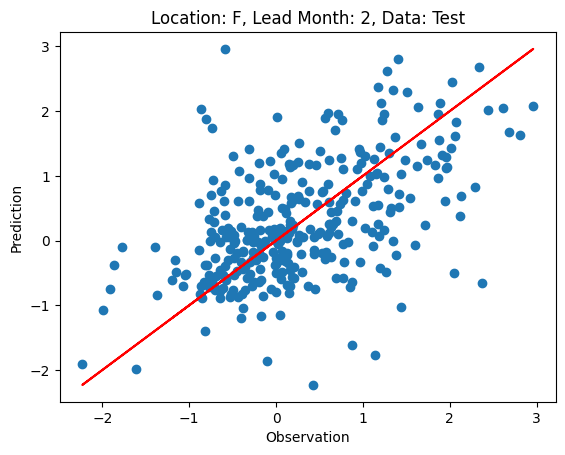}}
\caption{SSTA observation and prediction scatter plots of the two-lag model for two lead month forecasts at Fiordland, on the training and test data respectively.}
\end{figure}

\begin{figure}[H]
\centering
\subfloat{\includegraphics[scale=0.45]{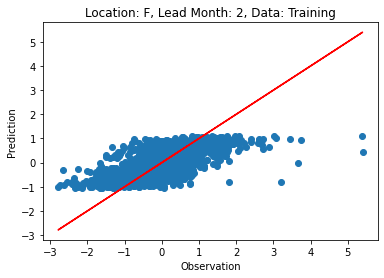}}
\subfloat{\includegraphics[scale=0.45]{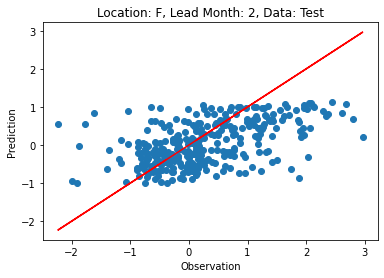}}\\
\subfloat{\includegraphics[scale=0.45]{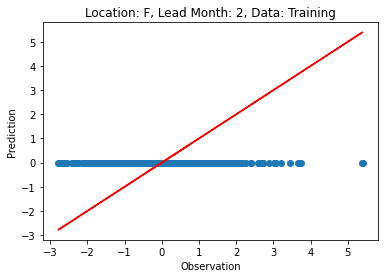}}
\subfloat{\includegraphics[scale=0.45]{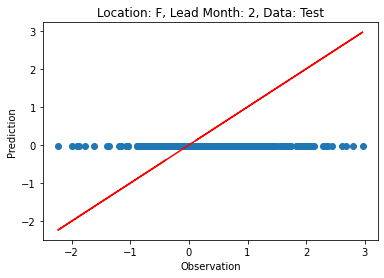}}\\
\subfloat{\includegraphics[scale=0.45]{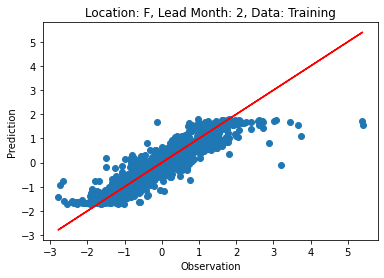}}
\subfloat{\includegraphics[scale=0.45]{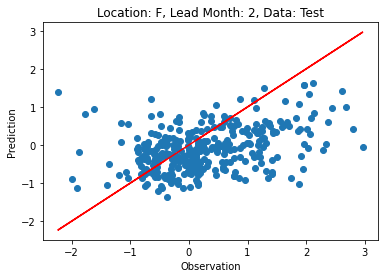}}
\caption{SSTA observation and prediction scatter plots of the last model for two lead month forecasts at Fiordland, on the training and test data respectively. The corresponding loss functions from top to bottom are the MSE, the MAE, and the Huber.}
\end{figure}

\begin{figure}[H]
\centering
\subfloat{\includegraphics[scale=0.45]{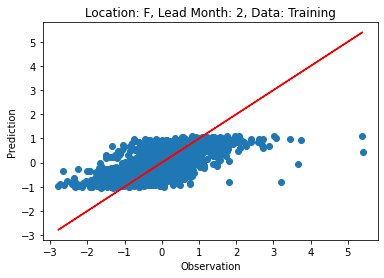}}
\subfloat{\includegraphics[scale=0.45]{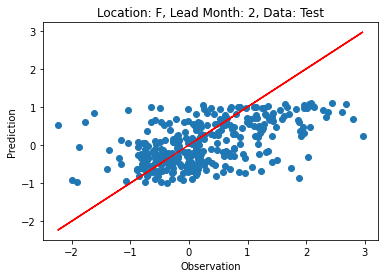}}\\
\subfloat{\includegraphics[scale=0.45]{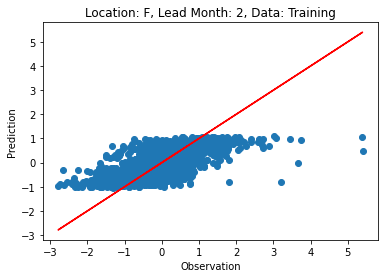}}
\subfloat{\includegraphics[scale=0.45]{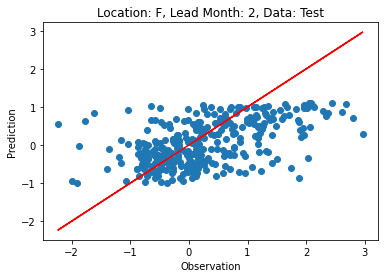}}\\
\subfloat{\includegraphics[scale=0.45]{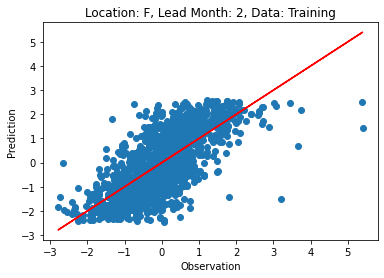}}
\subfloat{\includegraphics[scale=0.45]{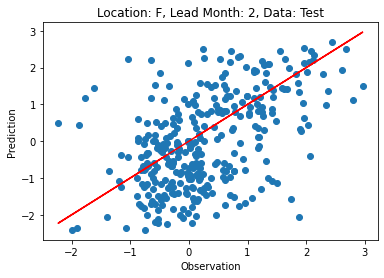}}
\caption{SSTA observation and prediction scatter plots of the last model for two lead month forecasts at Fiordland, on the training and test data respectively. The corresponding loss functions from top to bottom are the weighted MSE, the focal-R, and the balanced MSE.}
\end{figure}

\begin{figure}[H]
\centering
\subfloat{\includegraphics[scale=0.45]{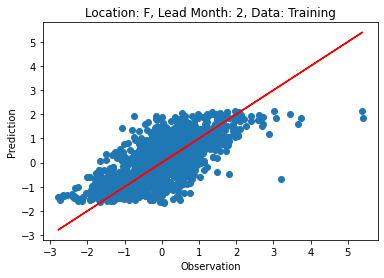}}
\subfloat{\includegraphics[scale=0.45]{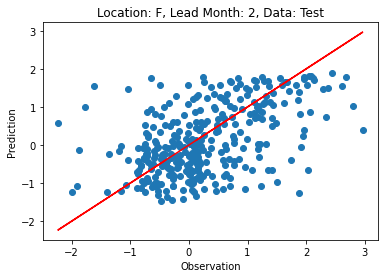}}\\
\subfloat{\includegraphics[scale=0.45]{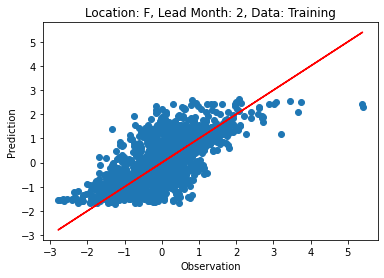}}
\subfloat{\includegraphics[scale=0.45]{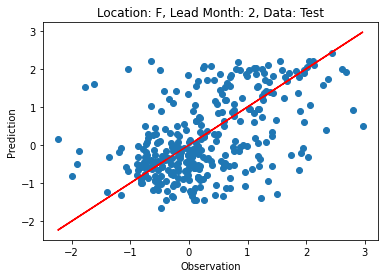}}\\
\subfloat{\includegraphics[scale=0.45]{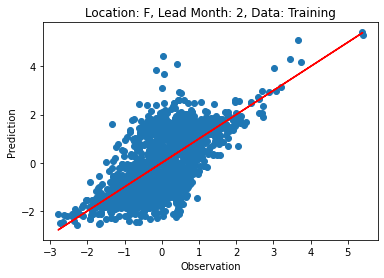}}
\subfloat{\includegraphics[scale=0.45]{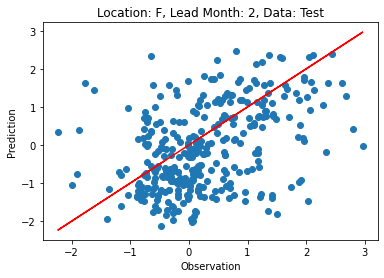}}
\caption{SSTA observation and prediction scatter plots of the last model for two lead month forecasts at Fiordland, on the training and test data respectively. The corresponding loss function is our proposed scaling-weighted MSE, from top to bottom with the hyperparameters $\alpha=1.5$, $\beta=0.5$, $\alpha=2$, $\beta=0.5$, and $\alpha=2$, $\beta=1$ respectively.}
\end{figure}

\subsubsection{Three Lead Month Forecasts}

\begin{figure}[H]
\centering
\subfloat{\includegraphics[scale=0.45]{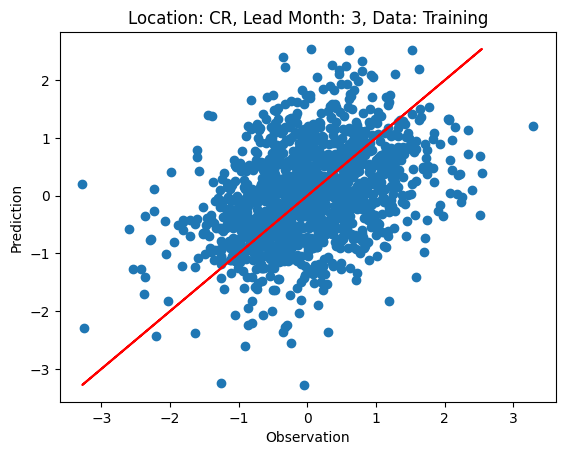}}
\subfloat{\includegraphics[scale=0.45]{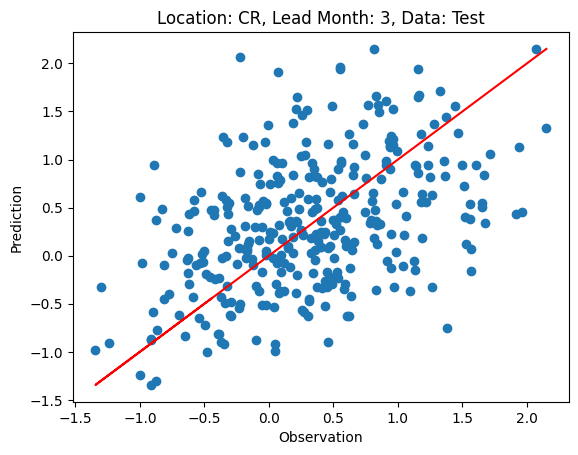}}
\caption{SSTA observation and prediction scatter plots of the three-lag model for three lead month forecasts at Cape Reinga, on the training and test data respectively.}
\end{figure}

\begin{figure}[H]
\centering
\subfloat{\includegraphics[scale=0.45]{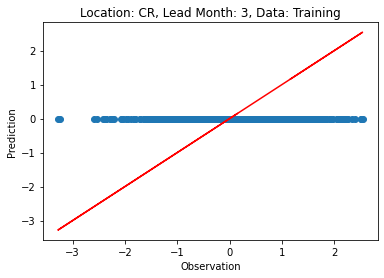}}
\subfloat{\includegraphics[scale=0.45]{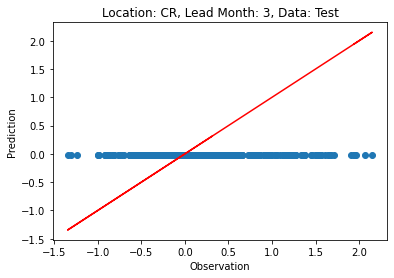}}\\
\subfloat{\includegraphics[scale=0.45]{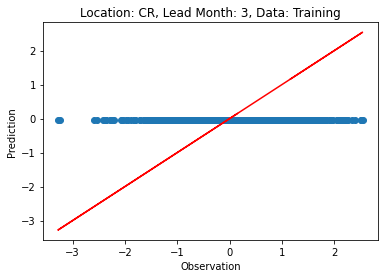}}
\subfloat{\includegraphics[scale=0.45]{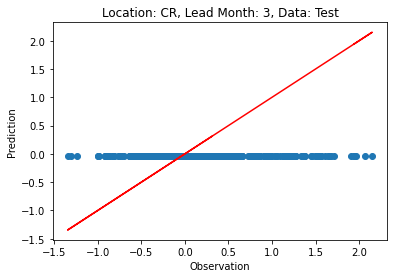}}\\
\subfloat{\includegraphics[scale=0.45]{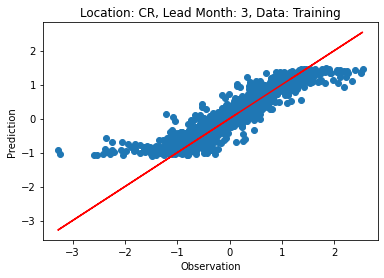}}
\subfloat{\includegraphics[scale=0.45]{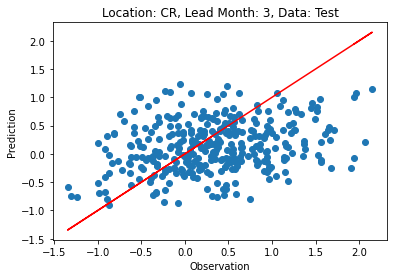}}
\caption{SSTA observation and prediction scatter plots of the last model for three lead month forecasts at Cape Reinga, on the training and test data respectively. The corresponding loss functions from top to bottom are the MSE, the MAE, and the Huber.}
\end{figure}

\begin{figure}[H]
\centering
\subfloat{\includegraphics[scale=0.45]{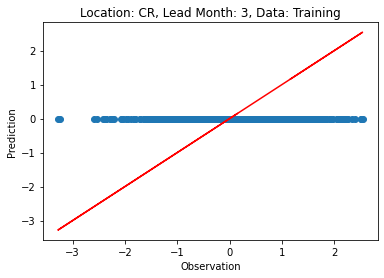}}
\subfloat{\includegraphics[scale=0.45]{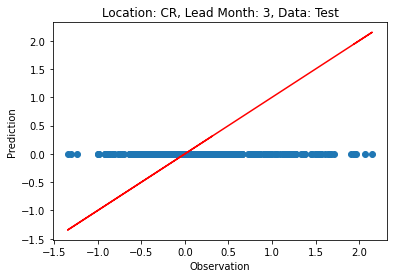}}\\
\subfloat{\includegraphics[scale=0.45]{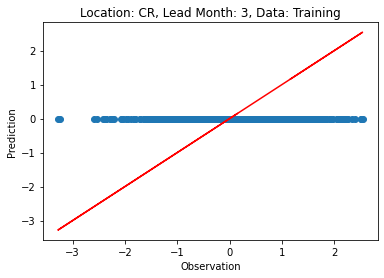}}
\subfloat{\includegraphics[scale=0.45]{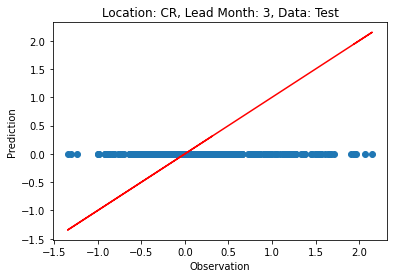}}\\
\subfloat{\includegraphics[scale=0.45]{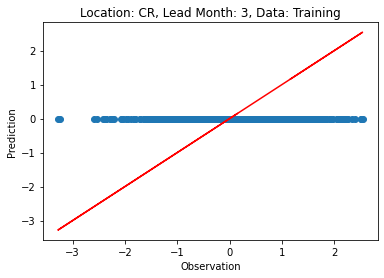}}
\subfloat{\includegraphics[scale=0.45]{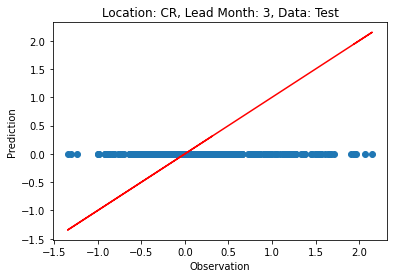}}
\caption{SSTA observation and prediction scatter plots of the last model for three lead month forecasts at Cape Reinga, on the training and test data respectively. The corresponding loss functions from top to bottom are the weighted MSE, the focal-R, and the balanced MSE.}
\end{figure}

\begin{figure}[H]
\centering
\subfloat{\includegraphics[scale=0.45]{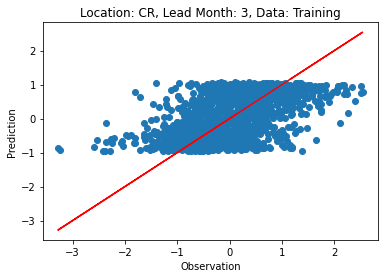}}
\subfloat{\includegraphics[scale=0.45]{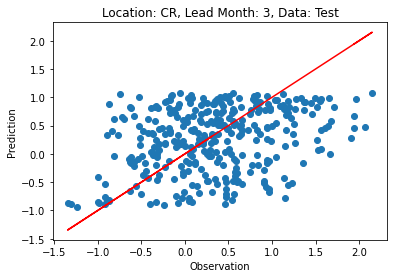}}\\
\subfloat{\includegraphics[scale=0.45]{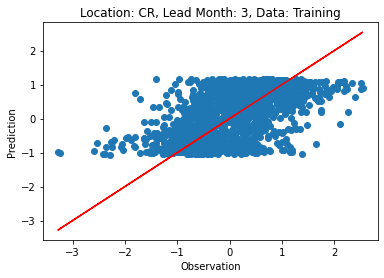}}
\subfloat{\includegraphics[scale=0.45]{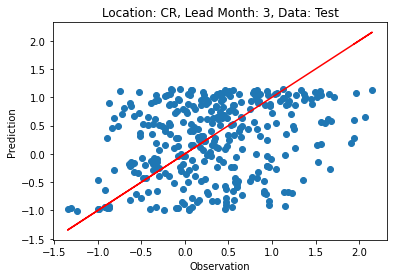}}\\
\subfloat{\includegraphics[scale=0.45]{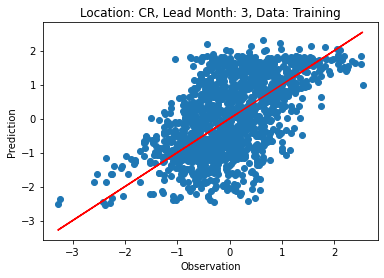}}
\subfloat{\includegraphics[scale=0.45]{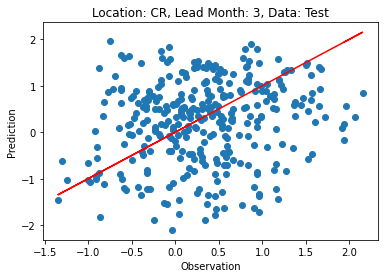}}
\caption{SSTA observation and prediction scatter plots of the last model for three lead month forecasts at Cape Reinga, on the training and test data respectively. The corresponding loss function is our proposed scaling-weighted MSE, from top to bottom with the hyperparameters $\alpha=1.5$, $\beta=0.5$, $\alpha=2$, $\beta=0.5$, and $\alpha=2$, $\beta=1$ respectively.}
\end{figure}

\begin{figure}[H]
\centering
\subfloat{\includegraphics[scale=0.45]{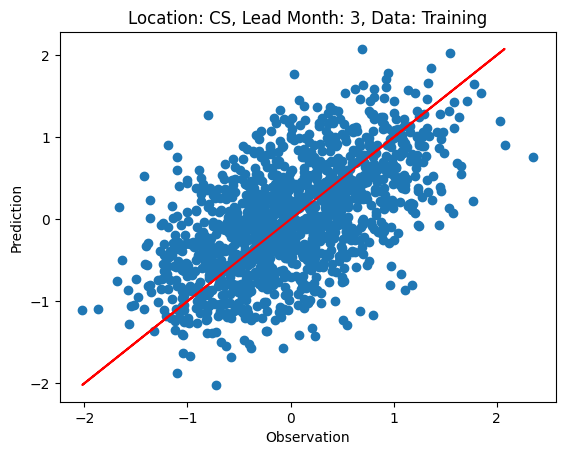}}
\subfloat{\includegraphics[scale=0.45]{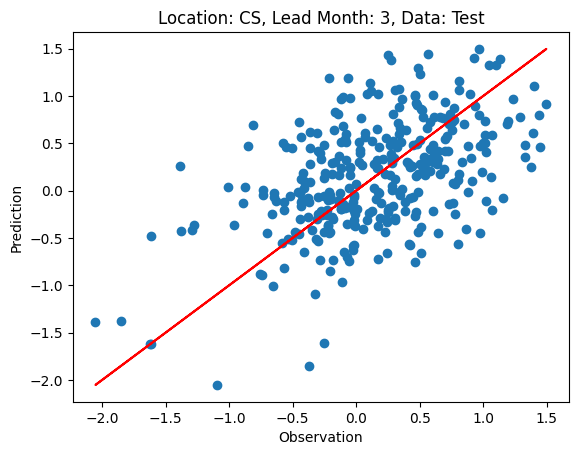}}
\caption{SSTA observation and prediction scatter plots of the three-lag model for three lead month forecasts at Cook Strait, on the training and test data respectively.}
\end{figure}

\begin{figure}[H]
\centering
\subfloat{\includegraphics[scale=0.45]{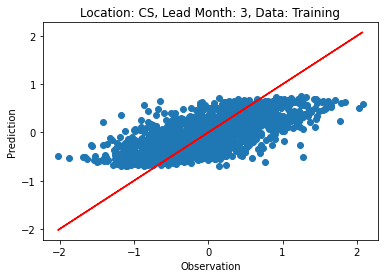}}
\subfloat{\includegraphics[scale=0.45]{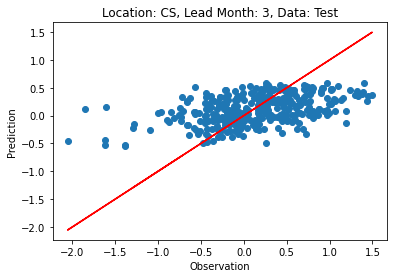}}\\
\subfloat{\includegraphics[scale=0.45]{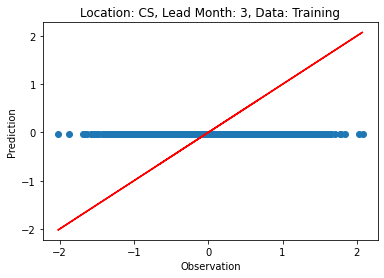}}
\subfloat{\includegraphics[scale=0.45]{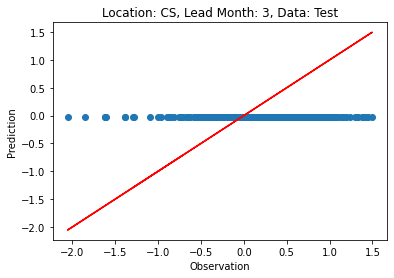}}\\
\subfloat{\includegraphics[scale=0.45]{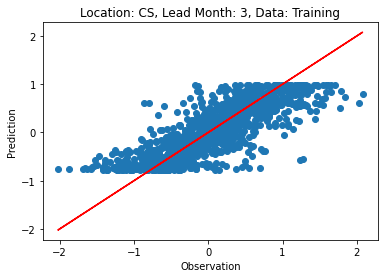}}
\subfloat{\includegraphics[scale=0.45]{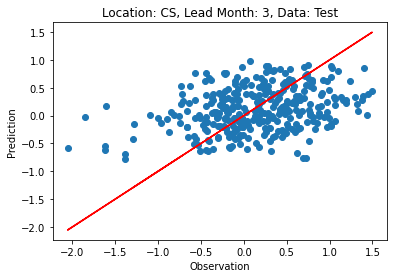}}
\caption{SSTA observation and prediction scatter plots of the last model for three lead month forecasts at Cook Strait, on the training and test data respectively. The corresponding loss functions from top to bottom are the MSE, the MAE, and the Huber.}
\end{figure}

\begin{figure}[H]
\centering
\subfloat{\includegraphics[scale=0.45]{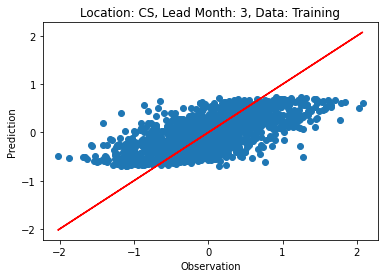}}
\subfloat{\includegraphics[scale=0.45]{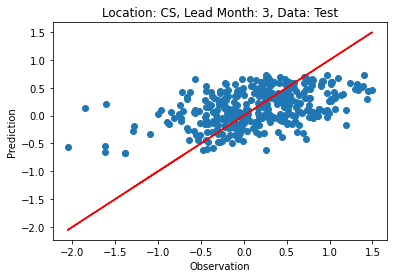}}\\
\subfloat{\includegraphics[scale=0.45]{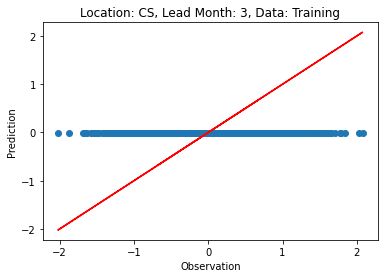}}
\subfloat{\includegraphics[scale=0.45]{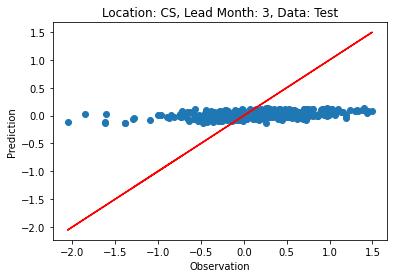}}\\
\subfloat{\includegraphics[scale=0.45]{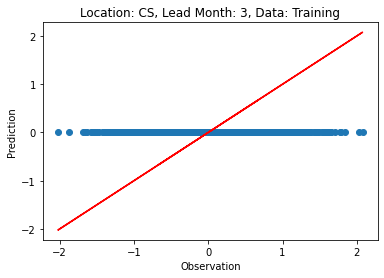}}
\subfloat{\includegraphics[scale=0.45]{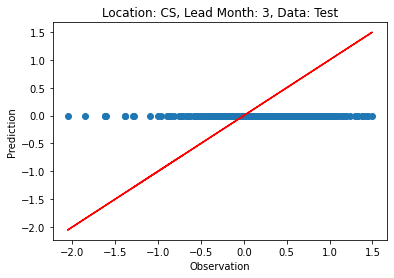}}
\caption{SSTA observation and prediction scatter plots of the last model for three lead month forecasts at Cook Strait, on the training and test data respectively. The corresponding loss functions from top to bottom are the weighted MSE, the focal-R, and the balanced MSE.}
\end{figure}

\begin{figure}[H]
\centering
\subfloat{\includegraphics[scale=0.45]{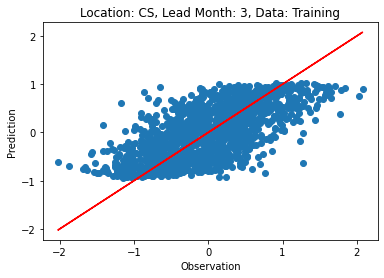}}
\subfloat{\includegraphics[scale=0.45]{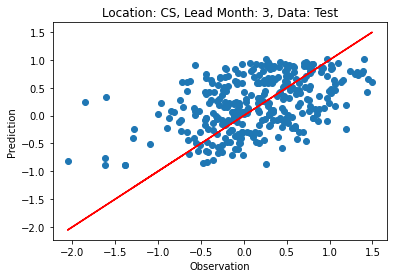}}\\
\subfloat{\includegraphics[scale=0.45]{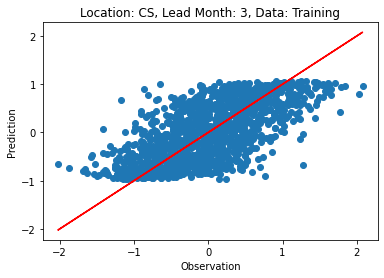}}
\subfloat{\includegraphics[scale=0.45]{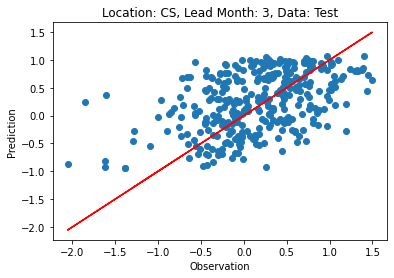}}\\
\subfloat{\includegraphics[scale=0.45]{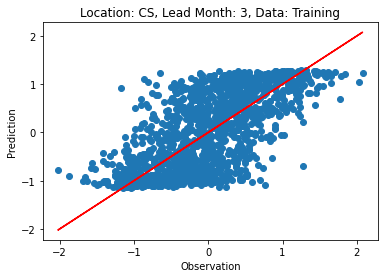}}
\subfloat{\includegraphics[scale=0.45]{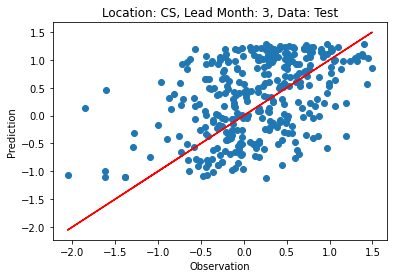}}
\caption{SSTA observation and prediction scatter plots of the last model for three lead month forecasts at Cook Strait, on the training and test data respectively. The corresponding loss function is our proposed scaling-weighted MSE, from top to bottom with the hyperparameters $\alpha=1.5$, $\beta=0.5$, $\alpha=2$, $\beta=0.5$, and $\alpha=2$, $\beta=1$ respectively.}
\end{figure}

\begin{figure}[H]
\centering
\subfloat{\includegraphics[scale=0.45]{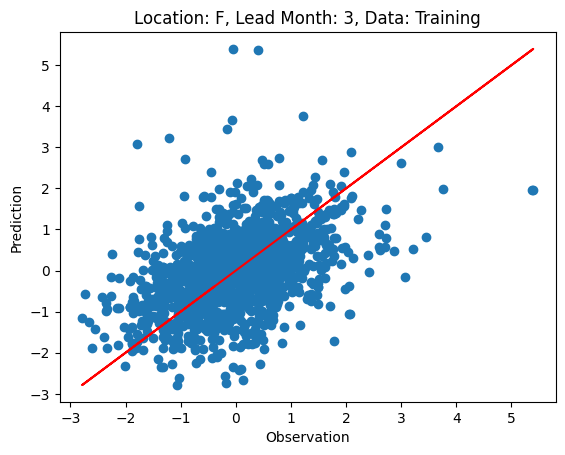}}
\subfloat{\includegraphics[scale=0.45]{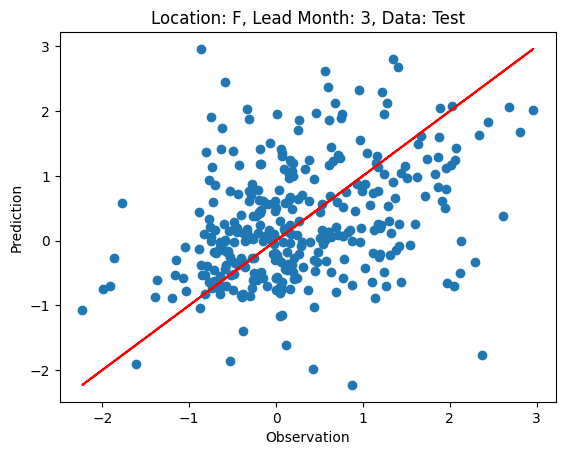}}
\caption{SSTA observation and prediction scatter plots of the three-lag model for three lead month forecasts at Fiordland, on the training and test data respectively.}
\end{figure}

\begin{figure}[H]
\centering
\subfloat{\includegraphics[scale=0.45]{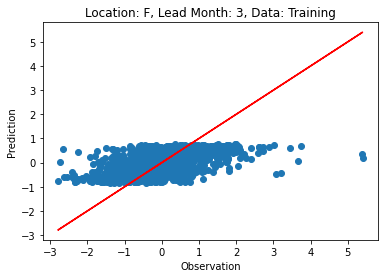}}
\subfloat{\includegraphics[scale=0.45]{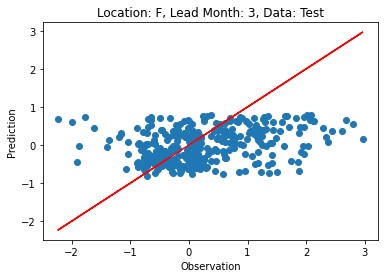}}\\
\subfloat{\includegraphics[scale=0.45]{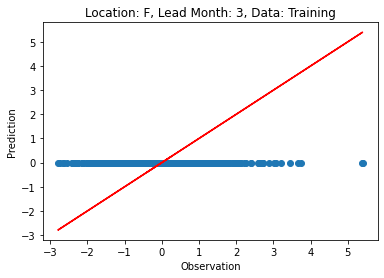}}
\subfloat{\includegraphics[scale=0.45]{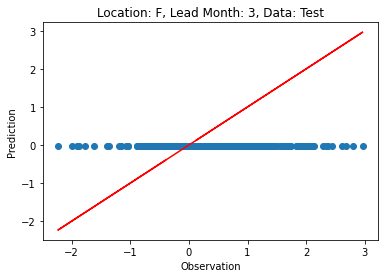}}\\
\subfloat{\includegraphics[scale=0.45]{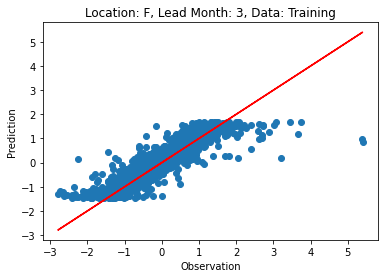}}
\subfloat{\includegraphics[scale=0.45]{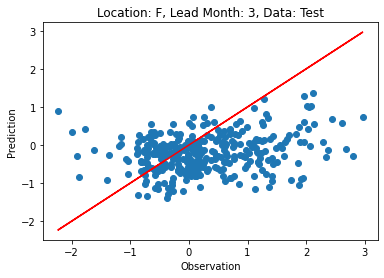}}
\caption{SSTA observation and prediction scatter plots of the last model for three lead month forecasts at Fiordland, on the training and test data respectively. The corresponding loss functions from top to bottom are the MSE, the MAE, and the Huber.}
\end{figure}

\begin{figure}[H]
\centering
\subfloat{\includegraphics[scale=0.45]{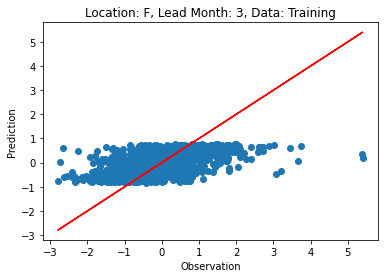}}
\subfloat{\includegraphics[scale=0.45]{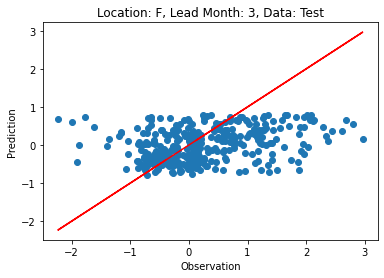}}\\
\subfloat{\includegraphics[scale=0.45]{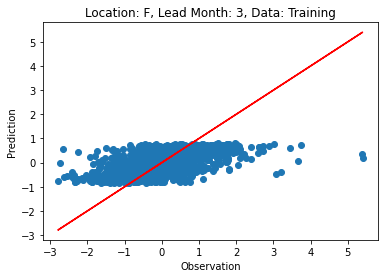}}
\subfloat{\includegraphics[scale=0.45]{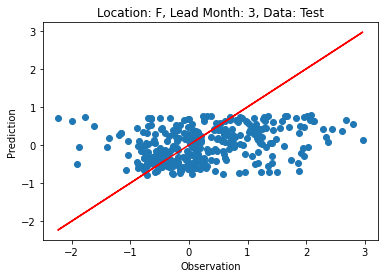}}\\
\subfloat{\includegraphics[scale=0.45]{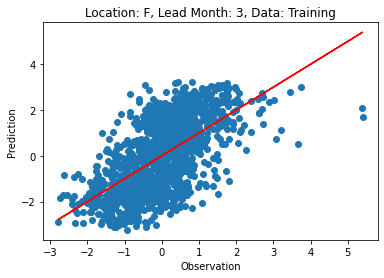}}
\subfloat{\includegraphics[scale=0.45]{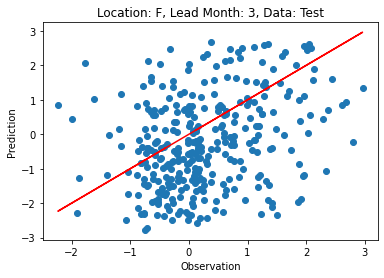}}
\caption{SSTA observation and prediction scatter plots of the last model for three lead month forecasts at Fiordland, on the training and test data respectively. The corresponding loss functions from top to bottom are the weighted MSE, the focal-R, and the balanced MSE.}
\end{figure}

\begin{figure}[H]
\centering
\subfloat{\includegraphics[scale=0.45]{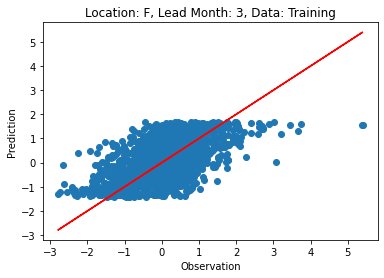}}
\subfloat{\includegraphics[scale=0.45]{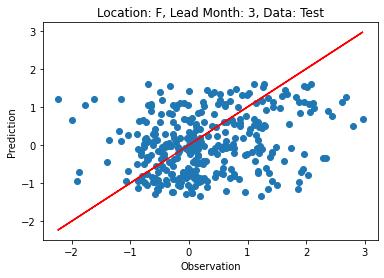}}\\
\subfloat{\includegraphics[scale=0.45]{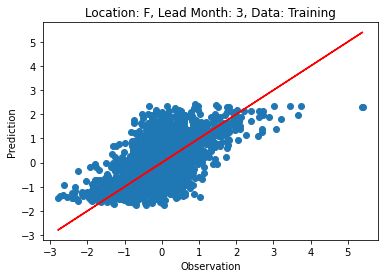}}
\subfloat{\includegraphics[scale=0.45]{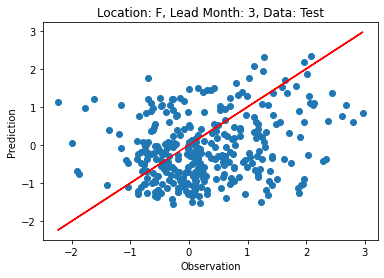}}\\
\subfloat{\includegraphics[scale=0.45]{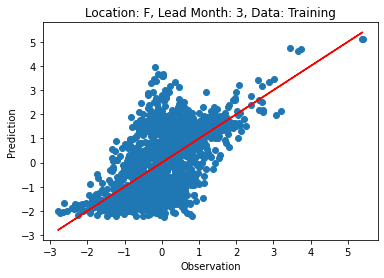}}
\subfloat{\includegraphics[scale=0.45]{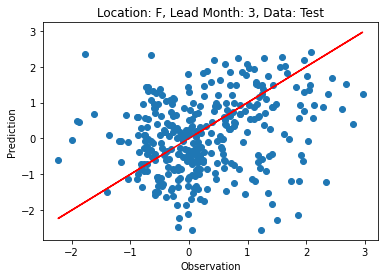}}
\caption{SSTA observation and prediction scatter plots of the last model for three lead month forecasts at Fiordland, on the training and test data respectively. The corresponding loss function is our proposed scaling-weighted MSE, from top to bottom with the hyperparameters $\alpha=1.5$, $\beta=0.5$, $\alpha=2$, $\beta=0.5$, and $\alpha=2$, $\beta=1$ respectively.}
\end{figure}

\subsubsection{Six Lead Month Forecasts}

\begin{figure}[H]
\centering
\subfloat{\includegraphics[scale=0.45]{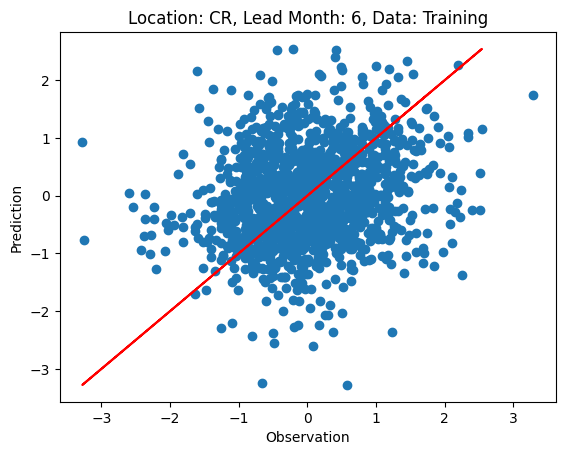}}
\subfloat{\includegraphics[scale=0.45]{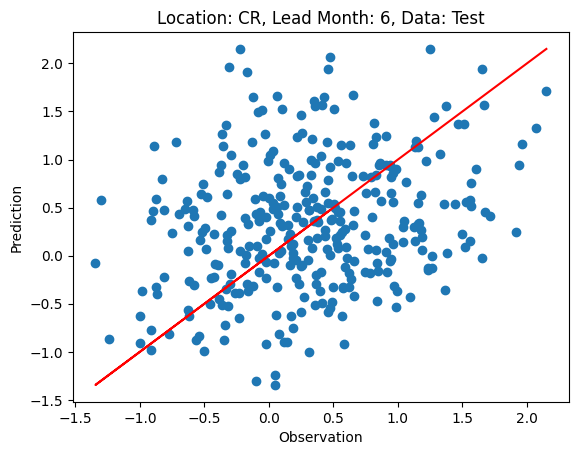}}
\caption{SSTA observation and prediction scatter plots of the six-lag model for six lead month forecasts at Cape Reinga, on the training and test data respectively.}
\end{figure}

\begin{figure}[H]
\centering
\subfloat{\includegraphics[scale=0.45]{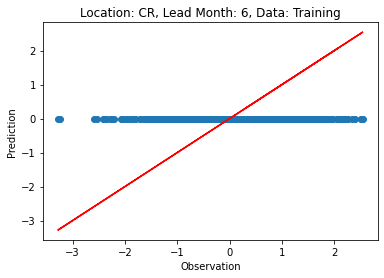}}
\subfloat{\includegraphics[scale=0.45]{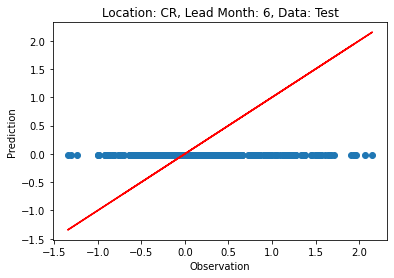}}\\
\subfloat{\includegraphics[scale=0.45]{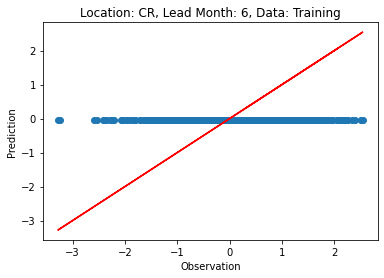}}
\subfloat{\includegraphics[scale=0.45]{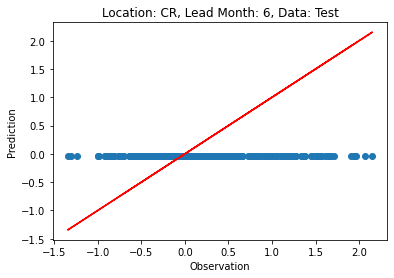}}\\
\subfloat{\includegraphics[scale=0.45]{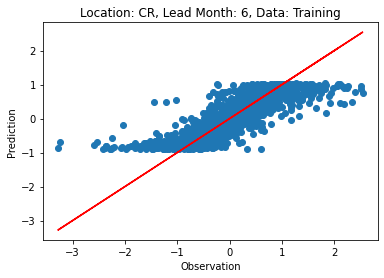}}
\subfloat{\includegraphics[scale=0.45]{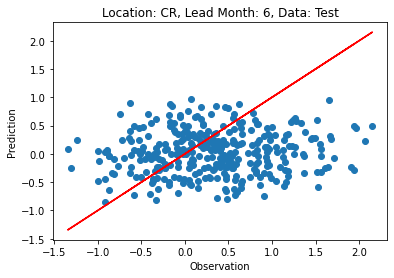}}
\caption{SSTA observation and prediction scatter plots of the last model for six lead month forecasts at Cape Reinga, on the training and test data respectively. The corresponding loss functions from top to bottom are the MSE, the MAE, and the Huber.}
\end{figure}

\begin{figure}[H]
\centering
\subfloat{\includegraphics[scale=0.45]{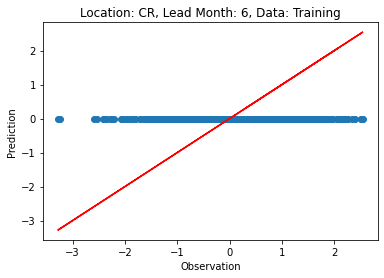}}
\subfloat{\includegraphics[scale=0.45]{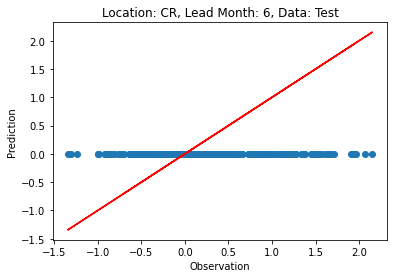}}\\
\subfloat{\includegraphics[scale=0.45]{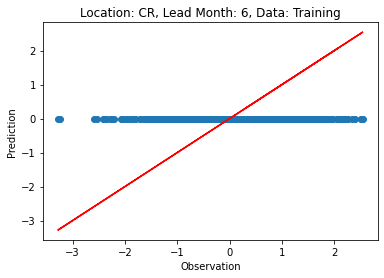}}
\subfloat{\includegraphics[scale=0.45]{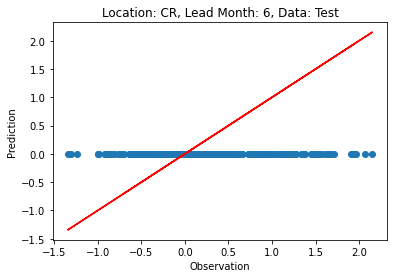}}\\
\subfloat{\includegraphics[scale=0.45]{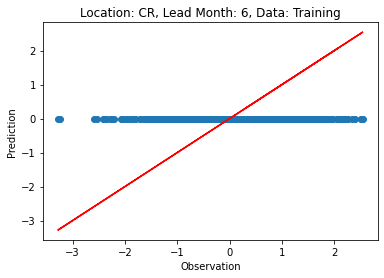}}
\subfloat{\includegraphics[scale=0.45]{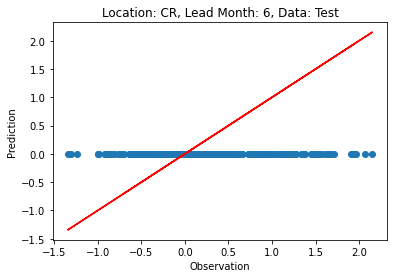}}
\caption{SSTA observation and prediction scatter plots of the last model for six lead month forecasts at Cape Reinga, on the training and test data respectively. The corresponding loss functions from top to bottom are the weighted MSE, the focal-R, and the balanced MSE.}
\end{figure}

\begin{figure}[H]
\centering
\subfloat{\includegraphics[scale=0.45]{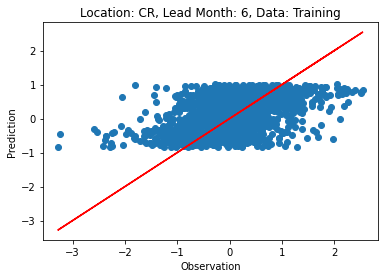}}
\subfloat{\includegraphics[scale=0.45]{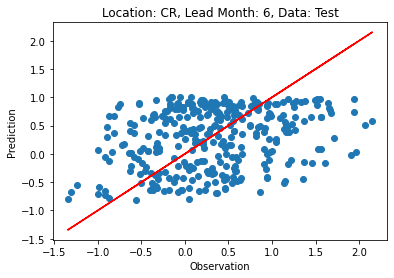}}\\
\subfloat{\includegraphics[scale=0.45]{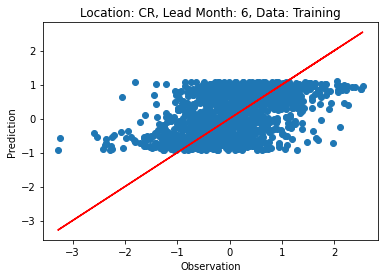}}
\subfloat{\includegraphics[scale=0.45]{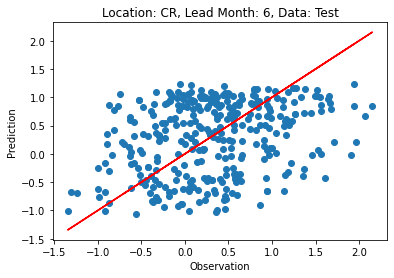}}\\
\subfloat{\includegraphics[scale=0.45]{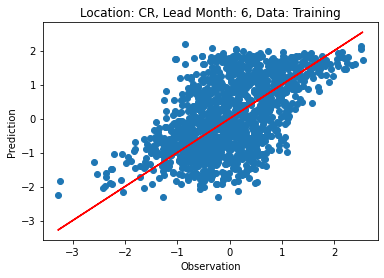}}
\subfloat{\includegraphics[scale=0.45]{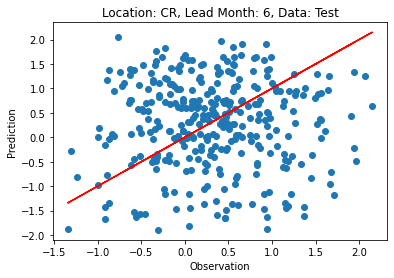}}
\caption{SSTA observation and prediction scatter plots of the last model for six lead month forecasts at Cape Reinga, on the training and test data respectively. The corresponding loss function is our proposed scaling-weighted MSE, from top to bottom with the hyperparameters $\alpha=1.5$, $\beta=0.5$, $\alpha=2$, $\beta=0.5$, and $\alpha=2$, $\beta=1$ respectively.}
\end{figure}

\begin{figure}[H]
\centering
\subfloat{\includegraphics[scale=0.45]{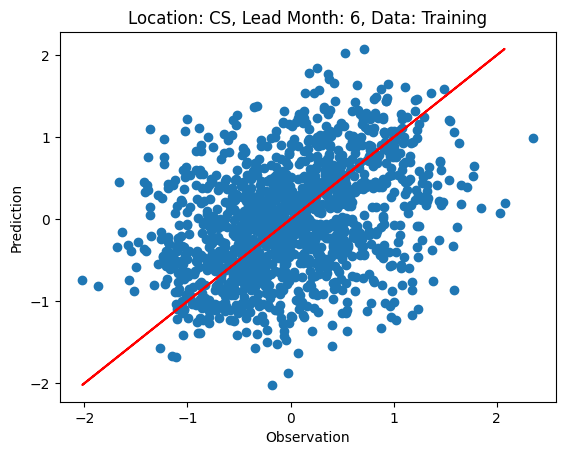}}
\subfloat{\includegraphics[scale=0.45]{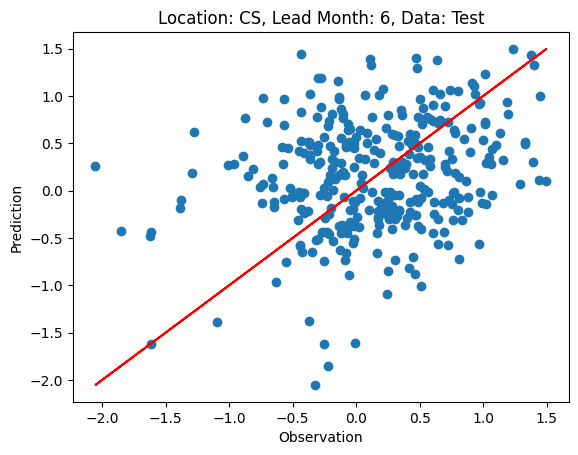}}
\caption{SSTA observation and prediction scatter plots of the six-lag model for six lead month forecasts at Cook Strait, on the training and test data respectively.}
\end{figure}

\begin{figure}[H]
\centering
\subfloat{\includegraphics[scale=0.45]{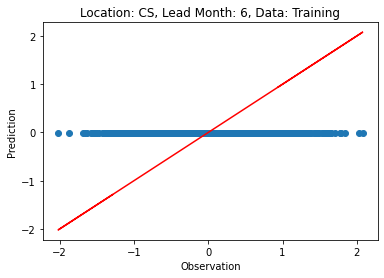}}
\subfloat{\includegraphics[scale=0.45]{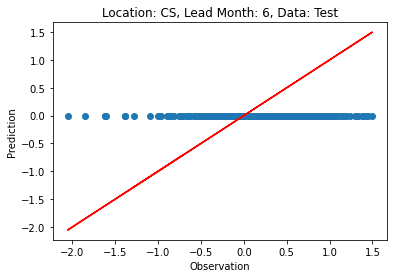}}\\
\subfloat{\includegraphics[scale=0.45]{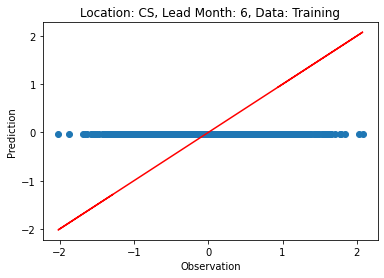}}
\subfloat{\includegraphics[scale=0.45]{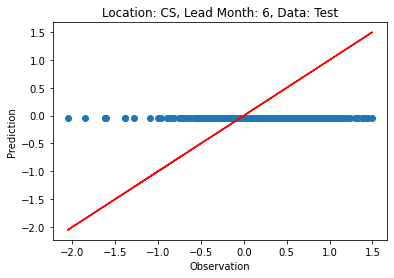}}\\
\subfloat{\includegraphics[scale=0.45]{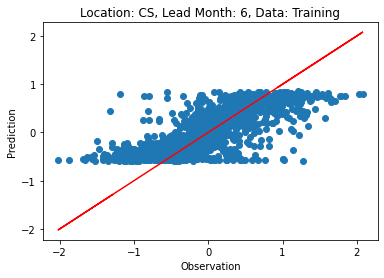}}
\subfloat{\includegraphics[scale=0.45]{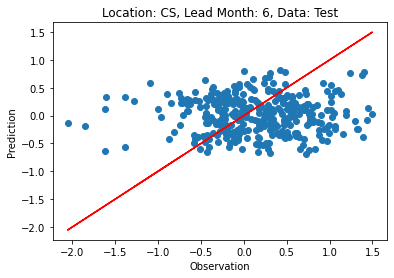}}
\caption{SSTA observation and prediction scatter plots of the last model for six lead month forecasts at Cook Strait, on the training and test data respectively. The corresponding loss functions from top to bottom are the MSE, the MAE, and the Huber.}
\end{figure}

\begin{figure}[H]
\centering
\subfloat{\includegraphics[scale=0.45]{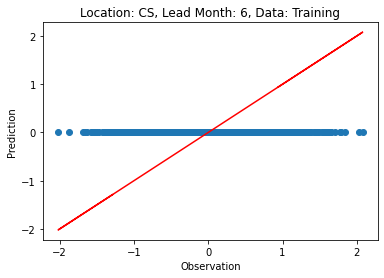}}
\subfloat{\includegraphics[scale=0.45]{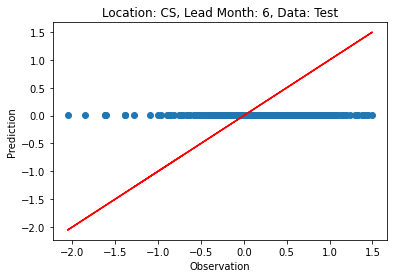}}\\
\subfloat{\includegraphics[scale=0.45]{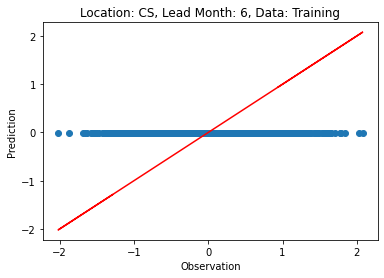}}
\subfloat{\includegraphics[scale=0.45]{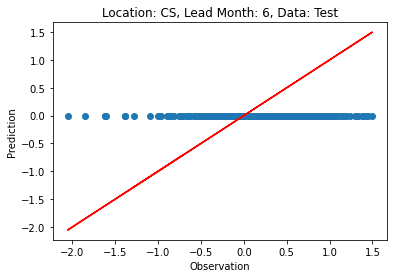}}\\
\subfloat{\includegraphics[scale=0.45]{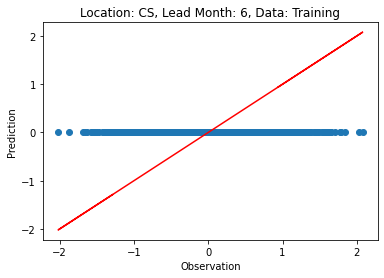}}
\subfloat{\includegraphics[scale=0.45]{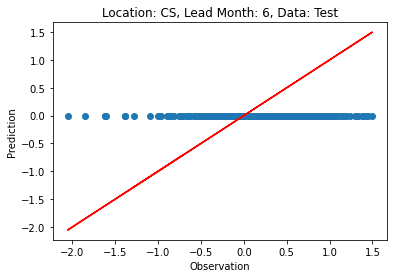}}
\caption{SSTA observation and prediction scatter plots of the last model for six lead month forecasts at Cook Strait, on the training and test data respectively. The corresponding loss functions from top to bottom are the weighted MSE, the focal-R, and the balanced MSE.}
\end{figure}

\begin{figure}[H]
\centering
\subfloat{\includegraphics[scale=0.45]{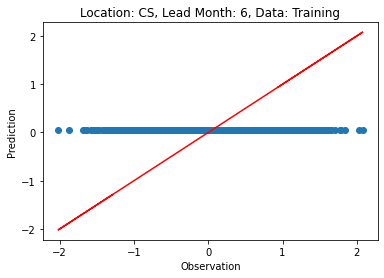}}
\subfloat{\includegraphics[scale=0.45]{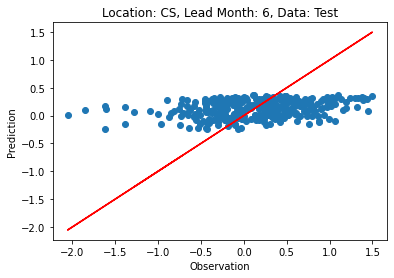}}\\
\subfloat{\includegraphics[scale=0.45]{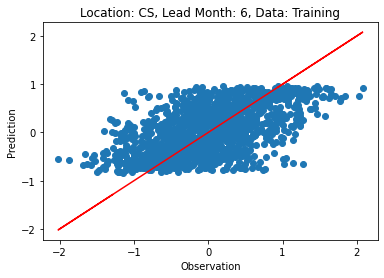}}
\subfloat{\includegraphics[scale=0.45]{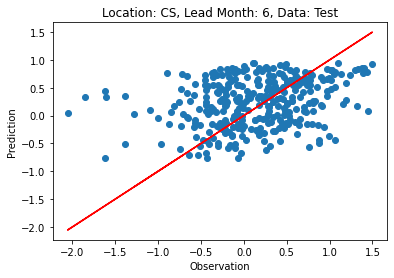}}\\
\subfloat{\includegraphics[scale=0.45]{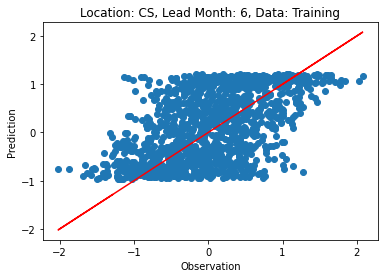}}
\subfloat{\includegraphics[scale=0.45]{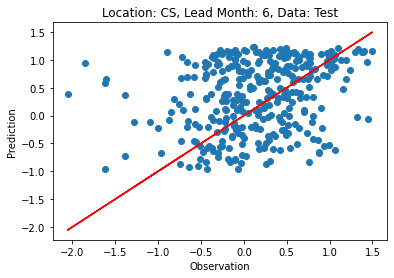}}
\caption{SSTA observation and prediction scatter plots of the last model for six lead month forecasts at Cook Strait, on the training and test data respectively. The corresponding loss function is our proposed scaling-weighted MSE, from top to bottom with the hyperparameters $\alpha=1.5$, $\beta=0.5$, $\alpha=2$, $\beta=0.5$, and $\alpha=2$, $\beta=1$ respectively.}
\end{figure}

\begin{figure}[H]
\centering
\subfloat{\includegraphics[scale=0.45]{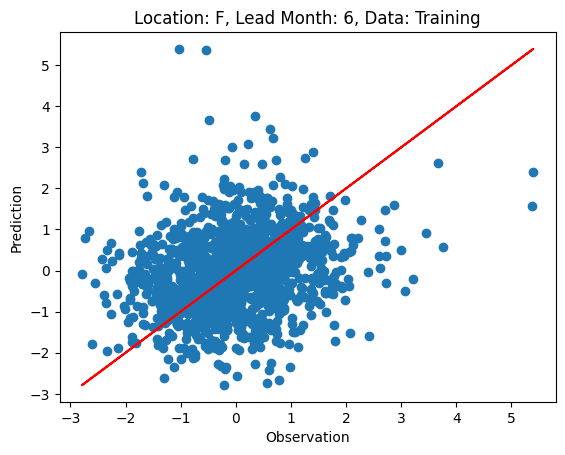}}
\subfloat{\includegraphics[scale=0.45]{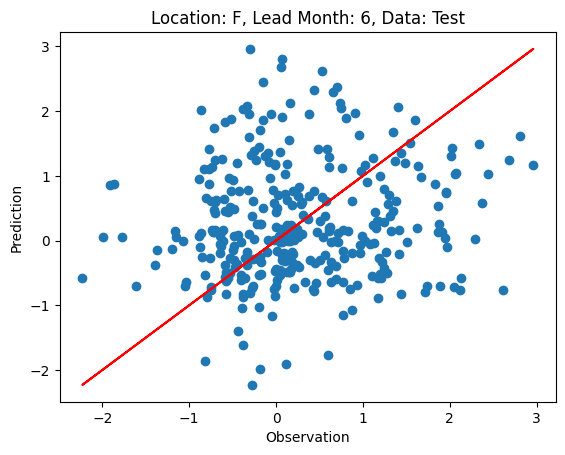}}
\caption{SSTA observation and prediction scatter plots of the six-lag model for six lead month forecasts at Fiordland, on the training and test data respectively.}
\end{figure}

\begin{figure}[H]
\centering
\subfloat{\includegraphics[scale=0.45]{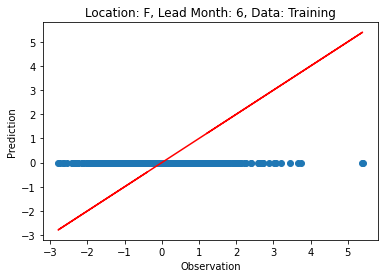}}
\subfloat{\includegraphics[scale=0.45]{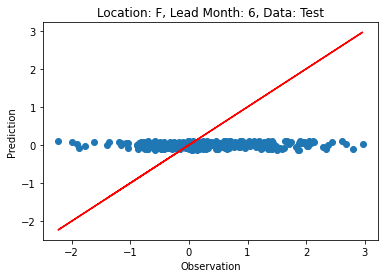}}\\
\subfloat{\includegraphics[scale=0.45]{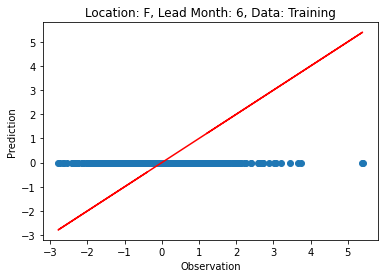}}
\subfloat{\includegraphics[scale=0.45]{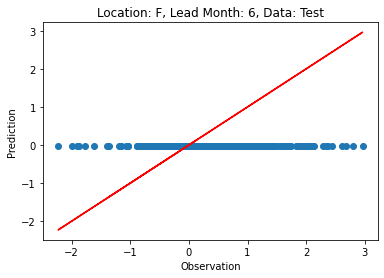}}\\
\subfloat{\includegraphics[scale=0.45]{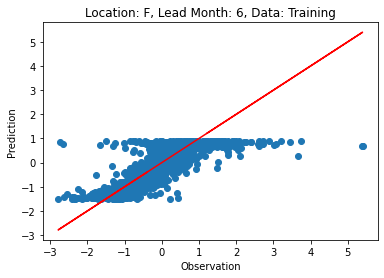}}
\subfloat{\includegraphics[scale=0.45]{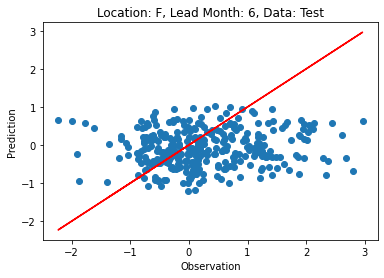}}
\caption{SSTA observation and prediction scatter plots of the last model for six lead month forecasts at Fiordland, on the training and test data respectively. The corresponding loss functions from top to bottom are the MSE, the MAE, and the Huber.}
\end{figure}

\begin{figure}[H]
\centering
\subfloat{\includegraphics[scale=0.45]{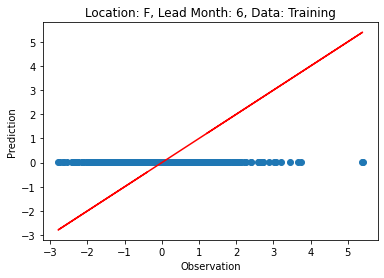}}
\subfloat{\includegraphics[scale=0.45]{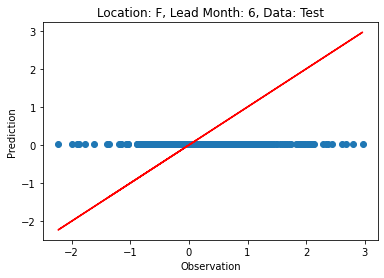}}\\
\subfloat{\includegraphics[scale=0.45]{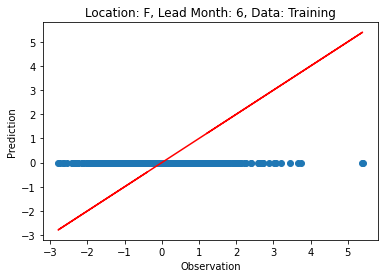}}
\subfloat{\includegraphics[scale=0.45]{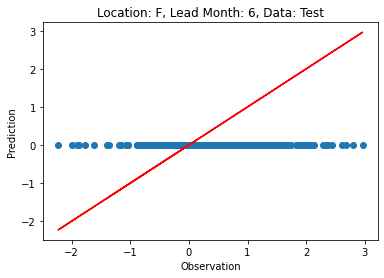}}\\
\subfloat{\includegraphics[scale=0.45]{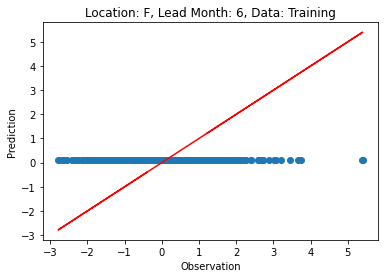}}
\subfloat{\includegraphics[scale=0.45]{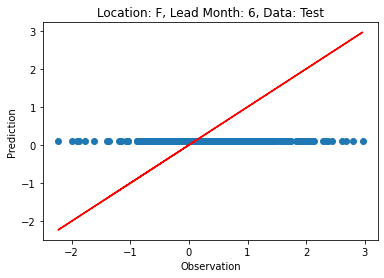}}
\caption{SSTA observation and prediction scatter plots of the last model for six lead month forecasts at Fiordland, on the training and test data respectively. The corresponding loss functions from top to bottom are the weighted MSE, the focal-R, and the balanced MSE.}
\end{figure}

\begin{figure}[H]
\centering
\subfloat{\includegraphics[scale=0.45]{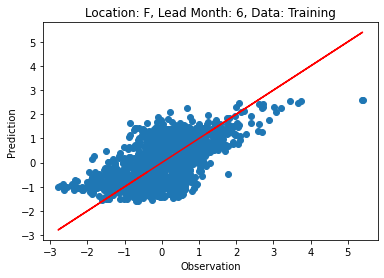}}
\subfloat{\includegraphics[scale=0.45]{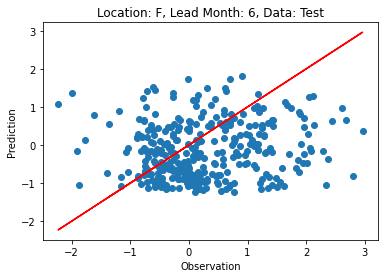}}\\
\subfloat{\includegraphics[scale=0.45]{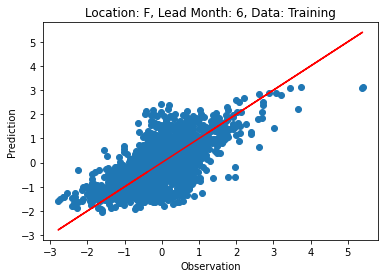}}
\subfloat{\includegraphics[scale=0.45]{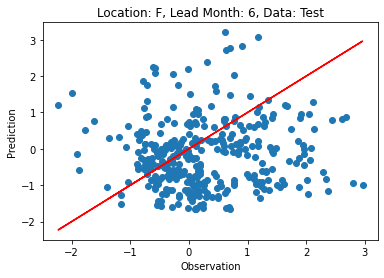}}\\
\subfloat{\includegraphics[scale=0.45]{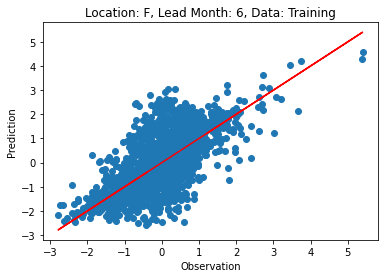}}
\subfloat{\includegraphics[scale=0.45]{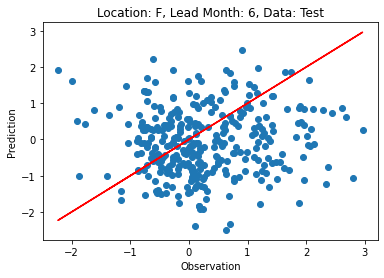}}
\caption{SSTA observation and prediction scatter plots of the last model for six lead month forecasts at Fiordland, on the training and test data respectively. The corresponding loss function is our proposed scaling-weighted MSE, from top to bottom with the hyperparameters $\alpha=1.5$, $\beta=0.5$, $\alpha=2$, $\beta=0.5$, and $\alpha=2$, $\beta=1$ respectively.}
\end{figure}

\subsection{Data Availability}\label{apd:data}

\def\UrlBreaks{\do\/\do-}
\sloppy

The preprocessed SSTA sequences are stored in the NumPy array files and accessible at \url{https://drive.google.com/drive/folders/1Nf2CyGK7LpgEgRpCBL0sGqcGcY3IbO4O?usp=sharing}.

\subsection{Code and Results Availability}\label{apd:code}

The code and results are presented in Jupyter Notebooks and accessible at \url{https://drive.google.com/drive/folders/1zOHx_BwZL45RnTWCMt3VNZ6bgMugLNQk?usp=sharing}.

\end{document}